\documentclass[twocolumn,10pt]{autart}

\usepackage{cite}
\usepackage{amsmath,amssymb,amsfonts,mathrsfs}
\usepackage{algorithm}
\usepackage{graphicx}
\usepackage{textcomp,enumitem}
\usepackage{array,booktabs}
\usepackage{float}
\usepackage{subfig}
\usepackage[justification=centering]{caption}
\usepackage{bm,bbm}
\usepackage{algpseudocode}
\usepackage{color}
\usepackage{threeparttable}
\usepackage{pifont}
\usepackage[round,authoryear]{natbib}
\usepackage{hyperref}
\hypersetup{colorlinks = true,citecolor = cyan}

\newtheorem{thm0}{Theorem}{}
\newtheorem{def0}{Definition}
\newtheorem{lemma}{Lemma}{}

\newtheorem{cor0}{Corollary}{}
\newtheorem{asm}{Assumption}
\newtheorem{rmk}{Remark}{}

\makeatletter
\def\@begintheorem#1#2{%
	\trivlist
	\item[\hskip \labelsep{\bfseries #1\ #2}]%
	\normalfont
}
\def\@opargbegintheorem#1#2#3{%
	\trivlist
	\item[\hskip \labelsep{\bfseries #1\ #2\ (#3)}]%
	\normalfont
}
\makeatother

\makeatletter
\newenvironment{breakablealgorithm}
{
	\begin{center}
		\refstepcounter{algorithm}
		\hrule height .8pt depth 0pt \kern 2pt
		\renewcommand{\caption}[2][\relax]{%
			{\raggedright\textbf{\ALG@name~\thealgorithm:} ##2\par}%
			\ifx\relax##1\relax
			\addcontentsline{loa}{algorithm}{\protect\numberline{\thealgorithm}##2}%
			\else
			\addcontentsline{loa}{algorithm}{\protect\numberline{\thealgorithm}##1}%
			\fi
			\kern 2pt\hrule\kern 2pt
		}
	}
	{
		\kern 2pt\hrule\relax
	\end{center}
}
\makeatother
\newcommand{\dg}{\mathcal{D}\hspace{-0.07em}\textit{G}(\sigma^2)}
\newcommand{\tdg}{\mathcal{D}\hspace{-0.07em}\textit{G}_\Gamma(\sigma^2)}
\newcommand{\hwt}{\mathcal{H}\hspace{-0.07em}\textit{WT}(h)}
\newcommand{\zo}{\mathcal{Z}\hspace{-0.07em}\textit{O}(\frac{1}{2})}

\newcommand{\p}{\mathbb{P}}
\newcommand{\E}{\mathbb{E}}
\newcommand{\iu}{\mathrm{i}}

\begin{document}
\begin{frontmatter}
\title{CKKS Cryptosystem-Based Secure Parameter Identification for Multi-Participant ARX Systems\thanksref{footnoteinfo}}
\thanks[footnoteinfo]{The work was supported by National Natural Science Foundation of China under Grants 62433020 and T2293770. The material in this paper was not presented at any conference. Corresponding author: Ji-Feng Zhang.}

\author[AMSS,UCAS]{Jialong Chen}\ead{chenjialong23@mails.ucas.ac.cn},  
\author[ZYUT,AMSS,UCAS]{Ji-Feng Zhang}\ead{jif@iss.ac.cn}, 
\author[USTB,USTBLAB]{Jimin Wang}\ead{jimwang@ustb.edu.cn}
\address[AMSS]{Key Laboratory of Systems and Control, Academy of Mathematics and Systems Science, Chinese Academy of Sciences, Beijing 100190, P. R. China.}
\address[UCAS]{School of Mathematical Sciences, University of Chinese Academy of Sciences, Beijing 100049, P. R. China.}
\address[ZYUT]{School of Automation and Electrical Engineering, Zhongyuan University of Technology, Zhengzhou 450007, P. R. China}
\address[USTB]{School of Automation and Electrical Engineering, University of Science and Technology Beijing, Beijing 100083, P. R. China.}
\address[USTBLAB]{Key Laboratory of Knowledge Automation for Industrial Processes, Ministry of Education, Beijing 100083, P. R. China.}

\begin{abstract}
This paper investigates parameter identification for multi-participant autoregressive systems with exogenous input (ARX systems) while protecting the system input and output. A novel Cheon-Kim-Kim-Song (CKKS) cryptosystem-based secure parameter identification algorithm is proposed. By combining a modified CKKS cryptosystem and a proxy re-encryption scheme, the algorithm enables the fusion center to perform homomorphic operations on the ciphertexts which are encrypted using different secret keys. A sufficient condition on the truncation value of the truncated discrete Gaussian noise is derived to ensure the indistinguishability under chosen-plaintext attack (IND-CPA) security of the algorithm under collusion and quantum attacks. For convergence analysis, an auxiliary plaintext sequence is constructed to characterize the encryption noise and quantization error in the encrypted estimate. Using this sequence, a criterion for avoiding plaintext overflow is given, based on which the mean square convergence and convergence rate of the algorithm are given. A numerical example demonstrates the effectiveness and superior performance of the algorithm. 
\end{abstract}
\end{frontmatter}

\section{Introduction}
Parameter identification for stochastic linear systems estimates the unknown parameters based on the system input and output \citep{ljung1999system}. As an important branch of stochastic linear system identification, parameter identification for autoregressive systems with exogenous input (ARX systems) focuses on estimating the unknown parameters in stochastic linear systems with auto-regressive outputs and exogenous inputs \citep{chen1990identification,chen1993identifiability,wang2022unified}, and has been extensively studied because of its wide applications \citep{deshpande2014optimized,griva2017commissioning}. Due to the limited computational capability of individual sensors \citep{akyildiz2002survey,lin2026contribution}, a fusion center is often required to process the system input and output collected from multiple sensors \citep{zhu2015parameter,sani2018distributed}. \vspace{-0.5em}

When sensors send the system input and output to the fusion center, the involved sensitive information may be leaked \citep{cock2015fast}. A common privacy-preserving approach is homomorphic encryption, which supports direct computation on ciphertexts without decryption \citep{zhang2021}. Homomorphic encryption has been successfully applied to feedback control \citep{murguia2020secure,teranishi2023designing,kim2023dynamic,feng2026asymptotic,tavazoei2026subtractionless}, distributed control \citep{wang2026privacy}, constrained quadratic optimization \citep{alexandru2021cloud}, distributed optimization \citep{lu2018privacy}, distributed consensus \citep{ruan2019secure,hadjicostis2020privacy}, discrete event system control \citep{fritz2026encrypted}, state estimation \citep{ristic2023distributed,fu2026privacy,zhang2025privacy}, and parameter identification \citep{tan2023cooperative}. In particular, \cite{tan2023cooperative} proposed a Paillier cryptosystem-based secure parameter identification algorithm. Since the Paillier cryptosystem only supports homomorphic addition, the multiplication is achieved using the pseudo-homomorphic multiplication property at the expense of increasing computational overhead. \vspace{-0.5em}

Ring learning with errors-based (RLWE-based) cryptosystems support both homomorphic addition and multiplication \citep{lyubashevsky2013ideal,lyubashevsky2013toolkit}, and then, have attracted much attention. Building on the hardness of the RLWE problem, the Cheon-Kim-Kim-Song (CKKS) cryptosystem provides the security against quantum attacks \citep{cheon2017homomorphic,cheon2018bootstrapping}, and has been applied to feedback control \citep{damera2026end,ikezaki2026encrypted} and parameter identification \citep{adamek2024encrypted}. Despite the advancements made in the above excellent works, research on ensuring the security against collusion and quantum attacks is relatively insufficient. Specifically, if sensors share the same secret key as in \cite{lu2018privacy,alexandru2021cloud,zhang2025privacy,hadjicostis2020privacy}, then a semi-honest sensor can collude with the fusion center to infer the sensitive information. Although using different secret keys for different sensors can ensure the security against collusion attacks, homomorphic operations cannot be performed on ciphertexts under different secret keys. The threshold encryption scheme is successfully employed in \cite{tan2023cooperative,ristic2023distributed} to address this issue. However, the Paillier cryptosystem therein is vulnerable to quantum attacks \citep{shor1999polynomial}. \vspace{-0.5em}

Another challenge arises in providing an indistinguishability under chosen-plaintext attack (IND-CPA) security proof of the CKKS cryptosystem with the truncated discrete Gaussian noise. Since generating a discrete Gaussian noise is infeasible in practice \citep{dwarakanath2014sampling}, the truncated discrete Gaussian noise is commonly used \citep{chen2017simple,albrecht2022homomorphic,dwarakanath2014sampling}. However, since the truncated discrete Gaussian noise changes the distributions of the public key and the ciphertexts, the IND-CPA security proof of the standard CKKS cryptosystem cannot be applied to the methods in \cite{adamek2024encrypted,damera2026end,ikezaki2026encrypted}. Although \cite{brakerski2014leveled} insightfully points out that the security level is not reduced, a rigorous proof is not provided. \vspace{-0.5em}

It is also non-trivial to provide a convergence analysis for CKKS cryptosystem-based secure parameter identification. \cite{adamek2024encrypted} first proposed a CKKS cryptosystem-based secure parameter identification algorithm, but did not provide a convergence analysis. Similarly, existing homomorphic encryption-based methods either do not provide a convergence analysis \citep{lu2018privacy,alexandru2021cloud} or provide one while neglecting the encryption noise and quantization error \citep{tan2023cooperative,ruan2019secure}. These errors are elegantly circumvented in \cite{hadjicostis2020privacy} by assuming rational-valued data. However, this assumption is hard to satisfy for CKKS cryptosystem-based secure parameter identification since approximate homomorphic operations introduce the encryption noise and quantization error. As these errors accumulate, the plaintext may exceed the range of the plaintext space, leading to plaintext overflow. In this case, the convergence of the algorithm may not be guaranteed. \vspace{-0.5em}

To address the above challenges, a novel CKKS cryptosystem-based secure parameter identification algorithm is proposed for multi-participant ARX systems. The main contribution is as follows: \vspace{-0.5em}

\begin{itemize}[leftmargin=*]
	\item The security of the algorithm under collusion and quantum attacks is ensured. A modified CKKS cryptosystem is given by replacing the discrete Gaussian noise with the truncated one. By combining the modified CKKS cryptosystem and a proxy re-encryption scheme, the security of the algorithm against collusion and quantum attacks is ensured. Compared to \cite{teranishi2023designing,feng2026asymptotic,kim2023dynamic,fritz2026encrypted,alexandru2021cloud,lu2018privacy,zhang2025privacy,fu2026privacy,ruan2019secure,murguia2020secure,tavazoei2026subtractionless,hadjicostis2020privacy,ristic2023distributed,adamek2024encrypted,tan2023cooperative,ikezaki2026encrypted,damera2026end}, a stronger adversary model is considered, and thus, the security level of the algorithm is enhanced.
	
	\item An IND-CPA security proof of the algorithm is given. The exponentially decreasing tail of the discrete Gaussian distribution is used to derive a sufficient condition on the truncation value. Based on this condition, an IND-CPA security proof of the algorithm is given. This shows advantage over \cite{adamek2024encrypted,kim2023dynamic,hadjicostis2020privacy,feng2026asymptotic,fritz2026encrypted,murguia2020secure,tavazoei2026subtractionless,ikezaki2026encrypted,damera2026end} since an IND-CPA security proof is not given therein. Compared to \cite{brakerski2014leveled}, an explicit sufficient condition on the truncation value is given.
	
	\item A convergence analysis of the algorithm is provided. An auxiliary plaintext sequence is constructed to characterize the encryption noise and quantization error in the encrypted estimate. By using this sequence, a criterion for avoiding plaintext overflow is given, based on which the mean square convergence and convergence rate of the algorithm are provided. Compared to \cite{lu2018privacy,alexandru2021cloud,adamek2024encrypted,tan2023cooperative,ruan2019secure}, the mean square convergence and convergence rate of the algorithm are provided.
\end{itemize}\vspace{-0.5em}

The remainder of this paper is organized as follows. Section 2 provides preliminaries and the problem formulation. Section 3 provides the proposed algorithm with the security and the convergence analysis. Section~4 presents a numerical example, and Section 5 concludes the paper. \vspace{-0.5em}

{\it Notations.} $\mathbb{C},\mathbb{R},\mathbb{Z},\mathbb{N}_+$ denote the sets of complex numbers, real numbers, integers, and positive integers, respectively. For a complex number $z\in\mathbb{C}$, $\bar{z}$ denotes its complex conjugate. $\mathbbm{1}_{\{\cdot\}}$ denotes the indicator function, whose value is 1 if its argument is true, and 0, otherwise. For $u\in\mathbb{R}$, $\lfloor u \rfloor$ denotes the largest integer no larger than $u$. $\mathbf{1}_N$ denotes an $N$-dimensional vector whose elements are all one, and $I_N$ represents the $N\times N$ identity matrix. For a vector $x\in\mathbb{R}^N$, $\|x\|_1,\|x\|,\|x\|_\infty$ denote the $1$-norm, $2$-norm and infinity norm of the vector $x$. For a positive integer $\mathfrak{q}$, we define $\mathbb{Z}_\mathfrak{q}=\mathbb{Z}\cap[-\frac{\mathfrak{q}}{2},\frac{\mathfrak{q}}{2})$ as the ring of integers modulo $\mathfrak{q}$. $\textsf{poly}(\lambda)$ denotes the set of polynomials of $\lambda$, and $\textsf{negl}(\lambda)$ denotes the set of negligible functions of $\lambda$, i.e., for any $f(\lambda)\in\textsf{negl}(\lambda)$ and $c>0$, there exists $\lambda_0\in\mathbb{N}_+$ such that for any $\lambda=\lambda_0+1,\lambda_0+2,\dots$, $|f(\lambda)|<\lambda^{-c}$.

\section{Preliminaries and problem formulation}
\subsection{Preliminaries on CKKS cryptosystem}
In this subsection, we first review the standard CKKS cryptosystem:
\begin{itemize}[leftmargin=*]
	\item $\textsf{KeyGen}$. Let $\lambda$$\in$$\mathbb{N}_+$ be the security parameter, $M$ be a power-of-two positive integer such that $M$$\in$$\textsf{poly}(\lambda)$, $M$$\geq$$8$, $N=\frac{M}{2}$ be the ring dimension, $\Phi_M(x)$ be the $M$-th cyclotomic polynomial, and $R=\mathbb{Z}[x]/(\Phi_M(x))$ be the cyclotomic ring. Then, $\textsf{e}\sim\dg$ denotes a discrete Gaussian noise over $R$ whose coefficients $\textsf{e}^{(1)},\dots,\textsf{e}^{(N)}$ are mutually independent such that
	\begin{align*}
		&\p(\textsf{e}^{(l)}=m)=\frac{\exp(-\frac{m^2}{2\sigma^2})}{\sum_{r\in\mathbb{Z}}\exp(-\frac{r^2}{2\sigma^2})},\cr
		&\forall\ m\in\mathbb{Z},\ l=1,\dots,N .
	\end{align*}
	Let	$\mathfrak{q}$ be a modulus, $h$ be a positive integer. Then, $R_{\mathfrak{q}}=R/(\mathfrak{q}R)$ is a quotient ring, $\mathcal{U}$ denotes the uniform distributions over $R_{\mathfrak{q}}$, and $\textsf{s}\sim\hwt$ denotes a random variable over $R_{\mathfrak{q}}$ whose coefficient vector is uniformly sampled from $\{\mathtt{s}$$\in$$\{-1,0,1\}^N$$:$$\|\mathtt{s}\|^2$$=$$h\}$. 
	
	\noindent Let $\textsf{s}\sim\hwt$, $\textsf{e},\textsf{e}^{\prime\prime}\sim\dg$, $\textsf{a},\textsf{a}^{\prime\prime}\sim\mathcal{U}$. Then, the secret key is $\textsf{sk}=(1,\textsf{s})$, the public key is $\textsf{pk}=(\textsf{b},\textsf{a})$, and the evaluation key is $\textsf{evk}=(\textsf{b}^{\prime\prime},\textsf{a}^{\prime\prime})$, where
	\begin{align*}
		\textsf{b}=-\textsf{a}\textsf{s}+\textsf{e}\bmod \mathfrak{q},\quad	\textsf{b}^{\prime\prime}=-\textsf{a}^{\prime\prime}\textsf{s}+\textsf{e}^{\prime\prime}+\mathfrak{p}\textsf{s}^2\bmod \mathfrak{p}\mathfrak{q}.
	\end{align*}
	
	\item $\textsf{Ecd}$. Let $\Delta$$>$$0$ be the scaling factor, $\textsf{CRT}$ be the canonical embedding that maps $\textsf{a}$ to $\textsf{CRT}(\textsf{a})$$=$ $(\textsf{a}(\exp(\frac{2\pi \iu}{M})),\dots, \textsf{a}(\exp(\frac{2(2N-1)\pi \iu}{M})))^\top$, $\Upsilon$ be the natural projection that maps $z$$\in$$\{(z^{(1)},\dots,z^{(N)})$$\in$$\mathbb{C}^N|$ $z^{(l)}=\overline{z^{(N-l+1)}}, l=1,\dots,N\}$ to $(z^{(1)},\dots,z^{(\frac{N}{2})})$$\in$$\mathbb{C}^{\frac{N}{2}}$, and $Q$ be the probabilistic quantizer that maps a vector $x$$=$$(x^{(1)},\dots,x^{(N)})$$\in$$\mathbb{R}^N$ to $Q(x)$$=$$(Q(x^{(1)}),\dots$, $Q(x^{(N)}))^\top$, where 
	\begin{align}\label{pq}
		&\begin{cases}
			\p(Q(x^{(l)})\!=\!\lfloor x^{(l)} \rfloor|x^{(l)})\!=\!1\!+\!\lfloor x^{(l)}\rfloor\!-\!x^{(l)};\\
			\p(Q(x^{(l)})\!=\!\lfloor x^{(l)} \rfloor\!+\!1|x^{(l)})\!=\!x^{(l)}\!-\!\lfloor x^{(l)}\rfloor,
		\end{cases}\cr
		&~~\forall\ l=1,\dots,N.
	\end{align}Then, a complex vector $z\in\mathbb{C}^{\frac{N}{2}}$ is encoded as  $\textsf{Ecd}_\Delta(z)$$=$$\textsf{CRT}^{-1}(Q(\Delta\cdot\Upsilon^{-1}(z))) \bmod \mathfrak{q}$.
	
	\item $\textsf{Dcd}$. A plaintext $\textsf{pt}\in R_\mathfrak{q}$ is decoded as  $\textsf{Dcd}_\Delta(\textsf{pt})$$=$ $\Upsilon(\textsf{CRT}(\Delta^{-1}\cdot\textsf{pt}))$.
	
	\item $\textsf{Enc}$. Let $\textsf{e}_1,\textsf{e}_2$$\sim$$\dg$, and $\textsf{v}\sim\zo$ be a random variable over $R_\mathfrak{q}$ whose coefficients $\textsf{v}^{(1)},\dots,\textsf{v}^{(N)}$ are mutually independent such that 
	\begin{align*}
		&\p(\textsf{v}^{(l)}=-1)=\p(\textsf{v}^{(l)}=1)=\frac{1}{4},\p(\textsf{v}^{(l)}=0)=\frac{1}{2},\cr 
		\noalign{\vskip -4pt}
		&\forall\ l=1,\dots,N.
	\end{align*}Then, a plaintext $\textsf{pt}$ is encrypted as $\textsf{Enc}_{\textsf{pk}}(\textsf{pt})=\textsf{v}\cdot\textsf{pk}+(\textsf{pt}+\textsf{e}_1,\textsf{e}_2)\bmod \mathfrak{q}$.
	
	\item $\textsf{Dec}$. A ciphertext $\textsf{ct}=(\textsf{ct}^{(1)},\textsf{ct}^{(2)})$ is decrypted as $\textsf{Dec}_{\textsf{sk}}(\textsf{ct})=\langle\textsf{ct},\textsf{sk}\rangle=\textsf{ct}^{(1)}+\textsf{ct}^{(2)} \textsf{s}\bmod \mathfrak{q}$.
\end{itemize}

The standard CKKS cryptosystem achieves the following IND-CPA security:
\begin{def0}\label{def1}
	\citep{katz2021introduction} If for any pair of plaintexts $(\textsf{pt}_0,\textsf{pt}_1)$ generated by an adversary $\mathcal{A}$ and any probabilistic polynomial-time (quantum) algorithm $\mathscr{B}$, the following holds:
	\begin{align}\label{ind-cpa}
		\smash{|\p(\mathscr{B}(\textsf{pt}_0,\textsf{pt}_1,\textsf{Enc}_{\textsf{pk}}(\textsf{pt}_b))=b)-\frac{1}{2}|\in\textsf{negl}(\lambda),}
	\end{align}
	then the cryptosystem $(\textsf{KeyGen},\textsf{Enc},\textsf{Dec})$ is said to achieve the IND-CPA security.
\end{def0}
\begin{rmk}
	Definition \ref{def1} is standard and commonly used in cryptography  \citep{cheon2017homomorphic,brakerski2014leveled,lyubashevsky2013toolkit}. The algorithm $\mathscr{B}$ in \eqref{ind-cpa} is used to infer the bit $b$ from the plaintexts $(\textsf{pt}_0,\textsf{pt}_1)$ and the ciphertext $\textsf{Enc}_{\textsf{pk}}(\textsf{pt}_b)$. If a cryptosystem $(\textsf{KeyGen},\textsf{Enc},\textsf{Dec})$ achieves the IND-CPA security, then the adversary $\mathcal{A}$ cannot obtain any information about the plaintexts from the ciphertexts \citep{teranishi2023designing}. 
\end{rmk}

The standard CKKS cryptosystem supports the following homomorphic operations on the ciphertexts  $\textsf{ct}_1=\textsf{Enc}_{\textsf{pk}}(\textsf{pt}_1)$, $\textsf{ct}_2=\textsf{Enc}_{\textsf{pk}}(\textsf{pt}_2)$:
\begin{itemize}[align=left,leftmargin=*]
	\item[H1 (Homomorphic addition):] There exists $\textsf{Add}$ such that $\textsf{Add}(\textsf{ct}_1,\textsf{ct}_2)=\textsf{Enc}_{\textsf{pk}}(\textsf{pt}_1+\textsf{pt}_2)\bmod \mathfrak{q}$.
	\item[H2 (Homomorphic multiplication):] There exists $\textsf{Mult}$ such that $\textsf{Mult}_{\textsf{evk}}(\textsf{ct}_1,\textsf{ct}_2)$$=$$\textsf{Enc}_{\textsf{pk}}(\textsf{pt}_1 \textsf{pt}_2)\bmod \mathfrak{q}$.
	\item[H3 (Homomorphic inner product):] For $\textsf{pt}_1=\textsf{Ecd}_\Delta(z_1)$ and $\textsf{pt}_2=\textsf{Ecd}_\Delta(z_2)$, there exists $\textsf{Dot}$ such that $\textsf{Dot}(\textsf{ct}_1,\textsf{ct}_2)=\textsf{Enc}_{\textsf{pk}}(\textsf{Ecd}_\Delta(z_1^\top z_2\cdot\mathbf{1}_{\frac{N}{2}}))\bmod \mathfrak{q}$.
	\item[H4 (Homomorphic rotation):] Let $\textsf{rot}(z_1;l)=(z_1^{(l+1)},\dots$, $z_1^{(\frac{N}{2})},z_1^{(1)},\dots,z_1^{(l)})$ be the rotation mapping for any $l=1,\dots,\frac{N}{2}-1$. Then, there exists $\textsf{Rot}$ such that $\textsf{Rot}(\textsf{ct}_1;l)=\textsf{Enc}_{\textsf{pk}}(\textsf{Ecd}_\Delta(\textsf{rot}(z_1;l)))\bmod \mathfrak{q}$.
\end{itemize} 

\subsection{Multi-participant ARX system}\label{section 2.2}
Consider the following parameter identification problem of a multi-participant ARX system:
\begin{align}\label{eq arx1}
	y_{i,k+1}=&\ a_1y_{i,k}+a_2y_{i,k-1}+\dots+a_py_{i,k-p+1}\cr
	&+b_1u_{i,k}+b_2u_{i,k-1}+\dots+b_qu_{i,k-q+1}\cr
	&+w_{i,k+1},\ i=1,\dots,n,\ k=0,\dots,K,
\end{align}
where $p,q\in\mathbb{N}_+$ are known system orders, $a_1,\dots,a_p,b_1$, $\dots,b_q\in\mathbb{R}$ are unknown parameters to be estimated, $u_{i,k},y_{i,k}\in\mathbb{R}$ are the system input and output measured by Sensor $i$, respectively, and $w_{i,k}\in\mathbb{R}$ is the system noise. Without loss of generality, we assume $y_{i,k}=0$ and $u_{i,k}=0$ for any $i=1,\dots,n$, $k=-1,\dots,-\max\{p,q\}$. For compact expression, let $\theta=(a_1, \dots , a_p, b_1,\dots,b_q)^\top$ and $\varphi_{i,k}=(y_{i,k},y_{i,k-1},\dots$, $y_{i,k-p+1},u_{i,k},u_{i,k-1},\dots,u_{i,k-q+1})^\top$. Then, the multi-participant ARX system \eqref{eq arx1} can be rewritten as
\begin{align}\label{eq arx2}
	y_{i,k+1}=\varphi_{i,k}^\top\theta+w_{i,k+1},i=1,\dots,n, k=0,\dots,K.
\end{align}

The problem architecture is that Sensors $1,\dots,n$ send the system input $\{u_{i,k}|i=1,\dots,n,k=0,\dots,K\}$ and output $\{y_{i,k}|i=1,\dots,n,k=0,\dots,K+1\}$ to the fusion center to estimate the unknown parameter $\theta$ in \eqref{eq arx2} while protecting the sensitive information $\{u_{i,k}|i=1,\dots,n,k=0,\dots,K\}$ and $\{y_{i,k}|i=1,\dots,n,k=0,\dots,K+1\}$. If the system input and output are directly sent to the fusion center, then the sensitive information may be inferred by the adversary $\mathcal{A}$ through the following attacks:
\begin{itemize}[leftmargin=*]
	\item \emph{Eavesdropping} on all communication channels between the sensors and the fusion center.
	\item \emph{Colluding} with the fusion center and the sensors in $\{1,\dots,n\}\setminus\{i_0\}$ to infer the sensitive information of an honest sensor $i_0$.
	\item Executes probabilistic polynomial-time \emph{quantum} algorithms to infer the sensitive information of Sensor~$i_0$.
\end{itemize}
\begin{rmk}\label{rmk: attack model}
	The attacks considered in this paper are stronger than those in the existing related works. Specifically, \cite{teranishi2023designing,feng2026asymptotic,fritz2026encrypted,lu2018privacy,zhang2025privacy,wang2026privacy,murguia2020secure,tavazoei2026subtractionless} only consider eavesdropping attacks. \cite{alexandru2021cloud,ruan2019secure,hadjicostis2020privacy,ristic2023distributed,tan2023cooperative,fu2026privacy} only consider eavesdropping and collusion attacks. \cite{kim2023dynamic,adamek2024encrypted,ikezaki2026encrypted,damera2026end} only consider eavesdropping and quantum attacks.
\end{rmk}

\begin{figure}[!htbp]
	\centering
	\includegraphics[width=0.42\textwidth]{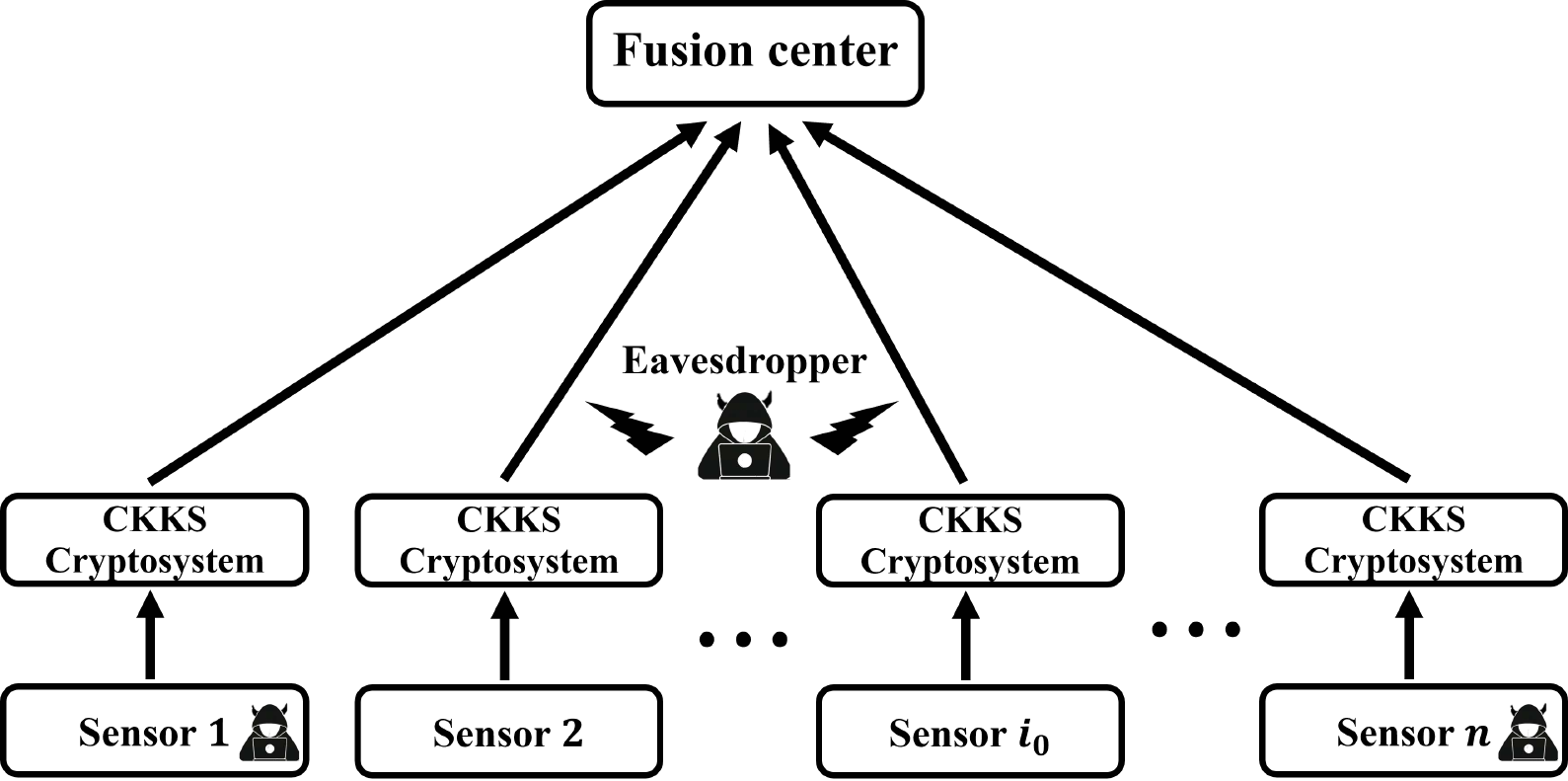}
	\caption{Problem architecture}
	\label{fig1}
\end{figure}

To protect the sensitive information against the adversary $\mathcal{A}$, different sensors use the CKKS cryptosystem to encrypt the system input and output with different secret keys before transmission. The problem architecture is given in Fig. \ref{fig1}.

The goal of this paper is to 1) propose a CKKS cryptosystem-based secure parameter identification algorithm, 2) give an IND-CPA security proof of the algorithm under collusion and quantum attacks, and 3) provide a convergence analysis of the algorithm.

\section{Main results}
\subsection{Algorithm design}
In this subsection, we will give the CKKS cryptosystem-based secure parameter identification algorithm. We first give the following modified CKKS cryptosystem by replacing the discrete Gaussian noise with the truncated one:
\begin{itemize}[leftmargin=*]
	\item $\textsf{KeyGen}$. Let $K\in\mathbb{N}_+$ be the maximum iteration number, $\Delta_K>0$ be the scaling factor, $\mathfrak{q}_{1,K},\mathfrak{q}_{2,K}\in\mathbb{N}_+$ the modulus, $d\in\mathbb{N}_+$ be the depth, $\mathfrak{q}_K=\mathfrak{q}_{1,K}\mathfrak{q}_{2,K}^d$, and $\mathfrak{p}_K$ be the auxiliary modulus. Then, $\textsf{e}^\dagger\sim\tdg$ denotes the truncated discrete Gaussian distribution with a truncation value $\Gamma>0$ over $R$, whose coefficients $\textsf{e}^{(1)\dagger},\dots,\textsf{e}^{(N)\dagger}$ are independently random variables such that\vspace{-1em}
	\begin{align*}
		&\p(\textsf{e}^{(l)\dagger}=m)=\frac{\mathbbm{1}_{\{|m|\leq \Gamma\}}\exp(-\frac{m^2}{2\sigma^2})}{\sum_{r\in\mathbb{Z},\,|r|\leq \Gamma}\exp(-\frac{r^2}{2\sigma^2})},\cr
		&\forall\ m\in\mathbb{Z},\ l=1,\dots,N .
	\end{align*}
	
	\noindent For any $i=1,\dots,n$, let $\textsf{s}_i\sim\hwt$, $\textsf{e}_i,\textsf{e}_i^{\prime\prime}\sim\tdg$, $\textsf{a}_i,\textsf{a}_i^{\prime\prime}\sim\mathcal{U}$. Then, Sensor $i$ generates the secret key 
	$\textsf{sk}_i=(1,\textsf{s}_i)$, the public key 
	$\textsf{pk}_i=(\textsf{b}_i,\textsf{a}_i)$, and the evaluation key 
	$\textsf{evk}_i=(\textsf{b}_i^{\prime\prime},\textsf{a}_i^{\prime\prime})$, where
	\begin{align*}
		\textsf{b}_i=&-\textsf{a}_i\textsf{s}_i+\textsf{e}_i\bmod \mathfrak{q}_K,\cr
		\textsf{b}_i^{\prime\prime}=&-\textsf{a}_i^{\prime\prime}\textsf{s}_i+\textsf{e}_i^{\prime\prime}+\mathfrak{p}_K\textsf{s}_i^2\bmod \mathfrak{p}_K\mathfrak{q}_K .
	\end{align*}
	
	\item $\textsf{Enc}$. For any $i=1,\dots,n$ and $k=0,\dots,K$, let $\textsf{v}_{i,k}\sim\zo$ and $\textsf{e}_{1,i,k},\textsf{e}_{2,i,k}$$\sim$ $\tdg$. Then, Sensor $i$ encrypts a plaintext $\textsf{pt}_{i,k}$ as $\textsf{Enc}_{\textsf{pk}_i}(\textsf{pt}_{i,k})=\textsf{v}_{i,k}\textsf{pk}_i+(\textsf{pt}_{i,k}+\textsf{e}_{1,i,k},\textsf{e}_{2,i,k})\bmod \mathfrak{q}_K$.
\end{itemize}

Since only the noise distribution is changed, the modified CKKS cryptosystem also supports the homomorphic operations H1-H4. However, the homomorphic operations H1-H4 cannot be performed on the ciphertexts under different secret keys $\textsf{sk}_i$. Then, the following proxy re-encryption scheme is given to address this issue:
\begin{itemize}[leftmargin=*]
	\item $\textsf{ReKeyGen}$. Let $\textsf{a}^\prime\sim \mathcal{U},\textsf{e}^\prime\sim\tdg$. Then, Sensor~$i_0$ sends $(\textsf{a}^\prime,\textsf{b}^\prime)$ to the fusion center, where $\textsf{b}^\prime=-\textsf{a}^\prime\textsf{s}_{i_0}+\textsf{e}^\prime\bmod \mathfrak{q}_K$. After that, the fusion center broadcasts $(\textsf{a}^\prime,\textsf{b}^\prime)$ to each Sensor $i\neq i_0$.
	
	\noindent For any $i\neq i_0$, let $\textsf{e}_{1,i\rightarrow i_0},\textsf{e}_{2,i\rightarrow i_0}$$\sim$$\tdg$, $\textsf{v}_{i\rightarrow i_0}$$\sim$ $\zo$. Then, Sensor $i$ generates the re-encryption key $\textsf{rk}_{i\rightarrow i_0}$ $=$$(\textsf{b}_{i\rightarrow i_0},\textsf{a}_{i\rightarrow i_0})$, where 
	\begin{align*}
		&\textsf{b}_{i\rightarrow i_0}=\textsf{v}_{i\rightarrow i_0}\textsf{b}^\prime+\mathfrak{p}_K \textsf{s}_i+\textsf{e}_{1,i\rightarrow i_0}\bmod \mathfrak{p}_K \mathfrak{q}_K,\cr &\textsf{a}_{i\rightarrow i_0}=\textsf{v}_{i\rightarrow i_0}\textsf{a}^\prime+\textsf{e}_{2,i\rightarrow i_0}\bmod\mathfrak{p}_K \mathfrak{q}_K.  
	\end{align*}
	Sensor $i$ sends the re-encryption key $\textsf{rk}_{i\rightarrow i_0}$ to the fusion center.
	
	\item $\textsf{ReEnc}$. For any $i\neq i_0$ and $k=0,\dots,K$, the fusion center re-encrypts a ciphertext $\textsf{ct}_{i,k}$$=$$(\textsf{ct}_{i,k}^{(1)},\textsf{ct}_{i,k}^{(2)})$ as $\textsf{ReEnc}_{\textsf{rk}_{i\rightarrow i_0}}(\textsf{ct}_{i,k})=(\textsf{ct}_{i,k}^{(1)},0)+Q(\mathfrak{p}_K^{-1}\textsf{ct}_{i,k}^{(2)}\cdot\textsf{rk}_{i\rightarrow i_0})$ $\bmod \mathfrak{q}_K$.
\end{itemize}

Combining the modified CKKS cryptosystem and the proxy re-encryption scheme, the CKKS-cryptosystem-based secure parameter identification algorithm is given as follows:\vspace{0.5em}
\begin{breakablealgorithm}
	\vspace{-1em}
	\caption{CKKS-cryptosystem-based secure parameter identification algorithm}
	\label{ckks pia}
	\renewcommand{\algorithmicensure}{\textbf{Setup:}}
	\begin{algorithmic}[1]\vspace{-0.2em}
		\Ensure The initial estimate $\hat{\theta}_0$ and the step-size $\alpha_K$.\vspace{0.2em}
		
		\State The fusion center encodes $\hat{\theta}_0$ as $\textsf{pt}_0^{\hat{\theta}}=\textsf{Ecd}_{\Delta_K}(\hat{\theta}_0)$, and sets $\textsf{ct}_0^{\hat{\theta}}=(\textsf{pt}_0^{\hat{\theta}},0)$.
		
		\hspace{-3.7em}{\bf for} $i=1,\dots,n$, {\bf do}
		
		\State Sensor $i$ uses the key generation oracle $\textsf{KeyGen}$ to generate the secret key $\textsf{sk}_i$, the public key $\textsf{pk}_i$, and the evaluation key $\textsf{evk}_i$. 
		
		\State Sensor $i$ uses the encoding oracle $\textsf{Ecd}_{\Delta_K}$ to encode $\{\varphi_{i,0}$, $\dots,\varphi_{i,K}\}$ and $\{y_{i,0},\dots,y_{i,K+1}\}$ as $\{\textsf{pt}_{i,0}^\varphi,\dots,\textsf{pt}_{i,K}^\varphi\}$ and $\{\textsf{pt}_{i,0}^y,\dots,\textsf{pt}_{i,K\!+\!1}^y\}$, respectively. 
		
		\State Sensor $i$ uses the encryption oracle $\textsf{Enc}$ to encrypt $\{\textsf{pt}_{i,0}^\varphi,\dots,\textsf{pt}_{i,K}^\varphi\}$ and $\{\textsf{pt}_{i,0}^y,\dots,\textsf{pt}_{i,K\!+\!1}^y\}$ as $\{\textsf{ct}_{i,0}^\varphi,\dots,\textsf{ct}_{i,K}^\varphi\}$ and $\{\textsf{ct}_{i,0}^y$, $\dots,\textsf{ct}_{i,K\!+\!1}^y\}$, respectively.
		
		\State Sensor $i$ sends $\{\textsf{ct}_{i,0}^\varphi,\dots,\textsf{ct}_{i,K}^\varphi\}$ and $\{\textsf{ct}_{i,0}^y,\dots$, $\textsf{ct}_{i,K\!+\!1}^y\}$ to the fusion center.
		
		\hspace{-3.7em}\textbf{end for}
		
		\hspace{-3.7em}{\bf for} $k=0,\dots,K$, {\bf do}
		
		\State For any $i=1,\dots,n$, the fusion center computes \vspace{-0.5em}
		\begin{align*}
			\textsf{ct}_{i,k}^{\text{res}}=&(\textsf{ct}_{i,k}^{\text{res}(1)},\textsf{ct}_{i,k}^{\text{res}(2)})\cr
			\noalign{\vskip -3pt}
			=&\textsf{Add}(\textsf{ct}_{i,k+1}^y,-\textsf{Dot}(\textsf{ct}_k^{\hat{\theta}},\textsf{ct}_{i,k}^\varphi))\bmod \mathfrak{q}_K,\cr
			\noalign{\vskip -3pt}
			\textsf{ct}_{i,k}=&(\textsf{ct}_{i,k}^{(1)},\textsf{ct}_{i,k}^{(2)})=\textsf{Mult}(\textsf{ct}_{i,k}^\varphi,\textsf{ct}_{i,k}^{\text{res}})\bmod \mathfrak{q}_K.
		\end{align*}
		
		\State The fusion center uses the re-encryption oracle $\textsf{ReEnc}$ to compute $\textsf{ReEnc}_{\textsf{rk}_{i\rightarrow i_0}}(\textsf{ct}_{i,k})$ for any $i\neq i_0$. Then, the fusion center aggregates the ciphertexts by\vspace{-0.5em}
		\begin{align*}
			\smash{\textsf{ct}_k=\textsf{Add}(\textsf{ct}_{i_0,k},\sum_{i\neq i_0}\textsf{ReEnc}_{\textsf{rk}_{i\rightarrow i_0}}(\textsf{ct}_{i,k}))\bmod \mathfrak{q}_K.}
		\end{align*}
		
		\State The fusion center updates the encrypted estimate by\vspace{-0.5em}
		\begin{align}\label{alg}
			\smash{\textsf{ct}_{k+1}^{\hat{\theta}}=\textsf{Add}(\textsf{ct}_k^{\hat{\theta}},\alpha_K\textsf{ct}_k)\bmod \mathfrak{q}_K.}
		\end{align}
		
		\hspace{-3.7em}{\bf end for}
		
		\State The fusion center sends $\textsf{ct}_{K+1}^{\hat{\theta}}$ to Sensor $i_0$. Then, Sensor $i_0$ computes and sends $\hat{\theta}_{K+1}$$=$$\textsf{Dcd}_{\Delta_K}($ $\textsf{Dec}_{\textsf{sk}_{i_0}}(\textsf{ct}_{K+1}^{\hat{\theta}}))$ to the fusion center. The fusion center broadcasts $\hat{\theta}_{K+1}$$=$$\textsf{Dcd}_{\Delta_K}(\textsf{Dec}_{\textsf{sk}_{i_0}}(\textsf{ct}_{K+1}^{\hat{\theta}}))$ to each Sensor $i\neq i_0$.
		
		\item[]\hspace{-1.7em}{\bf Return} $\hat{\theta}_{K+1}$
	\end{algorithmic}
\end{breakablealgorithm}

\subsection{Security analysis}
In this subsection, we will give the security analysis of Algorithm \ref{ckks pia}. First, we review the definition of the decision RLWE problem:
\begin{def0}\label{def2}
	\citep{brakerski2014leveled} For $\textsf{s}\in R_\mathfrak{q}$ and $T\in\mathbb{N}_+\cap\textsf{poly}(N)$, the decision RLWE problem over $\dg$ is to distinguish the following two sets:
	\begin{align*}
		D_0=&\{(\textsf{a}^{(l)},\textsf{b}^{(l)})|\textsf{a}^{(l)},\textsf{b}^{(l)}\sim\mathcal{U},l=1,\dots,T\},\cr
		\noalign{\vskip -3pt}
		D_1=&\{(\widetilde{\textsf{a}}^{(l)},\widetilde{\textsf{b}}^{(l)})|\widetilde{\textsf{a}}^{(l)}\!\sim\!\mathcal{U},\textsf{e}^{(l)}\!\sim\!\dg,\cr
		\noalign{\vskip -3pt}
		& \widetilde{\textsf{b}}^{(l)}\!=\! -\widetilde{\textsf{a}}^{(l)} \textsf{s}\! +\!\textsf{e}^{(l)} \bmod \mathfrak{q},l=1,\dots,T\}. 
	\end{align*}
	Let $D_0^\dagger=D_0$ and define $D_1^\dagger$ by replacing $\dg$ with $\tdg$. Then, the decision RLWE problem over $\tdg$ is to distinguish $D_0^\dagger$ and $D_1^\dagger$. For an adversary $\mathcal{A}$ executing an algorithm $\mathscr{B}$, the advantage of distinguishing $D_0$ and $D_1$ is defined as
	\begin{align*}
		\textsf{adv}^{\mathcal{A}}(\mathscr{B})=|\p(\mathscr{B}(D_b)=1|b=0)\!-\!\p(\mathscr{B}(D_b)=1|b=1)|.
	\end{align*}
	The advantage $\textsf{adv}_{\Gamma}^{\mathcal{A}}(\mathscr{B})$ of distinguishing $D_0^\dagger$ and $D_1^\dagger$ is defined by replacing $D_b$ with $D_b^\dagger$.
\end{def0}
\begin{rmk}
	Definition \ref{def2} is standard and commonly used in cryptography \citep{lyubashevsky2013ideal,brakerski2014leveled}. In particular, if $\textsf{adv}^{\mathcal{A}}(\mathscr{B})\in\textsf{negl}(N)$, then the adversary $\mathcal{A}$ cannot distinguish $D_0$ and $D_1$ in polynomial time.
\end{rmk}

Then, we give the following hardness assumption of the decision RLWE problem over $\dg$: 
\begin{asm}\label{asm1}
	For any $T\in\textsf{poly}(N)\cap\mathbb{N}_+$ and any probabilistic polynomial-time (quantum) algorithm $\mathscr{B}$, $\textsf{adv}^{\mathcal{A}}(\mathscr{B})\in\textsf{negl}(N)$.
\end{asm}

Assumption \ref{asm1} is standard and commonly used to ensure the IND-CPA security of RLWE-based cryptosystems \citep{lyubashevsky2013toolkit,brakerski2014leveled}. However, since the truncated discrete Gaussian distribution $\tdg$ in Algorithm \ref{ckks pia} changes the distributions of the public key and the ciphertexts, Assumption~\ref{asm1} is insufficient to ensure the IND-CPA security of Algorithm~\ref{ckks pia}. To address this issue, the following lemma provides a sufficient condition on the truncation value $\Gamma$ to ensure the hardness of the decision RLWE problem over $\tdg$:
\begin{lemma}\label{lemma 1}
	If Assumption \ref{asm1} and the following condition hold:
	\begin{align}\label{condition 1}
		\Gamma\geq \sqrt{2}\sigma\ln N,
	\end{align}
	then for any probabilistic polynomial-time (quantum) algorithm $\mathscr{B}$, $\textsf{adv}_{\Gamma}^{\mathcal{A}}(\mathscr{B})\in\textsf{negl}(N)$.
\end{lemma}
{\bf Proof.} See \ref{appendix a}. $\hfill\blacksquare$
\begin{rmk}
	In Lemma \ref{lemma 1}, an explicit sufficient condition \eqref{condition 1} on the truncation value $\Gamma$ is given for a rigorous hardness proof of the decision RLWE problem over $\tdg$. This shows advantage over \cite{brakerski2014leveled} that does not provide a rigorous hardness proof.
\end{rmk}
\begin{rmk}
	When the ring dimension $N=2^{14}$, \eqref{condition 1} leads to $\Gamma\geq 13.7\sigma$. In this case, \eqref{condition 1} is of the same order of magnitude as $\Gamma\geq 6\sigma$, which is commonly used \citep{cheon2017homomorphic,chen2017simple,dwarakanath2014sampling}. Then, \eqref{condition 1} is acceptable in practice.
\end{rmk}

Based on Lemma \ref{lemma 1}, the security analysis of Algorithm~\ref{ckks pia} is given as follows:
\begin{thm0}\label{thm1}
	Under the conditions of Lemma \ref{lemma 1}, Algorithm~\ref{ckks pia} achieves the IND-CPA security.
\end{thm0}
{\bf Proof.} See \ref{appendix b}. $\hfill\blacksquare$
\begin{rmk}
	By Theorem \ref{thm1}, Algorithm \ref{ckks pia} achieves the IND-CPA security. This shows advantage over \cite{adamek2024encrypted,kim2023dynamic,hadjicostis2020privacy,feng2026asymptotic,fritz2026encrypted,murguia2020secure,tavazoei2026subtractionless,ikezaki2026encrypted,damera2026end} which do not provide an IND-CPA security proof. Moreover, since Algorithm \ref{ckks pia} achieves the IND-CPA security against eavesdropping, collusion and quantum attacks, the security level of Algorithm \ref{ckks pia} is enhanced compared to the one of the methods mentioned in Remark \ref{rmk: attack model}.
\end{rmk}

\subsection{Convergence analysis}
In this subsection, we will give the convergence analysis of Algorithm \ref{ckks pia}. First, we  introduce the following standard assumptions:
\begin{asm}\label{asm2}
	System \eqref{eq arx2} is asymptotically stable, i.e., $1-a_1z-a_2z^2-\dots-a_pz^p\neq 0$, $\forall\ z\in\mathbb{C}, |z|\leq 1$.
\end{asm}
\begin{asm}\label{asm3}
	There exists $L>0$ such that $|u_{i,k}|\leq L$, $\forall\ i=1,\dots,n,k=0,\dots,K$. 
\end{asm}
\begin{asm}\label{asm4}
	System noises $\{w_{i,k}, i=1,\dots,n,k=0,\dots,K+1\}$ are independent and identically distributed, and there exists $L>0$ such that $\E w_{i,k}=0$, $|w_{i,k}|\leq L$, $\forall i=1,\dots,n$, $k=0,\dots,K+1$.
\end{asm}
\begin{rmk}
	Assumption \ref{asm2} is the stability condition on the autoregressive part of the multi-participant ARX system \eqref{eq arx2}. Assumption \ref{asm3} ensures that the system input is bounded. These assumptions are necessary, since the unbounded system input and output may exceed the range of the plaintext space $R_{\mathfrak{q}_K}$, leading to plaintext overflow. Moreover, Assumption \ref{asm4} covers commonly used noises, such as the unbiased uniform noise \citep{zhang2025privacy,tan2023cooperative} and the truncated (discrete) Gaussian distribution \citep{ikezaki2026encrypted,damera2026end,adamek2024encrypted}.
\end{rmk}

\begin{lemma}\label{lemma 2}
	If Assumptions \ref{asm2}-\ref{asm4} hold, then $|y_{i,k}|\leq\|\varphi_{i,k}\|_\infty\leq G_1$, $\forall\ i=1,\dots,n$, $k=0,\dots,K+1$, where $G_1=c\sqrt{p}\rho L+\frac{c\sqrt{(q+1)(1+\sum_{l=1}^q b_l^2)}L}{1-\rho}$, $c\geq 1$, $0<\rho<1$.
\end{lemma}
{\bf Proof.} See \ref{appendix c}. $\hfill\blacksquare$

Next, we give the following standard assumption on the regressors:
\begin{asm}\label{asm5}
	There exist $\varpi>0$ and $1\leq t\leq K$ such that the following inequality holds:
	\begin{align*}
		\smash{\frac{1}{t}\sum_{i=1}^n\sum_{l=k}^{k+t-1}\varphi_{i,l}\varphi_{i,l}^\top\geq\varpi I_{p+q}, \forall\ k=0,\dots,K-t+1.}
	\end{align*}
\end{asm}
\begin{rmk}
	Assumption \ref{asm5} is the cooperative excitation condition on all sensors, and is more general than requiring persistent excitation on each sensor \citep{wang2022unified,ljung1999system}.
\end{rmk}\vspace{-0.5em}

The parameter design of the step-size $\alpha_K$, the modulus $\mathfrak{q}_K,\mathfrak{q}_{1,K},\mathfrak{q}_{2,K}$, the auxiliary modulus $\mathfrak{p}_K$ and the scaling factor $\Delta_K$ in Algorithm \ref{ckks pia} is assumed as follows:
\begin{asm}\label{asm6}
	The step-size $\alpha_K$$=$$\frac{c_1}{(K+1)^{p_1}}$, the modulus $\mathfrak{q}_K$$=$$\mathfrak{q}_{1,K}\mathfrak{q}_{2,K}^d$, $\mathfrak{q}_{1,K}$$=$$\lfloor c_2(K+1)^{p_2}\rfloor$, $\mathfrak{q}_{2,K}$$=$$\lfloor c_3(K+1)^{p_3}\rfloor$, the scaling factor $\Delta_K$$=$$\lfloor c_3(K+1)^{p_3}\rfloor$, and the auxiliary modulus $\mathfrak{p}_K$$=$$\lfloor c_4(K+1)^{p_4}\rfloor$ satisfy the following conditions:\vspace{-1em}
	\begin{align*}
		&0<c_1<\min\{\frac{1}{ntG_1^2},\frac{1}{2t\varpi},G_2\},c_2\geq c_3\geq 1, c_4\geq c_2c_3^d,\cr
		\noalign{\vskip -3pt}
		&\frac{1}{2}<p_1< 1,p_2\geq p_3\geq 0, 3p_3\geq p_1, p_4\geq p_2+p_3d,
	\end{align*}
	{\vskip -20pt}\noindent where $G_2=G_1(\sqrt{p+q}+\frac{\sqrt{2}(N+1+\sqrt{Nh})N\Gamma}{2}+\frac{\sqrt{2}N^{\frac{3}{2}}(1+\sqrt{Nh})}{4}$ $+\frac{\sqrt{2\Gamma}N^3}{8})+(\sqrt{p+q}+\frac{\sqrt{2}(N+1+\sqrt{Nh})N\Gamma}{2})\cdot(G_1+\sqrt{p+q}+\frac{\sqrt{2}(N+1+\sqrt{Nh})N\Gamma}{2}+\frac{\sqrt{2}N^{\frac{3}{2}}(1+\sqrt{Nh})}{4}+\frac{\sqrt{2\Gamma}N^3}{8})$.
\end{asm}
\begin{rmk}
	By Assumption \ref{asm6}, $\mathfrak{p}_K>\mathfrak{q}_{1,K}\geq\mathfrak{q}_{2,K}$ and $\mathfrak{q}_{2,K}=\Delta_K$ hold. This condition is consistent with $\mathfrak{p}_K\gg\mathfrak{q}_{1,K}\geq\mathfrak{q}_{2,K}$ and $\mathfrak{q}_{2,K}\approx\Delta_K$, which are commonly used in practice \citep{cheon2017homomorphic,cheon2018bootstrapping}. 
\end{rmk}\vspace{-0.5em}

To characterize the encryption noise and quantization error in the encrypted estimate sequence $\{\textsf{ct}_k^{\hat{\theta}}\mid k=0,\dots,K+1\}$, we construct the following auxiliary plaintext sequence by Steps 6-8 of Algorithm \ref{ckks pia}:\vspace{-1em}
\begin{align}\label{l5 1}
	\hat{\theta}_{k+1}=&\ (I_{p+q}-\alpha_K\sum_{i=1}^n(\varphi_{i,k}\varphi_{i,k}^\top+M_{i,k}))\hat{\theta}_k\cr
	&+\alpha_KS_k,k=0,\dots,K,
\end{align}
{\vskip -20pt}\noindent where \vspace{-0.5em}
\begin{align*}
	&M_{i,k}=\varphi_{i,k}(\frac{\textsf{qe}_{i,k}^\varphi\!+\!\textsf{Dcd}_{\Delta_K}(\textsf{ee}_{i,k}^\varphi\!+\!\breve{\textsf{re}}_{i,k})}{\Delta_K})^\top\cr
	&~~~~~~~~~~+\frac{\textsf{qe}_{i,k}^\varphi+\textsf{Dcd}_{\Delta_K}(\textsf{ee}_{i,k}^\varphi)}{\Delta_K}(\varphi_{i,k}\!+\!\textsf{qe}_{i,k}^\varphi\cr
	&~~~~~~~~~~+\textsf{Dcd}_{\Delta_K}\!(\textsf{ee}_{i,k}^\varphi\!+\!\breve{\textsf{re}}_{i,k}))^\top,\cr
	&S_k=\sum_{i=1}^n\varphi_{i,k}(y_{i,k+1}+\frac{\textsf{Dcd}_{\Delta_K}(\textsf{ee}_{i,k+1}^y)+\textsf{qe}_{i,k+1}^y}{\Delta_K})\cr
	&~~~~~~+\sum_{i=1}^n\frac{\textsf{qe}_{i,k}^\varphi\!+\!\textsf{Dcd}_{\Delta_K}(\textsf{ee}_{i,k}^\varphi)}{\Delta_K}(y_{i,k+1}\!+\!\textsf{Dcd}_{\Delta_K}(\textsf{ee}_{i,k+1}^y)\cr
	&~~~~~~+\textsf{qe}_{i,k+1}^y)+\sum_{i=1}^n\textsf{Dcd}_{\Delta_K}(Q(\frac{\textsf{ct}_{i,k}}{\mathfrak{q}_{2,K}})-\frac{\textsf{ct}_{i,k}}{\mathfrak{q}_{2,K}})\cr
	&~~~~~~+\sum_{i=1}^n\textsf{Dcd}_{\Delta_K}(\textsf{me}_{i,k})+\sum_{i\neq i_0}\textsf{Dcd}_{\Delta_K}(\textsf{re}_{i,k}),\cr
	&\textsf{qe}_{i,k}^\varphi=\Delta_K^{-1}Q(\Delta_K\cdot\varphi_{i,k})-\varphi_{i,k},\cr
	&\textsf{ee}_{i,k}^\varphi=\textsf{Dec}_{\textsf{sk}_i}(\textsf{ct}_{i,k}^\varphi)-\textsf{pt}_{i,k}^\varphi,
\end{align*}
\begin{align*}
	&\breve{\textsf{re}}_{i,k}=\sum_{l=1}^{\frac{N}{2}-1} (\textsf{Dec}_{\textsf{sk}_i}(\textsf{Rot}(\textsf{ct}_{i,k}^\varphi;l))-\textsf{Dec}_{\textsf{sk}_i}(\textsf{ct}_{i,k}^\varphi)),\cr
	&\textsf{qe}_{i,k+1}^y=\Delta_K^{-1}Q(\Delta_K\cdot y_{i,k+1})-y_{i,k+1},\cr
	&\textsf{ee}_{i,k+1}^y=\textsf{Dec}_{\textsf{sk}_i}(\textsf{ct}_{i,k+1}^y)-\textsf{pt}_{i,k+1}^y,\cr
	&\textsf{me}_{i,k}=\langle Q(\mathfrak{p}_K^{-1}\textsf{ct}_{i,k}^{\varphi(2)}\textsf{ct}_{i,k}^{\text{res}(2)}\cdot\textsf{evk}_i)-\mathfrak{p}_K^{-1}\textsf{ct}_{i,k}^{\varphi(2)}\cr
	&~~~~~~~~~~~\textsf{ct}_{i,k}^{\text{res}(2)}\textsf{evk}_i,\textsf{sk}_i\rangle+\mathfrak{p}_K^{-1}\textsf{ct}_{i,k}^{\varphi(2)}\textsf{ct}_{i,k}^{\text{res}(2)}\textsf{e}^{\prime\prime}_i,\cr
	&\textsf{re}_{i,k}=\textsf{qe}_{i,k}^{\textsf{re}(1)}+\textsf{qe}_{i,k}^{\textsf{re}(2)}\textsf{s}_{i_0}+\mathfrak{p}_K^{-1}\textsf{ct}_{i,k}^{(2)}\textsf{e}_{i\rightarrow i_0},\cr	
	&(\textsf{qe}_{i,k}^{\textsf{re}(1)},\textsf{qe}_{i,k}^{\textsf{re}(2)})=Q(\mathfrak{p}_K^{-1}\textsf{ct}_{i,k}^{(2)}\cdot\textsf{rk}_{i\rightarrow i_0})-\mathfrak{p}_K^{-1}\textsf{ct}_{i,k}^{(2)}\cdot\textsf{rk}_{i\rightarrow i_0},\cr
	&\textsf{e}_{i\rightarrow i_0}=\textsf{v}_{i\rightarrow i_0}\textsf{e}^\prime+\textsf{e}_{1,i\rightarrow i_0}+\textsf{e}_{2,i\rightarrow i_0}\textsf{s}_{i_0}.
\end{align*}\vspace{-2em}

Note that $\textsf{Ecd}_{\Delta_K}(\hat{\theta}_k)\bmod \mathfrak{q}_K=\textsf{Dec}_{\textsf{sk}_{i_0}}(\textsf{ct}_k^{\hat{\theta}})$, $\forall\ k=0,\dots,K+1$. Then, \eqref{l5 1} is obtained by decrypting and decoding the encrypted estimate sequence modulo $\mathfrak{q}_K$. However, the cumulative encryption noise and quantization error in \eqref{l5 1} may cause the plaintext exceed the range of the plaintext space, leading to plaintext overflow. Then, the convergence of Algorithm \ref{ckks pia} cannot be ensured. To address this issue, the following lemma gives a criterion for avoiding overflow:
\begin{lemma}\label{lemma 3}
	Under Assumptions \ref{asm2}-\ref{asm6} and the following condition:
	\begin{align}\label{condition 2}
		&\hspace{-1em}(1\!+\!n\alpha_KG_{4,K})^{t-1}\|\hat{\theta}_0\|\!+\!\frac{G_{3,K}}{nG_{4,K}}((1\!+\!n\alpha_KG_{4,K})^{t-1}\!-\!1)\notag\\
		&\hspace{-1em}+\!\frac{2G_{3,K}}{nt\varpi\alpha_KG_{4,K}}((1\!+\!n\alpha_KG_{4,K})^t\!\!-\!1)\!\leq\!\frac{\sqrt{2N}(\mathfrak{q}_K\!-\!\!1)}{4\Delta_K},
	\end{align}
	plaintext overflow is avoided, i.e., $\textsf{Ecd}_{\Delta_K}(\hat{\theta}_k)$$=$$\textsf{Dec}_{\textsf{sk}_{i_0}}(\textsf{ct}_k^{\hat{\theta}})$, $\forall\ k=0,\dots,K+1$, where
	\begin{align*}
		G_{3,K}=&nG_1^2+2nG_1(\frac{\sqrt{2N}n(1+\sqrt{Nh})}{2\Delta_K^2}+\frac{\sqrt{p+q}}{\Delta_K^2})\cr
		&+\frac{n}{\Delta_K}(\frac{\sqrt{p+q}}{\Delta_K}+\frac{\sqrt{2N}n(1+\sqrt{Nh})}{2\Delta_K})^2+\frac{\sqrt{2N}n}{2\Delta_K}\cr
		&+\frac{\sqrt{2N}n(1+\sqrt{Nh})}{\Delta_K}+\frac{\sqrt{\Gamma\mathfrak{q}_K}nN^2(N\!+\!1\!+\!\sqrt{Nh})}{2\mathfrak{p}_K},\cr
		G_{4,K}=&\frac{G_1}{\Delta_K}(\frac{\sqrt{p+q}}{\Delta_K}+\frac{\sqrt{2}(N+1+\sqrt{Nh})N\Gamma}{2\Delta_K}\cr
		&+\frac{\sqrt{2}N^{\frac{3}{2}}(1+\sqrt{Nh})}{4\Delta_K}+\frac{\sqrt{\Gamma\mathfrak{q}_K}N^3}{4\mathfrak{p}_K\Delta_K})\cr
		&+(\frac{\sqrt{p+q}}{\Delta_K^2}+\frac{\sqrt{2}(N+1+\sqrt{Nh})N\Gamma}{2\Delta_K^2})\cdot\cr
		&(G_1+\frac{\sqrt{p+q}}{\Delta_K}+\frac{\sqrt{2}(N+1+\sqrt{Nh})N\Gamma}{2\Delta_K}\cr
		&+\frac{\sqrt{2}N^{\frac{3}{2}}(1+\sqrt{Nh})}{4\Delta_K}+\frac{\sqrt{\Gamma\mathfrak{q}_K}N^3}{4\mathfrak{p}_K\Delta_K}),
	\end{align*}
\end{lemma}\vspace{-2em}
{\bf Proof.} See \ref{appendix d}. $\hfill\blacksquare$

\begin{rmk}
	In practice, $N$$\geq$$ 2^{10}$, $\Delta_K$$\geq$$2^{30}$, $\mathfrak{q}_{1,K}$$>$$\mathfrak{q}_{2,K}$$\approx$ $\Delta_K$, and $\mathfrak{p}_K$$\approx$$\mathfrak{q}_K$ are commonly used \citep{albrecht2022homomorphic,cheon2017homomorphic,cheon2018bootstrapping}. In this case, $G_{3,K}\approx nG_1^2$, $G_{4,K}\approx0$, and then, \eqref{condition 2} can be approximately rewritten as
	\begin{align}\label{condition 2'}
		\smash{\|\hat{\theta}_0\|+nG_1^2K\alpha_K\leq\frac{\sqrt{2N}(\mathfrak{q}_K-1)}{4\Delta_K}.}
	\end{align}
	By \eqref{condition 2'}, if $K$$\leq$$O((\frac{\mathfrak{q}_K}{\Delta_K})^{\frac{1}{1-p_1}})$$=$$O(10^{18})$, then plaintext overflow is avoided. Thus, \eqref{condition 2'} is preferred for selecting the ring dimension $N$, the maximum iteration number $K$, the step-size $\alpha_K$, the scaling factor $\Delta_K$, and the modulus $\mathfrak{q}_K$ in practice.
\end{rmk}

Based on Lemma \ref{lemma 3}, the convergence analysis of Algorithm~\ref{ckks pia} is given as follows:
\begin{thm0}\label{thm2}
	Under the conditions of Lemma \ref{lemma 3}, the mean square convergence rate of Algorithm \ref{ckks pia} is given as follows:
	\begin{align}\label{conv rate}
		\smash{\E\|\hat{\theta}_{K+1}-\theta\|^2=O(\frac{1}{(K+1)^{p_1}}).}
	\end{align}
	Furthermore, Algorithm \ref{ckks pia} achieves the mean square convergence, i.e., $\lim_{K\to\infty}\E\|\hat{\theta}_{K+1}-\theta\|^2=0$.
\end{thm0}
{\bf Proof.} See \ref{appendix e}. $\hfill\blacksquare$
\begin{rmk}
	The key idea in proving Theorem \ref{thm2} is using the auxiliary plaintext sequence \eqref{l5 1} to characterize the encryption noise and quantization error $M_{i,k},S_k$ in the encrypted estimate $\textsf{ct}^{\hat{\theta}}_k$. Then, by using the decreasing step-size $\alpha_K$ to mitigate the encryption noise and quantization error, the mean square convergence and convergence rate of Algorithm \ref{ckks pia} are given. Compared to \cite{lu2018privacy,alexandru2021cloud,adamek2024encrypted,tan2023cooperative,ruan2019secure,zhang2025privacy}, a rigorous convergence analysis is provided.
\end{rmk}

Based on Theorems \ref{thm1} and \ref{thm2}, the trade-off between the IND-CPA security and the mean square convergence of Algorithm \ref{ckks pia} is as follows:
\begin{cor0}\label{cor 2}
	Under Assumptions \ref{asm1}-\ref{asm6} and \eqref{condition 1}, \eqref{condition 2}, Algorithm \ref{ckks pia} cannot achieve the IND-CPA security and the mean square convergence simultaneously.
\end{cor0}
{\bf Proof.} We prove this corollary by contradiction. 

Suppose that Algorithm \ref{ckks pia} achieves the IND-CPA security and the mean square convergence simultaneously. Then, by $3p_3\geq p_1$ and $p_1>\frac{1}{2}$ in Assumption \ref{asm6}, we have $\lim_{K\to\infty}\Delta_K=\infty$. Since \eqref{condition 2} is required to avoid plaintext overflow, we have $\lim_{K\to\infty}\mathfrak{q}_K=\infty$, and further, $\lim_{K\to\infty}\frac{\sigma}{\mathfrak{q}_K}=0$. Thus, by Lemma 4.1 in \cite{lindner2011better}, there exists a probabilistic polynomial-time algorithm to solve the RLWE problem over $\dg$. Then, Algorithm \ref{ckks pia} cannot achieve the IND-CPA security, which contradicts Theorem \ref{thm1}. $\hfill\blacksquare$
\begin{rmk}
	Corollary \ref{cor 2} shows a trade-off between the IND-CPA security and the mean square convergence, which is also observed in \cite{adamek2024encrypted}. This trade-off arises since achieving the mean square convergence of Algorithm \ref{ckks pia} requires the increasing scaling factor $\Delta_K$, while achieving the IND-CPA security of Algorithm \ref{ckks pia} requires the scaling factor $\Delta_K$ to be constant. 
\end{rmk}

\section{Numerical example}
In this section, we consider the parameter identification problem of the following multi-participant ARX systems:
\begin{align*}
	y_{i,k+1}=&\ 0.3y_{i,k}+0.5y_{i,k-1}-0.5y_{i,k-2}-0.4y_{i,k-3}\cr
	&+0.6y_{i,k-4}+0.7u_{i,k}+1.5u_{i,k-1}-0.3u_{i,k-2}\cr
	&-1.1u_{i,k-3}+w_{i,k+1},\cr
	&\ i=1,\dots,5,\ k=0,\dots,6000.
\end{align*}
where $w_{i,k}$ is drawn from the uniform distribution on $[-5,5]$, $u_{i,k}$ is drawn from the uniform distribution on $[1,15]$, and Sensor~$3$ is an honest sensor. Then, the system orders $p=5$, $q=4$, the system parameter $\theta=$ $(0.3, 0.5, -0.5, -0.4, 0.6,0.7, 1.5,-0.3, -1.1)^\top$, and the maximum iteration number $K=6000$.

\begin{figure}[!h]
	\centering
	\includegraphics[width=0.42\textwidth]{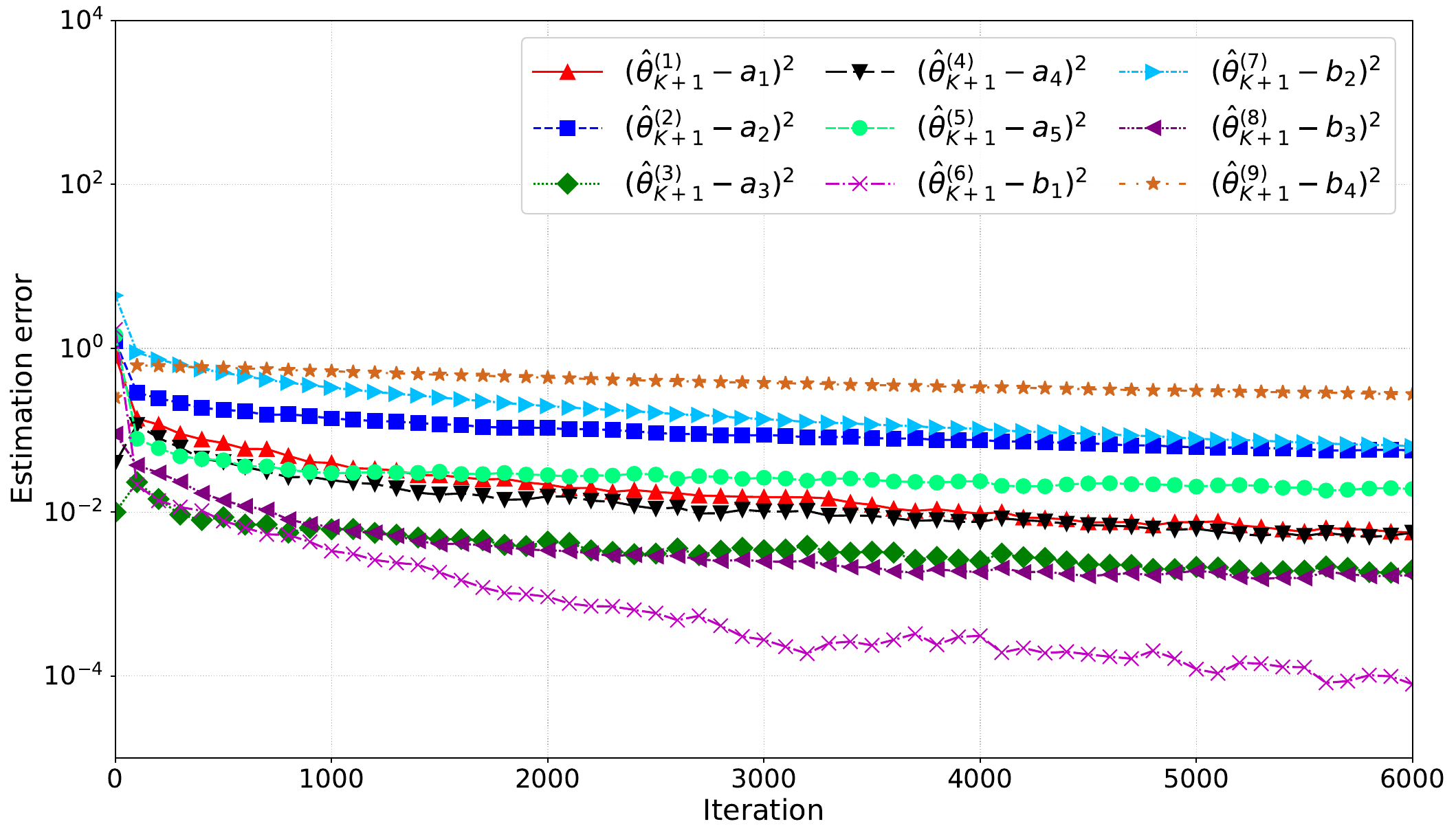}
	\caption{Estimation errors of Algorithm \ref{ckks pia}}
	\label{fig2}
\end{figure}

Let the step-size $\alpha_K=\frac{10^{-3}}{(K+1)^{0.6}}\approx6\cdot10^{-6}$, the initial estimate $\hat{\theta}_0=(-0.6,-0.6,-0.6,-0.6,-0.6,-0.6,-0.6$, $-0.6,-0.6)^\top$, the ring dimension $N=2^{14}$, the scaling factor $\Delta_K$$=$$\lfloor2\cdot10^{11}\cdot(K+1)^{0.2}\rfloor$$\approx 2^{40}$, the noise parameter $\sigma=3.2$, the truncation value $\Gamma=44$, $h=64$, the modulus $\mathfrak{q}_K=\mathfrak{q}_{1,K}\mathfrak{q}_{2,K}^3$, and the auxiliary modulus $\mathfrak{p}_K=\mathfrak{q}_K$, where the modulus $\mathfrak{q}_{1,K},\mathfrak{q}_{2,K}$ are primes satisfying $\mathfrak{q}_{1,K}=\lfloor2\cdot10^{17}\cdot(K+1)^{0.2}\rfloor\approx 2^{60}$, $\mathfrak{q}_{2,K}=\lfloor2\cdot10^{11}\cdot(K+1)^{0.2}\rfloor\approx 2^{40}$. Then, Algorithm \ref{ckks pia} is implemented based on the SEAL library \citep{chen2017simple}, and the modulus $\mathfrak{q}_K$ is formulated as the bit-lengths $(60,40,40,40)$ of the primes $(\mathfrak{q}_{1,K},\mathfrak{q}_{2,K},\mathfrak{q}_{2,K},\mathfrak{q}_{2,K})$. The trajectories of estimation errors are given in Fig.~\ref{fig2}, from which one can see that the estimation error is in a small neighborhood of zero. This result is consistent with Theorem \ref{thm2}.

\begin{figure}[!h]
	\centering
	\includegraphics[width=0.42\textwidth]{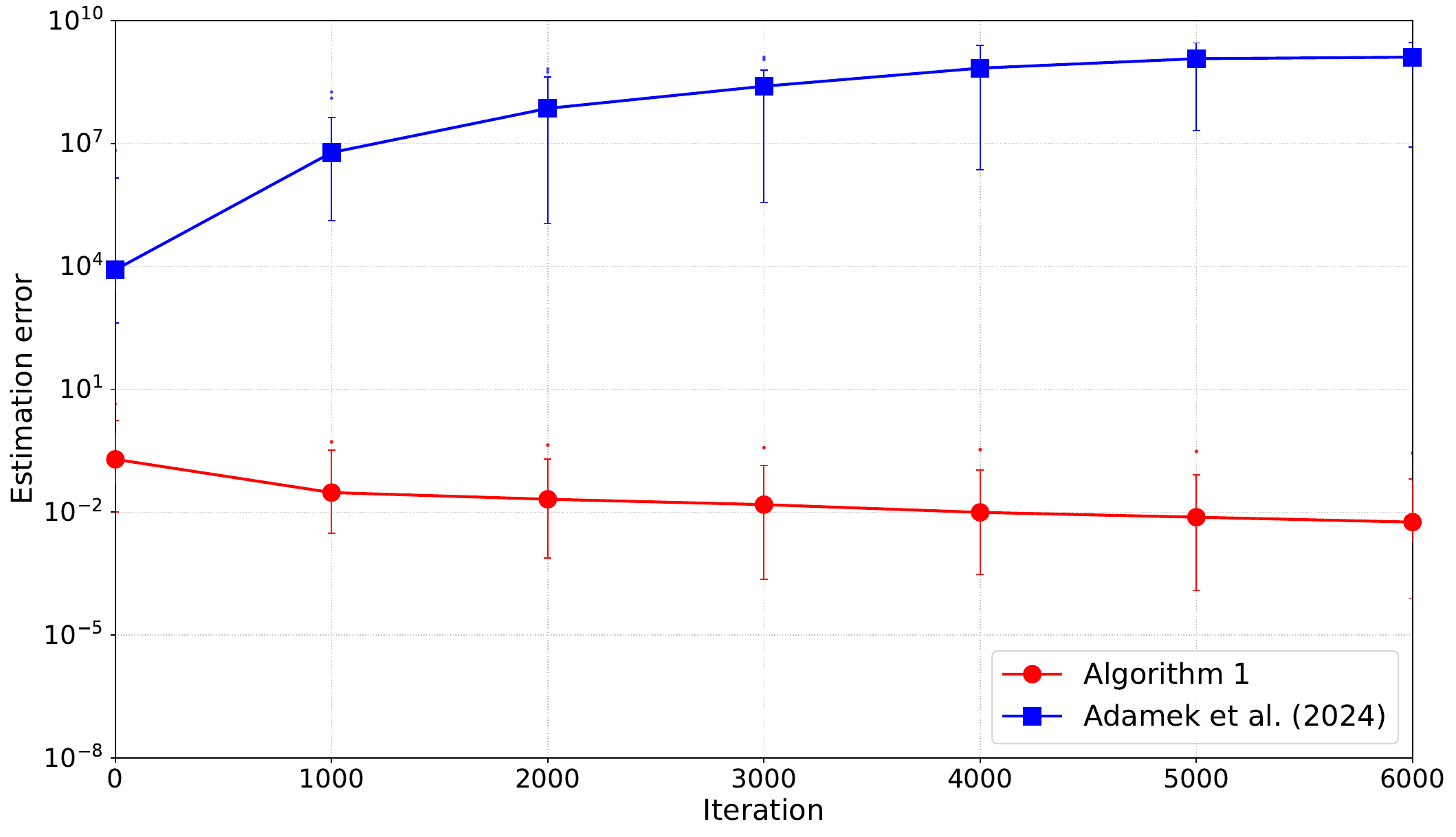}
	\caption{Comparison of estimation errors}
	\label{fig3}
\end{figure}

Next, we show the superior performance of Algorithm \ref{ckks pia} over the method in \cite{adamek2024encrypted}. To ensure a fair comparison, let the ring dimension $N$, the scaling factor $\Delta_K$, the noise parameter $\sigma$, the truncation value $\Gamma$, the modulus $\mathfrak{q}_K$, and the auxiliary modulus $\mathfrak{p}_K$ in \cite{adamek2024encrypted} be the same as Algorithm \ref{ckks pia}. Then, the comparison of estimation error between Algorithm~\ref{ckks pia} and the method in \cite{adamek2024encrypted} is given in Fig. \ref{fig3}. From Fig. \ref{fig3}, one can see that Algorithm \ref{ckks pia} has a smaller estimation error than the method in \cite{adamek2024encrypted}.

\section{Conclusion}
This paper has proposed a CKKS cryptosystem-based secure identification algorithm for multi-participant ARX systems. A sufficient condition on the truncation value has been derived to ensure the IND-CPA security of the algorithm under collusion and quantum attacks. An auxiliary plaintext sequence has been constructed to derive a criterion for avoiding plaintext overflow, based on which the mean square convergence and convergence rate of the algorithm have been established. By combining the security and convergence analysis, the trade-off between the IND-CPA security and the mean square convergence of the algorithm has been revealed. A numerical example has demonstrated the effectiveness and superior performance of the algorithm. Future work will focus on incorporating differential privacy into the proposed algorithm to further enhance the privacy protection level.

\appendix
\renewcommand{\thesection}{Appendix~\Alph{section}}
\renewcommand{\theequation}{\Alph{section}.\arabic{equation}}
\section{Proof of Lemma \ref{lemma 1}}\label{appendix a}
Let $\delta_{\text{SD}}$ be the statistical distance, and $A,A^\dagger$ be the probability distributions of $D_1,D_1^\dagger$, respectively. Then, the proof proceeds in two steps. 

{\bf Step 1.} First, we prove $\delta_{\text{SD}}(A,A^\dagger)\in\textsf{negl}(N)$. 

Note that
\begin{align}\label{eq1}
	&\delta_{\text{SD}}(\dg,\tdg)\cr
	=&\frac{1}{2}\sum_{l=1}^{N}\sum_{m\in\mathbb{Z}}\left|\frac{\exp(-\frac{m^2}{2\sigma^2})}{\sum_{r\in\mathbb{Z}}\exp(-\frac{r^2}{2\sigma^2})}-\frac{\mathbbm{1}_{\{|m|\leq \Gamma\}}\exp(-\frac{m^2}{2\sigma^2})}{\sum_{r\in\mathbb{Z},|r|\leq \Gamma}\exp(-\frac{r^2}{2\sigma^2})}\right|\cr
	=&\frac{N}{2}\sum_{m\in\mathbb{Z},|m|\leq \Gamma}(\frac{\exp(-\frac{m^2}{2\sigma^2})}{\sum_{r\in\mathbb{Z},|r|\leq \Gamma}\exp(-\frac{r^2}{2\sigma^2})}-\frac{\exp(-\frac{m^2}{2\sigma^2})}{\sum_{r\in\mathbb{Z}}\exp(-\frac{r^2}{2\sigma^2})})\cr
	&+\frac{N}{2}\frac{\sum_{m\in\mathbb{Z},|m|>\Gamma}\exp(-\frac{m^2}{2\sigma^2})}{\sum_{r\in\mathbb{Z}}\exp(-\frac{r^2}{2\sigma^2})}\cr
	=&\frac{N\sum_{m\in\mathbb{Z},|m|>\Gamma}\exp(-\frac{m^2}{2\sigma^2})}{\sum_{r\in\mathbb{Z}}\exp(-\frac{r^2}{2\sigma^2})}.
\end{align}
Then, by Lemma 2.4 in \cite{banaszczyk1995inequalities}, we have
\begin{align}\label{eq2}
	\frac{\sum_{m\in\mathbb{Z},|m|>\Gamma}\exp(-\frac{m^2}{2\sigma^2})}{\sum_{r\in\mathbb{Z}}\exp(-\frac{r^2}{2\sigma^2})}\leq 2\exp(-\frac{\Gamma^2}{2\sigma^2}).
\end{align}
By $\Gamma\geq \sqrt{2}\sigma\ln N$, \eqref{eq2} can be rewritten as
\begin{align}\label{eq3}
	\frac{\sum_{m\in\mathbb{Z},|m|>\Gamma}\exp(-\frac{m^2}{2\sigma^2})}{\sum_{r\in\mathbb{Z}}\exp(-\frac{r^2}{2\sigma^2})}\leq 2N^{-\ln N}.
\end{align}
Thus, substituting \eqref{eq3} into \eqref{eq1} implies
\begin{align}\label{eq4}
	\delta_{\text{SD}}(\dg,\tdg)\leq 2N^{1-\ln N}\in\textsf{negl}(N).
\end{align}
Note that $T\in\textsf{poly}(N)\cap\mathbb{N}_+$. Then, by Proposition 8.10 in \cite{micciancio2002complexity} and \eqref{eq4}, we have
\begin{align}\label{eq5}
	\hspace{-1em}\delta_{\text{SD}}(A,A^\dagger)\leq T\delta_{\text{SD}}(\dg,\tdg)\in\textsf{negl}(N).
\end{align}

{\bf Step 2.} In this step, we prove Lemma \ref{lemma 1}. 

For any probabilistic polynomial-time (quantum) algorithm $\mathscr{B}$, by \eqref{eq5} and Proposition 8.10 in \cite{micciancio2002complexity}, we have
\begin{align}\label{eq6}
	|\textsf{adv}^{\mathcal{A}}(\mathscr{B})-\textsf{adv}_{\Gamma}^{\mathcal{A}}(\mathscr{B})|\leq\delta_{\text{SD}}(A,A^\dagger)\in\textsf{negl}(N).
\end{align}
Note that by Assumption \ref{asm1}, $\textsf{adv}^{\mathcal{A}}(\mathscr{B})\in\textsf{negl}(N)$ holds for any probabilistic polynomial-time (quantum) algorithm $\mathscr{B}$. Then, by \eqref{eq6}, we have
\begin{align}\label{eq7}
	\textsf{adv}_{\Gamma}^{\mathcal{A}}(\mathscr{B})\leq&|\textsf{adv}_{\Gamma}^{\mathcal{A}}(\mathscr{B})-\textsf{adv}^{\mathcal{A}}(\mathscr{B})|+|\textsf{adv}^{\mathcal{A}}(\mathscr{B})|\cr
	\in& \textsf{negl}(N).
\end{align}
Thus, this lemma is proved. $\hfill\blacksquare$

\section{Proof of Theorem \ref{thm1}}\label{appendix b}
{\bf Proof.} The proof proceeds in two steps.

\textbf{Step 1}. First, we prove that the modified CKKS cryptosystem achieves the IND-CPA security.

Let $(\textsf{pt}_0,\textsf{pt}_1)$ be a pair of plaintexts generated by the adversary $\mathcal{A}$, and $\textsf{c}_{1,i},\textsf{c}_{2,i},\textsf{d}_{1,i},\textsf{d}_{2,i}$ be independent random variables sampled from $\mathcal{U}$ for any $i=1,\dots,n$. Then, by Proposition 8.10 in \cite{micciancio2002complexity} and Lemma \ref{lemma 1}, we have
\begin{align}\label{t1 1}
	&|\p(\mathscr{B}(\textsf{pt}_0,\textsf{pt}_1,\textsf{Enc}_{\textsf{pk}_i}(\textsf{pt}_b))=b)\cr
	&-\p(\mathscr{B}(\textsf{pt}_0,\textsf{pt}_1,\textsf{Enc}_{(\textsf{c}_{1,i},\textsf{c}_{2,i})}(\textsf{pt}_b))=b)|\cr
	\leq&\textsf{adv}_{\Gamma}^{\mathcal{A}}(\mathscr{B})\in\textsf{negl}(N).
\end{align}
Note that there exists $\textsf{v}\sim\zo,\textsf{e}_1,\textsf{e}_2\sim\tdg$ such that
\begin{align}\label{t1 2}
	\textsf{Enc}_{(\textsf{c}_{1,i},\textsf{c}_{2,i})}(\textsf{pt}_b)\!=\!(\textsf{pt}_b\!+\!\textsf{v}\!\cdot\!\textsf{c}_{2,i}\!+\!\textsf{e}_2,\textsf{v}\!\cdot\!\textsf{c}_{1,i}\!+\!\textsf{e}_1).
\end{align}
Then, by Proposition 8.10 in \cite{micciancio2002complexity} and Lemma \ref{lemma 1}, we have
\begin{align}\label{t1 3}
	&|\p(\mathscr{B}(\textsf{pt}_0,\textsf{pt}_1,\textsf{Enc}_{(\textsf{c}_{1,i},\textsf{c}_{2,i})}(\textsf{pt}_b))=b)\cr
	&-\p(\mathscr{B}(\textsf{pt}_0,\textsf{pt}_1,(\textsf{pt}_b+\textsf{d}_{2,i},\textsf{d}_{1,i}))=b)|\cr
	\leq&\textsf{adv}_{\Gamma}^{\mathcal{A}}(\mathscr{B})\in\textsf{negl}(N).
\end{align}
Since $\textsf{d}_{2,i}\sim\mathcal{U}$, $\textsf{pt}_b+\textsf{d}_{2,i}\sim\mathcal{U}$ holds for any $b\in\{0,1\}$. Then, it can be seen that
\begin{align}\label{t1 4}
	&\p(\mathscr{B}(\textsf{pt}_0,\textsf{pt}_1,(\textsf{pt}_b+\textsf{d}_{2,i},\textsf{d}_{1,i}))=b)\cr
	=&\p(\mathscr{B}(\textsf{pt}_0,\textsf{pt}_1,(\textsf{pt}_b+\textsf{d}_{2,i},\textsf{d}_{1,i}))=b|b=0)\p(b=0)\cr
	&+\p(\mathscr{B}(\textsf{pt}_0,\textsf{pt}_1,(\textsf{pt}_b+\textsf{d}_{2,i},\textsf{d}_{1,i}))=b|b=1)\p(b=1)\cr
	=&\frac{1}{2}(\p(\mathscr{B}(\textsf{pt}_0,\textsf{pt}_1,(\textsf{pt}_0+\textsf{d}_{2,i},\textsf{d}_{1,i}))=0)\cr
	&+\p(\mathscr{B}(\textsf{pt}_0,\textsf{pt}_1,(\textsf{pt}_1+\textsf{d}_{2,i},\textsf{d}_{1,i}))=1))\cr
	=&\frac{1}{2}(\p(\mathscr{B}(\textsf{pt}_0,\textsf{pt}_1,(\textsf{d}_{2,i},\textsf{d}_{1,i}))=0)\cr
	&+\p(\mathscr{B}(\textsf{pt}_0,\textsf{pt}_1,(\textsf{d}_{2,i},\textsf{d}_{1,i}))=1))\cr
	=&\frac{1}{2}.
\end{align}
Combining \eqref{t1 1}, \eqref{t1 3}, and \eqref{t1 4} gives
\begin{align}\label{t1 5}
	&|\p(\mathscr{B}(\textsf{pt}_0,\textsf{pt}_1,\textsf{Enc}_{\textsf{pk}_i}(\textsf{pt}_b))=b)-\frac{1}{2}|\cr
	\leq&|\p(\mathscr{B}(\textsf{pt}_0,\textsf{pt}_1,\textsf{Enc}_{\textsf{pk}_i}(\textsf{pt}_b))=b)\cr
	&-\p(\mathscr{B}(\textsf{pt}_0,\textsf{pt}_1,\textsf{Enc}_{(\textsf{c}_{1,i},\textsf{c}_{2,i})}(\textsf{pt}_b))=b)|\cr
	&+|\p(\mathscr{B}(\textsf{pt}_0,\textsf{pt}_1,\textsf{Enc}_{(\textsf{c}_{1,i},\textsf{c}_{2,i})}(\textsf{pt}_b))=b)\cr
	&-\p(\mathscr{B}(\textsf{pt}_0,\textsf{pt}_1,(\textsf{pt}_b+\textsf{d}_{2,i},\textsf{d}_{1,i}))=b)|\cr
	&+|\p(\mathscr{B}(\textsf{pt}_0,\textsf{pt}_1,(\textsf{pt}_b+\textsf{d}_{2,i},\textsf{d}_{1,i}))=b)-\frac{1}{2}|\cr
	\leq&\textsf{adv}_{\Gamma}^{\mathcal{A}}(\mathscr{B})+\textsf{adv}_{\Gamma}^{\mathcal{A}}(\mathscr{B})\in\textsf{negl}(N).
\end{align}
Since $M\in\textsf{poly}(\lambda), N=\frac{M}{2}$, we have $N\in\textsf{poly}(\lambda)$. Then, \eqref{t1 5} implies
\begin{align*}
	|\p(\mathscr{B}(\textsf{pt}_0,\textsf{pt}_1,\textsf{Enc}_{\textsf{pk}_i}(\textsf{pt}_b))=b)-\frac{1}{2}|\in\textsf{negl}(\lambda).
\end{align*}
Thus, by Definition \ref{def1}, the modified CKKS cryptosystem achieves the IND-CPA security. 

\textbf{Step 2}. In this step, we prove that Algorithm \ref{ckks pia} achieves the IND-CPA security.

Under collusion attacks, let $(g(\textsf{Enc}_{\textsf{pk}_i}(\textsf{pt}_0)),g(\textsf{Enc}_{\textsf{pk}_i}$ $(\textsf{pt}_1)))$ be the post-processed ciphertexts in Algorithm~\ref{ckks pia} for any $i=1,\dots,n$. Then, it can be seen that $\mathscr{B}(\textsf{pt}_0,\textsf{pt}_1,g(\cdot))$ is a probabilistic polynomial-time (quantum) algorithm. Thus, replacing $\mathscr{B}$ with $\mathscr{B}(\textsf{pt}_0,\textsf{pt}_1,g(\cdot))$ in \eqref{t1 5} gives
\begin{align}\label{t1 6}
	&|\p(\mathscr{B}(\textsf{pt}_0,\textsf{pt}_1,g(\textsf{Enc}_{\textsf{pk}_i}(\textsf{pt}_b)))=b)-\frac{1}{2}|\cr
	\leq&\textsf{adv}_{\Gamma}^{\mathcal{A}}(\mathscr{B})+\textsf{adv}_{\Gamma}^{\mathcal{A}}(\mathscr{B})\in\textsf{negl}(N).
\end{align}
Since $N\in\textsf{poly}(\lambda)$, \eqref{t1 6} implies
\begin{align}\label{t1 7}
	|\p(\mathscr{B}(\textsf{pt}_0,\textsf{pt}_1,g(\textsf{Enc}_{\textsf{pk}_i}(\textsf{pt}_b)))\!=\!b)\!-\!\frac{1}{2}|\!\in\!\textsf{negl}(\lambda).
\end{align}
Then, by Definition \ref{def1}, this theorem is proved. $\hfill\blacksquare$

\section{Proof of Lemma \ref{lemma 3}}\label{appendix c}
For notational convenience, define
\begin{align*}
	Y_{i,k}=&(y_{i,k},\dots,y_{i,k-p+1})^\top, \cr
	U_{i,k}=&(u_{i,k},\dots,u_{i,k-q+1},w_{i,k+1})^\top,\cr
	A=&\left[\begin{matrix}
		a_1&\dots&a_{p-1}&a_p\\
		1&\dots&0&0\\
		\vdots&\dots&\vdots&\vdots\\
		0&\dots&1&0
	\end{matrix}\right],B=\left[\begin{matrix}
		b_1&b_2&\dots&b_q&1\\
		0&0&\dots&0&0\\
		\vdots&\vdots&\dots&\vdots&\vdots\\
		0&0&\dots&0&0
	\end{matrix}\right].
\end{align*}
Then the proof proceeds in two steps.

\textbf{Step 1}. First, we prove that there exists $c\geq 1,0<\rho<1$ such that $\|A^k\|\leq c\rho^k$, $\forall k=0,\dots,K+1$.

Let $z_1,\dots,z_p$ be the roots of the equation $1-a_1z-a_2z^2-\dots-a_pz^p=0$ and $\rho_A$ be the spectral radius of the matrix $A$. Then, by Assumption \ref{asm2}, $|z_l|>1$ holds for any $l=1,\dots,p$. Note that $\det(zI_p-A)=z^p-a_1z^{p-1}-a_2z^{p-2}-\dots-a_p$. Then, $\frac{1}{z_1},\dots,\frac{1}{z_p}$ are the roots of the equation $z^p-a_1z^{p-1}-a_2z^{p-2}-\dots-a_p=0$, and thus, $\frac{1}{z_1},\dots,\frac{1}{z_p}$ are the eigenvalues of the matrix $A$ and $|\frac{1}{z_l}|<1$ holds for any $l=1,\dots,p$. Hence, we have $\rho_A<1$. 

By Gelfand's formula (Corollary 5.6.16 in \cite{horn2012matrix}), $\lim_{k\to\infty}\|A^k\|^{\frac{1}{k}}=\rho_A$. Then, there exists $\rho_A<\rho<1$ and $r\in\mathbb{N}_+$ such that for any $k=r+1,r+2,\dots$, $\|A^k\|\leq \rho^k$. Let $c=\max\{1,\frac{\|A\|}{\rho},\dots,\frac{\|A^r\|}{\rho^r}\}$. Then, for any $k=0,1,\dots$, $\|A^k\|\leq c\rho^k$. 

\textbf{Step 2}. In this step, we prove that Lemma \ref{lemma 2}.

By \eqref{eq arx2} we have
\begin{align}\label{eqx4}
	Y_{i,k+1}=&AY_{i,k}+BU_{i,k}\cr
	\noalign{\vskip -3pt}
	=&A^{k+1}Y_{i,0}+\sum_{l=0}^k A^l B U_{i,k-l}.
\end{align}
By Assumptions \ref{asm3} and \ref{asm4}, $\|Y_{i,0}\|\leq \sqrt{p}L,\|U_{i,k}\|\leq\sqrt{q+1}L$ holds for any $i=1,\dots,n,k=0,1,\dots$. Then, taking the $2$-norm of \eqref{eqx4} implies
\begin{align}\label{eqx6}
	\hspace{-1.5em}\|Y_{i,k+1}\|=&\|A^{k+1}Y_{i,0}+\sum_{l=0}^k A^l B U_{i,k-l}\|\cr
	\noalign{\vskip -3pt}
	\leq&\|A^{k+1}\|\sqrt{p}L+\sum_{l=0}^k\sqrt{q+1}\|A^l\|\|B\|L.
\end{align}
By Step 1, $\|A^k\|\leq c\rho^k$ holds for any $k=0,1,\dots$. Then, \eqref{eqx6} can be rewritten as
\begin{align}\label{eqx6.5}
	\|Y_{i,k+1}\|\leq& c\sqrt{p}\rho^{k+1}L+\sum_{l=0}^kc\sqrt{q+1}\rho^l\|B\|L\notag\\
	\leq& c\sqrt{p}\rho^{k+1}L+\frac{c\sqrt{q+1}\|B\|L}{1-\rho}.
\end{align}
Since $0<\rho<1$ and $\|B\|=\sqrt{1+\sum_{l=1}^q b_l^2}$, \eqref{eqx6.5} can be rewritten as
\begin{align*}
	\smash{\|Y_{i,k+1}\|\leq c\sqrt{p}\rho L+\frac{c\sqrt{(q+1)(1+\sum_{l=1}^q b_l^2)}L}{1-\rho}.}
\end{align*}
Let $G_1=c\sqrt{p}\rho L+\frac{c\sqrt{(q+1)(1+\sum_{l=1}^q b_l^2)}L}{1-\rho}$. Then, by $0<\rho<1$, $c\geq 1$, we have $\|Y_{i,0}\|\leq L\leq G_1$. Thus, for any $i=1,\dots,n,k=0,1,\dots$, $\|Y_{i,k}\|\leq G_1$, and hence, we have
\begin{align}\label{eqx7}
	\hspace{-1em}|y_{i,k}|\leq\|\varphi_{i,k}\|_\infty\leq\max\{\|Y_{i,k}\|_\infty,\|U_{i,k}\|_\infty\}\leq G_1.
\end{align}
Therefore, this lemma is proved. $\hfill\blacksquare$

\section{Proof of Lemma \ref{lemma 3}}\label{appendix d}
The proof proceeds in six steps.

{\bf Step 1}. First, we give an upper bound of the re-encryption error and the multiplication error.

For any $\mathsf{a}\in R_{\mathfrak{q}_K}$, let $\|\mathsf{a}\|^{\textsf{can}}=\|\textsf{CRT}(\mathsf{a})\|$ be the canonical embedding $2$-norm. Then, taking the canonical embedding $2$-norm of $\textsf{re}_{i,k}$ gives
\begin{align}\label{l5 2}
	\|\textsf{re}_{i,k}\|^{\textsf{can}}=&\|\textsf{qe}_{i,k}^{\textsf{re}(1)}+\textsf{qe}_{i,k}^{\textsf{re}(2)}\textsf{s}_{i_0}+\mathfrak{p}^{-1}\textsf{ct}_{i,k}^{(2)}\textsf{e}_{i\rightarrow i_0}\|^{\textsf{can}}\notag\\
	\leq&\|\textsf{qe}_{i,k}^{\textsf{re}(1)}\|^{\textsf{can}}+\|\textsf{qe}_{i,k}^{\textsf{re}(2)}\|^{\textsf{can}}\|\textsf{s}_{i_0}\|^{\textsf{can}}\notag\\
	&+\|\mathfrak{p}_K^{-1}\textsf{ct}_{i,k}^{(2)}\textsf{e}_{i\rightarrow i_0}\|^{\textsf{can}}.
\end{align}
Note that $\textsf{s}_{i_0}$$\sim$$\hwt$. Then, we have\vspace{-0.7em}
\begin{align}\label{l5 3}
	\hspace{-1em}\|\textsf{s}_{i_0}\|^{\textsf{can}}=\|\textsf{CRT}(\textsf{s}_{i_0})\|\leq\sqrt{Nh}.
\end{align}
{\vskip -15pt}\noindent Since $\textsf{v}_{i\rightarrow i_0}\sim\zo$, $\textsf{e}^\prime,\textsf{e}_{1,i\rightarrow i_0},\textsf{e}_{2,i\rightarrow i_0}\sim\tdg$, it can be seen that\vspace{-0.7em}
\begin{align}\label{l5 4.2}
	&\|\textsf{v}_{i\rightarrow i_0}\|^{\textsf{can}}\leq N,\|\textsf{e}^\prime\|^{\textsf{can}}\leq N\sqrt{\Gamma},\cr
	&\|\textsf{e}_{1,i\rightarrow i_0}\|^{\textsf{can}}\leq N\sqrt{\Gamma},\|\textsf{e}_{2,i\rightarrow i_0}\|^{\textsf{can}}\leq N\sqrt{\Gamma}.
\end{align}
{\vskip -15pt}\noindent Since $\textsf{e}_{i\rightarrow i_0}=\textsf{v}_{i\rightarrow i_0}\textsf{e}^\prime+\textsf{e}_{1,i\rightarrow i_0}+\textsf{e}_{2,i\rightarrow i_0}\textsf{s}_{i_0}$, by \eqref{l5 3} and \eqref{l5 4.2}, we have\vspace{-0.7em}
\begin{align}\label{l5 4.1}
	\|\textsf{e}_{i\rightarrow i_0}\|^{\textsf{can}}\leq(N+1+\sqrt{Nh})N\sqrt{\Gamma}.
\end{align}
{\vskip -15pt}\noindent Since $\textsf{ct}_{i,k}^{(2)}$$\in$$R_{\mathfrak{q}_K}$, by \eqref{l5 4.1} we have\vspace{-0.7em}
\begin{align}\label{l5 5}
	\hspace{-1em}\|\mathfrak{p}_K^{-1}\textsf{ct}_{i,k}^{(2)}\textsf{e}_{i\rightarrow i_0}\|^{\textsf{can}}\leq\frac{\sqrt{2\Gamma\mathfrak{q}_K}N^2(N+1+\sqrt{Nh})}{2\mathfrak{p}_K}.
\end{align}
{\vskip -15pt}\noindent By the probabilistic quantizer \eqref{pq}, we have\vspace{-0.7em}
\begin{align}\label{l5 4}
	&\|\textsf{qe}_{i,k}^{\textsf{re}(1)}\|^{\textsf{can}}\leq \sqrt{N},\|\textsf{qe}_{i,k}^{\textsf{re}(2)}\|^{\textsf{can}}\leq \sqrt{N}.
\end{align}
{\vskip -15pt}\noindent Substituting \eqref{l5 3}, \eqref{l5 5}, and \eqref{l5 4} into \eqref{l5 2} implies\vspace{-0.7em}
\begin{align}\label{l5 6}
	\|\textsf{re}_{i,k}\|^{\textsf{can}}\leq&\sqrt{N}(1+\sqrt{Nh})\cr
	&+\frac{\sqrt{2\Gamma\mathfrak{q}_K}N^2(N+1+\sqrt{Nh})}{2\mathfrak{p}_K}.
\end{align}

{\vskip -15pt}Note that for any $\mathsf{a}\in R_{\mathfrak{q}_K}$, we have
\begin{align}\label{crt}
	\|\textsf{Dcd}_{\Delta_K}(\textsf{a})\|=\|\Upsilon(\textsf{CRT}(\Delta_K^{-1}\textsf{a}))\|=\frac{\sqrt{2}\|\textsf{a}\|^{\textsf{can}}}{2\Delta_K}.
\end{align}
Then, combining \eqref{l5 6} and \eqref{crt} gives
\begin{align}\label{l5 6.1}
	&\|\textsf{Dcd}_{\Delta_K}(\textsf{re}_{i,k})\|\cr
	\leq&\frac{\sqrt{2N}(1\!+\!\sqrt{Nh})}{2\Delta_K}\!+\!\frac{\sqrt{\Gamma\mathfrak{q}_K}N^2(N\!+\!1\!+\!\sqrt{Nh})}{2\mathfrak{p}_K\Delta_K}.
\end{align}
By \eqref{l5 6} and \eqref{l5 6.1}, we have
\begin{align}\label{l5 7}
	&\|\sum_{i\neq i_0}\textsf{Dcd}_{\Delta_K}(\textsf{re}_{i,k})\|\cr
	\hspace{-1em}\leq&\frac{\sqrt{2N}n(1\!+\!\sqrt{Nh})}{2\Delta_K}\!+\!\frac{\sqrt{\Gamma\mathfrak{q}_K}nN^2(N\!+\!1\!+\!\sqrt{Nh})}{2\mathfrak{p}_K\Delta_K}.
\end{align}

By \eqref{crt}, taking the $2$-norm of $\sum_{i=1}^n$ $\textsf{Dcd}_{\Delta_K}(\textsf{me}_{i,k})$ gives\vspace{-1em}
\begin{align}\label{l5 8}
	&\|\sum_{i=1}^n\textsf{Dcd}_{\Delta_K}(\textsf{me}_{i,k})\|\cr
	\leq&\frac{\sqrt{2}}{2\Delta_K}\sum_{i=1}^n\|\mathfrak{p}_K^{-1}\textsf{ct}_{i,k}^{\varphi(2)}\textsf{ct}_{i,k}^{\text{res}(2)}\textsf{e}^{\prime\prime}_i\|^{\textsf{can}}\cr
	&+\frac{\sqrt{2}}{2\Delta_K}\sum_{i=1}^n\|\langle Q(\mathfrak{p}_K^{-1}\textsf{ct}_{i,k}^{\varphi(2)}\textsf{ct}_{i,k}^{\text{res}(2)}\!\cdot\!\textsf{evk}_i)\cr
	&-\mathfrak{p}_K^{-1}\textsf{ct}_{i,k}^{\varphi(2)}\textsf{ct}_{i,k}^{\text{res}(2)}\!\cdot\!\textsf{evk}_i,\textsf{sk}_i\rangle\|^{\textsf{can}}.
\end{align}
By the probabilistic quantizer \eqref{pq}, we have\vspace{-0.5em}
\begin{align}\label{l5 9}
	&\hspace{-1em}\|\langle Q(\mathfrak{p}_K^{-1}\textsf{ct}_{i,k}^{\varphi(2)}\textsf{ct}_{i,k}^{\text{res}(2)}\!\hspace{-0.5em}\cdot\!\textsf{evk}_i)\notag\\
	&\hspace{-1em}-\mathfrak{p}_K^{-1}\textsf{ct}_{i,k}^{\varphi(2)}\textsf{ct}_{i,k}^{\text{res}(2)}\hspace{-0.5em}\!\cdot\!\textsf{evk}_i,\textsf{sk}_i\rangle\|^{\textsf{can}}\!\leq\!\sqrt{N}(1\!+\!\sqrt{Nh}).
\end{align} 
Since $\textsf{ct}_{i,k}^{\varphi(2)}$$\textsf{ct}_{i,k}^{\text{res}(2)}$$\in$$R_{\mathfrak{q}_K}$ and $\textsf{e}^{\prime\prime}_i$$\sim$$\tdg$, we have\vspace{-0.5em}
\begin{align}\label{l5 10}
	\|\mathfrak{p}_K^{-1}\textsf{ct}_{i,k}^{\varphi(2)}\textsf{ct}_{i,k}^{\text{res}(2)}\textsf{e}^{\prime\prime}_i\|^{\textsf{can}}\leq\frac{\sqrt{2\Gamma\mathfrak{q}_K}N^2}{2\mathfrak{p}_K}.
\end{align}
Substituting \eqref{l5 9} and \eqref{l5 10} into \eqref{l5 8} implies
\begin{align}\label{l5 11}
	&\|\sum_{i=1}^n\textsf{Dcd}_{\Delta_K}(\textsf{me}_{i,k})\|\cr
	\leq&\frac{\sqrt{2N}n(1\!+\!\sqrt{Nh})}{2\Delta_K}\!+\!\frac{\sqrt{\Gamma\mathfrak{q}_K}nN^2}{2\mathfrak{p}_K\Delta_K}.
\end{align}

{\vskip -20pt}\indent{\bf Step 2}. In this step, we prove that $\|S_k\|\leq G_{3,K}, \forall\ k=0,\dots,K$.

Taking the $2$-norm of $S_k$ gives\vspace{-0.7em}
\begin{align}\label{l5 12}
	&\|S_k\|\notag\\
	\hspace{-0.5em}\leq&\sum_{i=1}^n\|\varphi_{i,k}\|(\|y_{i,k+1}\|+\frac{\|\textsf{Dcd}_{\Delta_K}(\textsf{ee}_{i,k+1}^y)\|+\|\textsf{qe}_{i,k+1}^y\|}{\Delta_K})\notag\\
	&+\sum_{i=1}^n\frac{\|\textsf{Dcd}_{\Delta_K}(\textsf{ee}_{i,k}^\varphi)\|+\|\textsf{qe}_{i,k}^\varphi\|}{\Delta_K}(\|y_{i,k+1}\|+\|\textsf{qe}_{i,k+1}^y\|\notag\\
	&+\|\textsf{Dcd}_{\Delta_K}(\textsf{ee}_{i,k+1}^y)\|)+\sum_{i=1}^n\|\textsf{Dcd}_{\Delta_K}(Q(\frac{\textsf{ct}_{i,k}}{\mathfrak{q}_{2,K}})-\frac{\textsf{ct}_{i,k}}{\mathfrak{q}_{2,K}})\|\notag\\
	&+\sum_{i=1}^n\|\textsf{Dcd}_{\Delta_K}(\textsf{me}_{i,k})\|+\sum_{i\neq i_0}\|\textsf{Dcd}_{\Delta_K}(\textsf{re}_{i,k})\|.
\end{align}
{\vskip -10pt}\noindent Similar to \eqref{l5 4.1}, we have\vspace{-0.7em}
\begin{align}\label{l5 13}
	\|\textsf{ee}_{i,k}^\varphi\|^{\textsf{can}}\leq(N+1+\sqrt{Nh})N\sqrt{\Gamma}.
\end{align}
{\vskip -10pt}\noindent Thus, combining \eqref{crt} and \eqref{l5 13} gives\vspace{-0.7em}
\begin{align}\label{l5 17}
	\|\textsf{Dcd}_{\Delta_K}(\textsf{ee}_{i,k}^\varphi)\|\leq&\frac{\sqrt{2}(N+1+\sqrt{Nh})N\sqrt{\Gamma}}{2\Delta_K}.
\end{align}
{\vskip -20pt}\noindent Similar to \eqref{l5 17}, we have\vspace{-0.7em}
\begin{align}\label{l5 18}
	\hspace{-1em}\|\textsf{Dcd}_{\Delta_K}(\textsf{ee}_{i,k+1}^y)\|\leq \frac{\sqrt{2}(N+1+\sqrt{Nh})N\sqrt{\Gamma}}{2\Delta_K}.
\end{align}

{\vskip -10pt}\indent By the probabilistic quantizer \eqref{pq}, it can be seen that\vspace{-0.7em}
\begin{align}
	&\hspace{-0.5em}\|Q(\frac{\textsf{ct}_{i,k}}{\mathfrak{q}_{2,K}})-\frac{\textsf{ct}_{i,k}}{\mathfrak{q}_{2,K}}\|^{\textsf{can}}\leq\sqrt{N},\label{l5 20.1}\\
	&\hspace{-0.5em}\|\textsf{qe}_{i,k}^\varphi\|\!=\!\|\frac{Q(\Delta_K\!\cdot\!\varphi_{i,k})\!-\!\Delta_K\varphi_{i,k}}{\Delta_K}\|\!\leq\!\frac{\sqrt{p\!+\!q}}{\Delta_K},\label{l5 19}\\
	&\hspace{-0.5em}\|\textsf{qe}_{i,k+1}^y\|\!=\!\|\frac{Q(\Delta_K\!\cdot\! y_{i,k+1})\!-\!\Delta_K y_{i,k+1}}{\Delta_K}\|\!\leq\!\frac{1}{\Delta_K}.\label{l5 20}
\end{align}
{\vskip -15pt}\noindent Then, by \eqref{l5 20.1}, we have\vspace{-0.7em}
\begin{align}\label{l5 20.2}
	\sum_{i=1}^n\|\textsf{Dcd}_{\Delta_K}(Q(\frac{\textsf{ct}_{i,k}}{\mathfrak{q}_{2,K}})-\frac{\textsf{ct}_{i,k}}{\mathfrak{q}_{2,K}})\|\leq\frac{\sqrt{2N}n}{2\Delta_K}.
\end{align}
{\vskip -15pt}\noindent For any $k=0,\dots,K$, substituting \eqref{l5 7}, \eqref{l5 11}, \eqref{l5 17}, \eqref{l5 19}-\eqref{l5 20.2} into \eqref{l5 12} implies\vspace{-0.7em}
\begin{align}\label{l5 21}
	\|S_k\|\leq&nG_1^2+nG_1(\frac{\sqrt{2N}n(1+\sqrt{Nh})}{2\Delta_K^2}+\frac{\sqrt{p+q}}{\Delta_K^2})\notag\\
	&+nG_1(\frac{\sqrt{2N}n(1+\sqrt{Nh})}{2\Delta_K^2}+\frac{1}{\Delta_K^2})\notag\\
	&+\frac{n}{\Delta_K}(\frac{\sqrt{2N}n(1+\sqrt{Nh})}{2\Delta_K}+\frac{1}{\Delta_K})\cdot\notag\\
	&(\frac{\sqrt{2N}n(1+\sqrt{Nh})}{2\Delta_K}+\frac{\sqrt{p+q}}{\Delta_K})+\frac{\sqrt{2N}n}{2\Delta_K}\notag\\
	&+\frac{\sqrt{2N}n(1+\sqrt{Nh})}{\Delta_K}+\frac{\sqrt{\Gamma\mathfrak{q}_K}nN^2}{\mathfrak{p}_K\Delta_K}\notag\\
	=&G_{3,K}.
\end{align}

{\vskip -10pt}\indent {\bf Step 3}. In this step, we prove that $\|M_{i,k}\|\leq G_{4,K}, \forall\ k=0,\dots,K$. \vspace{-0.7em}

Similar to \eqref{l5 7}, we have\vspace{-0.7em}
\begin{align}\label{l5 23}
	\hspace{-1em}\|\textsf{Dcd}_{\Delta_K}(\breve{\textsf{re}}_{i,k})\|\leq\frac{\sqrt{2}N^{\frac{3}{2}}(1\!+\!\sqrt{Nh})}{4\Delta_K}\!+\!\frac{\sqrt{\Gamma\mathfrak{q}_K}N^3}{4\mathfrak{p}_K\Delta_K}.
\end{align}
{\vskip -13pt}\noindent Taking the $2$-norm of $M_{i,k}$ gives\vspace{-0.5em}
\begin{align}\label{l5 22}
	&\|M_{i,k}\|\notag\\
	=&\|\varphi_{i,k}(\frac{\textsf{qe}_{i,k}^\varphi\!+\!\textsf{Dcd}_{\Delta_K}(\textsf{ee}_{i,k}^\varphi\!+\!\breve{\textsf{re}}_{i,k})}{\Delta_K})^\top\notag\\
	&+\frac{\textsf{qe}_{i,k}^\varphi+\textsf{Dcd}_{\Delta_K}(\textsf{ee}_{i,k}^\varphi)}{\Delta_K}(\varphi_{i,k}\!+\!\textsf{qe}_{i,k}^\varphi\notag\\
	&+\textsf{Dcd}_{\Delta_K}\!(\textsf{ee}_{i,k}^\varphi\!+\!\breve{\textsf{re}}_{i,k}))^\top\|\notag\\
	\leq&\frac{\|\varphi_{i,k}\|}{\Delta_K}(\|\textsf{qe}_{i,k}^\varphi\|+\|\textsf{Dcd}_{\Delta_K}(\textsf{ee}_{i,k}^\varphi)\|+\|\textsf{Dcd}_{\Delta_K}(\breve{\textsf{re}}_{i,k})\|)\notag\\
	&+\frac{\|\textsf{qe}_{i,k}^\varphi\|+\|\textsf{Dcd}_{\Delta_K}(\textsf{ee}_{i,k}^\varphi)\|}{\Delta_K}(\|\varphi_{i,k}\|+\|\textsf{qe}_{i,k}^\varphi\|\notag\\
	&+\|\textsf{Dcd}_{\Delta_K}(\textsf{ee}_{i,k}^\varphi)\|+\|\textsf{Dcd}_{\Delta_K}(\breve{\textsf{re}}_{i,k})\|).
\end{align}
{\vskip -13pt}\noindent For any $k=0,\dots,K$, substituting \eqref{l5 17}-\eqref{l5 20} and \eqref{l5 23} into \eqref{l5 22} implies\vspace{-0.3em}
\begin{align}\label{l5 24}
	\|M_{i,k}\|\leq&\frac{G_1}{\Delta_K}(\frac{\sqrt{p+q}}{\Delta_K}+\frac{\sqrt{2}(N+1+\sqrt{Nh})N\Gamma}{2\Delta_K}\notag\\
	&+\frac{\sqrt{2}N^{\frac{3}{2}}(1+\sqrt{Nh})}{4\Delta_K}+\frac{\sqrt{\Gamma\mathfrak{q}_K}N^3}{4\mathfrak{p}_K\Delta_K})\notag\\
	&+(\frac{\sqrt{p+q}}{\Delta_K^2}+\frac{\sqrt{2}(N+1+\sqrt{Nh})N\Gamma}{2\Delta_K^2})\cdot\notag\\
	&(G_1+\frac{\sqrt{p+q}}{\Delta_K}+\frac{\sqrt{2}(N+1+\sqrt{Nh})N\Gamma}{2\Delta_K}\notag\\
	&+\frac{\sqrt{2}N^{\frac{3}{2}}(1+\sqrt{Nh})}{4\Delta_K}+\frac{\sqrt{\Gamma\mathfrak{q}_K}N^3}{4\mathfrak{p}_K\Delta_K})\notag\\
	=&G_{4,K}.
\end{align}\vspace{-2em}

{\bf Step 4}. In this step, we prove that\vspace{-0.5em}
\begin{align}\label{l5 29}
	&\sum_{l=k}^{k+t-2}\!\alpha_K\|S_l\|\!\prod_{m=l+1}^{k+t-1}\|I_{p+q}\!-\!\alpha_K\!\sum_{i=1}^n\!(\varphi_{i,m}\varphi_{i,m}^\top\!\!+\!\!M_{i,m})\|\notag\\
	&+\!\alpha_K\|S_{k+t-1}\|\leq\frac{G_{3,K}((1+n\alpha_KG_{4,K})^{t}-1)}{nG_{4,K}},\notag\\
	&\forall\ k=0,\dots,K-t+1.
\end{align}

{\vskip -15pt}Since $\|S_k\|\leq G_{3,K}$, we have\vspace{-0.5em}
\begin{align}\label{l5 27}
	&\sum_{l=k}^{k+t-2}\!\alpha_K\|S_l\|\!\prod_{m=l+1}^{k+t-1}\|I_{p+q}\!-\!\alpha_K\!\sum_{i=1}^n\!(\varphi_{i,m}\varphi_{i,m}^\top\!\!+\!\!M_{i,m})\| \notag\\
	&+\!\alpha_K\|S_{k+t-1}\|\notag\\
	\leq&\sum_{l=k}^{k+t-2}\!\alpha_KG_{3,K}\!\prod_{m=l+1}^{k+t-1}\!\|(I_{p+q}\!-\!\alpha_K\sum_{i=1}^n\!(\varphi_{i,m}\varphi_{i,m}^\top\!+\!\!M_{i,m}))\|\notag\\
	&+\alpha_KG_{3,K}.
\end{align}
Note that by Lemma \ref{lemma 2}, we have $\|\alpha_K\sum_{i=1}^n\varphi_{i,k}\varphi_{i,k}^\top\|\leq n\alpha_KG_1^2$. Then, by $c_1$$<$$\frac{1}{ntG_1^2}$ in Assumption \ref{asm6}, $\|I_{p+q}$$-$$\alpha_K$ $\sum_{i=1}^n\varphi_{i,k}$ $\varphi_{i,k}^\top\|\leq 1$ holds for any $k=0,\dots,K$. Thus, by \eqref{l5 24}, it can be seen that
\begin{align}\label{l5 28}
	\hspace{-0.5em}\|I_{p+q}\!-\!\alpha_K\sum_{i=1}^n(\varphi_{i,k}\varphi_{i,k}^\top\!+\!M_{i,k})\|\!\leq\! 1\!+\!n\alpha_K G_{4,K}.
\end{align}
Then, substituting \eqref{l5 28} into \eqref{l5 27} proves \eqref{l5 29}.

{\bf Step 5}. In this step, we prove that
\begin{align}\label{l5 34.9}
	&\|\prod_{l=k}^{k+t-1}\!\!\!(I_{p+q}\!-\!\alpha_K\sum_{i=1}^n(\varphi_{i,l}\varphi_{i,l}^\top+M_{i,l}))\|\notag\\
	\leq& 1-\frac{t\varpi\alpha_K}{2},\forall\ k=0,\dots,K-t+1.
\end{align}

Note that
\begin{align}\label{l5 35}
	&\|\!\!\prod_{l=k}^{k+t-1}\!\!\!(I_{p+q}\!-\!\alpha_K\!\sum_{i=1}^n(\varphi_{i,l}\varphi_{i,l}^\top\!+\!M_{i,l}))\|\notag\\
	\leq&\|I_{p+q}\!-\!\sum_{l=k}^{k+t-1}\sum_{i=1}^n\alpha_K\varphi_{i,l}\varphi_{i,l}^\top\|\notag\\
	&+\!\!\sum_{m=2}^t\sum_{k\leq l_1<\dots< l_m\leq k+t-1}\hspace{-1em}\alpha_K^m\prod_{r=1}^m(\|\sum_{i=1}^n\varphi_{i,l_r}\varphi_{i,l_r}^\top\|\notag\\
	&+\|\sum_{i=1}^nM_{i,l_r}\|)+\alpha_K\sum_{l=k}^{k+t-1}\sum_{i=1}^n\|M_{i,l}\|.
\end{align}
Then, by \eqref{l5 24} and Lemma \ref{lemma 2}, \eqref{l5 35} can be rewritten as
\begin{align}\label{l5 36}
	&\|\!\!\prod_{l=k}^{k+t-1}\!\!\!(I_{p+q}\!-\!\alpha_K\!\sum_{i=1}^n(\varphi_{i,l}\varphi_{i,l}^\top\!+\!M_{i,l}))\|\cr
	\leq&\|I_{p+q}\!-\!\sum_{l=k}^{k+t-1}\sum_{i=1}^n\alpha_K\varphi_{i,l}\varphi_{i,l}^\top\|+(1+n\alpha_K(G_1^2+G_{4,K}))^t\cr
	&-1-nt(G_1^2+G_{4,K})\alpha_K+ntG_{4,K}\alpha_K.
\end{align}
By Taylor's formula (Corollary 5.3.2 in \cite{zorich2015analysis}), \eqref{l5 36} can be rewritten as
\begin{align}\label{l5 37}
	&\|\prod_{l=k}^{k+t-1}(I_{p+q}-\alpha_K\sum_{i=1}^n(\varphi_{i,l}\varphi_{i,l}^\top+M_{i,l}))\|\cr
	\leq&\|I_{p+q}-\sum_{l=k}^{k+t-1}\sum_{i=1}^n\alpha_K\varphi_{i,l}\varphi_{i,l}^\top\|+\frac{n^2t^2(G_1^2+G_{4,K})^2\alpha_K^2}{2}\cdot\cr
	&(1+n\alpha_K(G_1^2+G_{4,K}))^{t-2}+ntG_{4,K}\alpha_K.
\end{align}
Note that $c_1\leq \frac{1}{ntG_1^2}$ holds by Assumption \ref{asm6}. Then, \eqref{l5 37} can be rewritten as
\begin{align}\label{l5 37.1}
	&\|\prod_{l=k}^{k+t-1}(I_{p+q}-\alpha_K\sum_{i=1}^n(\varphi_{i,l}\varphi_{i,l}^\top+M_{i,l}))\|\cr
	\leq&\|I_{p+q}-\sum_{l=k}^{k+t-1}\sum_{i=1}^n\alpha_K\varphi_{i,l}\varphi_{i,l}^\top\|+ntG_{4,K}\alpha_K\cr
	&+\frac{n^2t^2(G_1^2\!+\!G_{4,K})^2\alpha_K^2}{2}(1\!+\!\frac{G_1^2\!+\!G_{4,K}}{tG_1^2})^{t-2}.
\end{align}

{\vskip -15pt}By Lemma \ref{lemma 2} and Assumption \ref{asm6}, we have $\|\sum_{l=k}^{k+t-1}$ $\sum_{i=1}^n\alpha_K\varphi_{i,l}\varphi_{i,l}^\top\|\leq\alpha_KntG_1^2\leq 1$. Then, by Assumption~\ref{asm5}, it can be seen that
\begin{align}\label{l5 37.2}
	\|I_{p+q}-\sum_{l=k}^{k+t-1}\sum_{i=1}^n\alpha_K\varphi_{i,l}\varphi_{i,l}^\top\|\leq 1-\varpi t\alpha_K.
\end{align}
Substituting \eqref{l5 37.2} into \eqref{l5 37.1} implies\vspace{-0.7em}
\begin{align}\label{l5 38}
	&\|\prod_{l=k}^{k+t-1}(I_{p+q}-\alpha_K\sum_{i=1}^n(\varphi_{i,l}\varphi_{i,l}^\top+M_{i,l}))\|\cr
	\leq&1-\varpi t\alpha_K+ntG_{4,K}\alpha_K\cr
	&+\frac{n^2t^2(G_1^2\!+\!G_{4,K})^2\alpha_K^2}{2}(1\!+\!\frac{G_1^2+G_{4,K}}{tG_1^2})^{t-2}.
\end{align}
{\vskip -15pt}\noindent Then, substituting $c_1\leq G_2$ in Assumption~\ref{asm6} into \eqref{l5 38} proves \eqref{l5 34.9}.

{\bf Step 6}. In this step, we prove that $\textsf{Dec}_{\textsf{sk}_{i_0}}(\textsf{ct}_k^{\hat{\theta}})\bmod \mathfrak{q}_K=\textsf{Dec}_{\textsf{sk}_{i_0}}(\textsf{ct}_k^{\hat{\theta}})$, $\forall\ k=0,\dots,K+1$.

{\vskip -5pt} When $k=1,\dots,t-1$, taking the $2$-norm of \eqref{l5 1} gives\vspace{-0.7em}
\begin{align}\label{l5 40}
	\|\hat{\theta}_k\|\leq&\|I_{p+q}-\alpha_K\sum_{i=1}^n(\varphi_{i,k-1}\varphi_{i,k-1}^\top+M_{i,k-1})\|\|\hat{\theta}_{k-1}\|\notag\\
	&+\alpha_K\|S_{k-1}\|.
\end{align}
{\vskip -15pt}\noindent By \eqref{l5 21} and \eqref{l5 24}, \eqref{l5 40} can be rewritten as\vspace{-0.7em}
\begin{align}\label{l5 40.1}
	\|\hat{\theta}_k\|\leq&(1+n\alpha_KG_{4,K})^k\|\hat{\theta}_0\|\cr
	&+\alpha_KG_{3,K}\sum_{l=0}^{k-1}(1+n\alpha_KG_{4,K})^l\cr
	=&(1+n\alpha_KG_{4,K})^k\|\hat{\theta}_0\|\cr
	&+\frac{G_{3,K}((1+n\alpha_KG_{4,K})^k-1)}{nG_{4,K}}.
\end{align}
{\vskip -15pt}\noindent Then, for any $k=0,\dots,t-1$, by \eqref{l5 40.1} we have\vspace{-0.7em}
\begin{align}\label{l5 40.2}
	\|\hat{\theta}_k\|\leq&(1+n\alpha_KG_{4,K})^{t-1}\|\hat{\theta}_0\|\cr
	&+\frac{G_{3,K}((1+n\alpha_KG_{4,K})^{t-1}-1)}{nG_{4,K}}.
\end{align}

{\vskip -15pt} When $k=t,\dots,K+1$, taking the $2$-norm of \eqref{l5 1} gives\vspace{-0.7em}
\begin{align}\label{l5 40.4}
	\|\hat{\theta}_k\|\leq&\|\prod_{l=k-t}^{k-1}(I_{p+q}-\alpha_K\sum_{i=1}^n(\varphi_{i,l}\varphi_{i,l}^\top+M_{i,l}))\|\|\hat{\theta}_{k-t}\|\notag\\
	\noalign{\vskip -4pt}
	&+\alpha_KG_{3,K}\sum_{l=0}^{t-1}(1+n\alpha_KG_{4,K})^l\notag\\
	\noalign{\vskip -4pt}
	=&\|\prod_{l=k-t}^{k}(I_{p+q}-\alpha_K\sum_{i=1}^n(\varphi_{i,l}\varphi_{i,l}^\top+M_{i,l}))\|\|\hat{\theta}_{k-t}\|\notag\\
	&+\frac{G_{3,K}((1+n\alpha_KG_{4,K})^t-1)}{nG_{4,K}}.
\end{align}
By \eqref{l5 34.9}, \eqref{l5 40.4} can be rewritten as
\begin{align}\label{l5 40.5}
	\|\hat{\theta}_k\|\leq&(1-\frac{t\varpi\alpha_K}{2})\|\hat{\theta}_{k-t}\|\notag\\
	&+\frac{G_{3,K}((1+n\alpha_KG_{4,K})^t-1)}{nG_{4,K}}.
\end{align}
Recursively computing \eqref{l5 40.5} gives\vspace{-0.7em}
\begin{align}\label{l5 40.6}
	&\|\hat{\theta}_k\|\notag\\
	\leq&(1-\frac{t\varpi\alpha_K}{2})^{\lfloor\frac{k}{t}\rfloor}\|\hat{\theta}_{k-t\lfloor\frac{k}{t}\rfloor}\|\notag\\
	&+\frac{G_{3,K}((1+n\alpha_KG_{4,K})^t\!-\!1)}{nG_{4,K}}\!\sum_{l=0}^{\lfloor\frac{k}{t}\rfloor-1}\!(1\!-\!\frac{t\varpi\alpha_K}{2})^l\notag\\
	\leq&(1-\frac{t\varpi\alpha_K}{2})^{\lfloor\frac{k}{t}\rfloor}\|\hat{\theta}_{k-t\lfloor\frac{k}{t}\rfloor}\|\notag\\
	&+\frac{2G_{3,K}((1+n\alpha_KG_{4,K})^t\!-\!1)}{nt\varpi\alpha_KG_{4,K}}.
\end{align}
Note that $0\leq k-t\lfloor\frac{k}{t}\rfloor\leq t-1$. Then, substituting \eqref{l5 40.2} into \eqref{l5 40.6} implies\vspace{-0.7em}
\begin{align}\label{l5 40.7}
	\|\hat{\theta}_k\|\leq&(1-\frac{t\varpi \alpha_K}{2})^{\lfloor\frac{k}{t}\rfloor}(1+n\alpha_KG_{4,K})^{t-1}\|\hat{\theta}_0\|\cr
	&+\frac{G_{3,K}}{nG_{4,K}}(1-\frac{t\varpi \alpha_K}{2})^{\lfloor\frac{k}{t}\rfloor}((1+n\alpha_KG_{4,K})^{t-1}-1)\cr
	&+\frac{2G_{3,K}}{nt\varpi\alpha_KG_{4,K}}((1+n\alpha_KG_{4,K})^t-1)\cr
	\leq&(1+n\alpha_KG_{4,K})^{t-1}\|\hat{\theta}_0\|\cr
	&+\frac{G_{3,K}((1+n\alpha_KG_{4,K})^{t-1}-1)}{nG_{4,K}}\cr
	&+\frac{2G_{3,K}((1+n\alpha_KG_{4,K})^t-1)}{nt\varpi\alpha_KG_{4,K}}.
\end{align}
Thus, substituting \eqref{condition 2} into \eqref{l5 40.7} implies \vspace{-0.7em}
\begin{align}\label{l5 41}
	\|\hat{\theta}_k\|\leq \frac{\sqrt{2N}(\mathfrak{q}_K-1)}{4\Delta_K}, \forall\ k=0,\dots,K+1.
\end{align}

{\vskip -12pt} Let $\tau$ be the mapping that maps $\textsf{a}\in R$ to the coefficient vector  $(\textsf{a}^{(1)},\dots,\textsf{a}^{(N)})\in\mathbb{Z}^N$. Then, we have\vspace{-0.5em}
\begin{align}\label{l5 46}
	\|\tau(\textsf{Ecd}_{\Delta_K}(\hat{\theta}_k))\|_\infty\leq\frac{\sqrt{2N}}{N}\|Q(\Delta_K\cdot\hat{\theta}_k)\|.
\end{align}
Note that $\hat{\theta}_k$ is obtained by decoding the plaintexts in $R_{\mathfrak{q}_K}$. Then, $Q(\Delta_K\hat{\theta}_k)=\Delta_K\hat{\theta}_k$. Thus, substituting \eqref{l5 41} into \eqref{l5 46} implies\vspace{-0.7em}
\begin{align}\label{l5 47}
	\|\tau(\textsf{Ecd}_{\Delta_K}(\hat{\theta}_k))\|_\infty\leq\frac{\sqrt{2N}\Delta_K}{N}\|\hat{\theta}_k\|\leq\frac{\mathfrak{q}_K-1}{2}.
\end{align}
By \eqref{l5 47}, each coefficient of the plaintext $\textsf{Ecd}_{\Delta_K}(\hat{\theta}_k)$ is in $[-\frac{\mathfrak{q}_K}{2},\frac{\mathfrak{q}_K}{2})$. Then, $\textsf{Dec}_{\textsf{sk}_{i_0}}(\textsf{ct}_k^{\hat{\theta}})\bmod \mathfrak{q}_K$$=$ $\textsf{Dec}_{\textsf{sk}_{i_0}}(\textsf{ct}_k^{\hat{\theta}})$. Thus, this lemma is proved. $\hfill\blacksquare$

\section{Proof of Theorem \ref{thm2}}\label{appendix e}
For convenience of analysis, define
\begin{align*}
	\mathcal{F}_k=&\sigma(\{\varphi_{i,l},y_{i,l},w_{i,l},\hat{\theta}_l|l=0,\dots,k,i=1,\dots,n\}),\cr
	\mathcal{H}_k=&\sigma(\mathcal{F}_k\cup\{\textsf{qe}_{i,k}^\varphi,\textsf{ee}_{i,k}^\varphi,\breve{\textsf{re}}_{i,k}|i=1,\dots,n\}).
\end{align*}
Then, the proof of Theorem \ref{thm2} is given in the following six steps.

{\bf Step 1}. First, we prove that\vspace{-0.7em}
\begin{align}\label{t2 0}
	&\E(\|\hat{\theta}_{k+1}-\theta\|^2|\mathcal{H}_k)\notag\\
	\leq&\|(I_{p+q}\!-\!\alpha_K\sum_{i=1}^n(\varphi_{i,k}\!+\!\frac{\textsf{Dcd}_{\Delta_K}(\textsf{ee}_{i,k}^\varphi)\!+\!\textsf{qe}_{i,k}^\varphi}{\Delta_K})\varphi_{i,k}^\top)(\hat{\theta}_k\!-\!\theta)\notag\\
	&-\alpha_K\sum_{i=1}^n\frac{\varphi_{i,k}}{\Delta_K}(\textsf{Dcd}_{\Delta_K}(\textsf{ee}_{i,k}^\varphi+\breve{\textsf{re}}_{i,k})+\textsf{qe}_{i,k}^\varphi)^\top\hat{\theta}_k\notag\\
	&-\alpha_K\sum_{i=1}^n\frac{\textsf{Dcd}_{\Delta_K}(\textsf{ee}_{i,k}^\varphi)\!+\!\textsf{qe}_{i,k}^\varphi}{\Delta_K}(\textsf{Dcd}_{\Delta_K}(\textsf{ee}_{i,k}^\varphi\!+\!\breve{\textsf{re}}_{i,k})\!\notag\\
	&+\textsf{qe}_{i,k}^\varphi)^\top\hat{\theta}_k\|^2+O(\alpha_K^2).
\end{align}

{\vskip -17pt}\indent Note that by Assumptions \ref{asm2}-\ref{asm6} and \eqref{condition 2}, Lemma \ref{lemma 3} holds. Then, $\hat{\theta}_k=\textsf{Dcd}_{\Delta_K}(\textsf{Dec}_{\textsf{sk}_{i_0}}(\textsf{ct}_k^{\hat{\theta}}))$ for any $k=0,\dots,K+1$. Thus, subtracting $\theta$ from both sides of \eqref{l5 1} and taking its squared $2$-norm leads to\vspace{-0.7em}
\begin{align}\label{t2 0.2}
	&\|\hat{\theta}_{k+1}-\theta\|^2\notag\\
	=&\|\hat{\theta}_k-\theta+\alpha_K\sum_{i=1}^n\varphi_{i,k}(y_{i,k+1}-\varphi_{i,k}^\top\hat{\theta}_k)\notag\\
	&+\alpha_K\sum_{i=1}^n\frac{\textsf{Dcd}_{\Delta_K}(\textsf{ee}_{i,k}^\varphi)+\textsf{qe}_{i,k}^\varphi}{\Delta_K}(y_{i,k+1}-\varphi_{i,k}^\top\hat{\theta}_k)\notag\\
	&+\alpha_K\sum_{i=1}^n\frac{\varphi_{i,k}}{\Delta_K}(\textsf{Dcd}_{\Delta_K}(\textsf{ee}_{i,k+1}^y)+\textsf{qe}_{i,k+1}^y\notag\\
	&-(\textsf{Dcd}_{\Delta_K}(\textsf{ee}_{i,k}^\varphi+\breve{\textsf{re}}_{i,k})+\textsf{qe}_{i,k}^\varphi)^\top\hat{\theta}_k)\notag\\
	&+\alpha_K\sum_{i=1}^n\frac{\textsf{Dcd}_{\Delta_K}(\textsf{ee}_{i,k}^\varphi)+\textsf{qe}_{i,k}^\varphi}{\Delta_K}(\textsf{Dcd}_{\Delta_K}(\textsf{ee}_{i,k\!+\!1}^y)\!+\!\textsf{qe}_{i,k\!+\!1}^y\notag\\
	&-(\textsf{Dcd}_{\Delta_K}(\textsf{ee}_{i,k}^\varphi+\breve{\textsf{re}}_{i,k})+\textsf{qe}_{i,k}^\varphi)^\top\hat{\theta}_k)\notag\\
	&+\alpha_K\!\sum_{i=1}^n\textsf{Dcd}_{\Delta_K}(Q(\frac{\textsf{ct}_{i,k}}{\mathfrak{q}_{2,K}})\!-\!\frac{\textsf{ct}_{i,k}}{\mathfrak{q}_{2,K}})\notag\\
	&+\alpha_K\!\sum_{i\neq i_0}\textsf{Dcd}_{\Delta_K}(\textsf{re}_{i,k})\!+\!\alpha_K\sum_{i=1}^n\textsf{Dcd}_{\Delta_K}(\textsf{me}_{i,k})\|^2\!\!.
	\raisetag{23pt}
\end{align}
{\vskip -7pt}\noindent Substituting \eqref{eq arx2} into \eqref{t2 0.2} implies\vspace{-0.7em}
\begin{align*}
	&\|\hat{\theta}_{k+1}-\theta\|^2\cr
	=&\|(I_{p+q}-\alpha_K\sum_{i=1}^n\varphi_{i,k}\varphi_{i,k}^\top)(\hat{\theta}_k-\theta)\cr
	&-\alpha_K\sum_{i=1}^n\frac{\textsf{Dcd}_{\Delta_K}(\textsf{ee}_{i,k}^\varphi)+\textsf{qe}_{i,k}^\varphi}{\Delta_K}\varphi_{i,k}^\top(\hat{\theta}_k-\theta)\cr
	&+\alpha_K\sum_{i=1}^n(\varphi_{i,k}+\frac{\textsf{Dcd}_{\Delta_K}(\textsf{ee}_{i,k}^\varphi)+\textsf{qe}_{i,k}^\varphi}{\Delta_K})w_{i,k+1}\cr
	&+\alpha_K\sum_{i=1}^n\frac{\varphi_{i,k}}{\Delta_K}(\textsf{Dcd}_{\Delta_K}(\textsf{ee}_{i,k+1}^y)+\textsf{qe}_{i,k+1}^y
\end{align*}
\begin{align}\label{t2 1}
	&-(\textsf{Dcd}_{\Delta_K}(\textsf{ee}_{i,k}^\varphi+\breve{\textsf{re}}_{i,k})+\textsf{qe}_{i,k}^\varphi)^\top\hat{\theta}_k)\cr
	&+\alpha_K\sum_{i=1}^n\frac{\textsf{Dcd}_{\Delta_K}(\textsf{ee}_{i,k}^\varphi)+\textsf{qe}_{i,k}^\varphi}{\Delta_K}(\textsf{Dcd}_{\Delta_K}(\textsf{ee}_{i,k\!+\!1}^y)\!+\!\textsf{qe}_{i,k\!+\!1}^y\cr
	&-(\textsf{Dcd}_{\Delta_K}(\textsf{ee}_{i,k}^\varphi+\breve{\textsf{re}}_{i,k})+\textsf{qe}_{i,k}^\varphi)^\top\hat{\theta}_k)\cr
	&+\!\alpha_K\!\sum_{i=1}^n\textsf{Dcd}_{\Delta_K}(Q(\frac{\textsf{ct}_{i,k}}{\mathfrak{q}_{2,K}})\!-\!\frac{\textsf{ct}_{i,k}}{\mathfrak{q}_{2,K}})\cr
	\noalign{\vskip -3pt}
	&+\!\alpha_K\!\sum_{i\neq i_0}\textsf{Dcd}_{\Delta_K}(\textsf{re}_{i,k})\!+\!\alpha_K\sum_{i=1}^n\textsf{Dcd}_{\Delta_K}(\textsf{me}_{i,k})\|^2.
\end{align}
{\vskip -15pt}\noindent Taking the conditional mathematical expectation on \eqref{t2 1} with respect to $\mathcal{H}_k$ implies\vspace{-1em}
\begin{align}\label{t2 2}
	&\E(\|\hat{\theta}_{i,k+1}-\theta\|^2|\mathcal{H}_k)\notag\\
	=&\|(I_{p+q}\!-\!\alpha_K\sum_{i=1}^n(\varphi_{i,k}\!+\!\frac{\textsf{Dcd}_{\Delta_K}(\textsf{ee}_{i,k}^\varphi)\!+\!\textsf{qe}_{i,k}^\varphi}{\Delta_K})\varphi_{i,k}^\top)(\hat{\theta}_k\!-\!\theta)\notag\\
	&-\alpha_K\sum_{i=1}^n\frac{\varphi_{i,k}}{\Delta_K}(\textsf{Dcd}_{\Delta_K}(\textsf{ee}_{i,k}^\varphi+\breve{\textsf{re}}_{i,k})+\textsf{qe}_{i,k}^\varphi)^\top\hat{\theta}_k\notag\\
	&-\alpha_K\sum_{i=1}^n\frac{\textsf{Dcd}_{\Delta_K}(\textsf{ee}_{i,k}^\varphi)\!+\!\textsf{qe}_{i,k}^\varphi}{\Delta_K}(\textsf{Dcd}_{\Delta_K}(\textsf{ee}_{i,k}^\varphi\!+\!\breve{\textsf{re}}_{i,k})\notag\\
	&+\textsf{qe}_{i,k}^\varphi)^\top\hat{\theta}_k\|^2+\alpha_K^2\E(\|\sum_{i=1}^n\frac{\varphi_{i,k}}{\Delta_K}(\textsf{Dcd}_{\Delta_K}(\textsf{ee}_{i,k+1}^y)\notag\\
	&+\!\textsf{qe}_{i,k+1}^y)+\sum_{i=1}^n(\varphi_{i,k}\!+\!\frac{\textsf{Dcd}_{\Delta_K}(\textsf{ee}_{i,k}^\varphi)\!+\!\textsf{qe}_{i,k}^\varphi}{\Delta_K})w_{i,k+1}\notag\\
	&+\sum_{i=1}^n\textsf{Dcd}_{\Delta_K}(Q(\frac{\textsf{ct}_{i,k}}{\mathfrak{q}_{2,K}})\!-\!\frac{\textsf{ct}_{i,k}}{\mathfrak{q}_{2,K}})\!+\!\sum_{i\neq i_0}\textsf{Dcd}_{\Delta_K}(\textsf{re}_{i,k})\notag\\
	&\!+\!\sum_{i=1}^n\textsf{Dcd}_{\Delta_K}(\textsf{me}_{i,k})\|^2|\mathcal{H}_k)+2\alpha_K\langle(I_{p+q}\!-\!\alpha_K\sum_{i=1}^n(\varphi_{i,k}\notag\\
	&\!+\!\frac{\textsf{Dcd}_{\Delta_K}(\textsf{ee}_{i,k}^\varphi)\!+\!\textsf{qe}_{i,k}^\varphi}{\Delta_K})\varphi_{i,k}^\top)(\hat{\theta}_k\!-\!\theta)\notag\\
	&-\alpha_K\sum_{i=1}^n\frac{\varphi_{i,k}}{\Delta_K}(\textsf{Dcd}_{\Delta_K}(\textsf{ee}_{i,k}^\varphi+\breve{\textsf{re}}_{i,k})+\textsf{qe}_{i,k}^\varphi)^\top\hat{\theta}_k\notag\\
	&-\alpha_K\sum_{i=1}^n\frac{\textsf{Dcd}_{\Delta_K}(\textsf{ee}_{i,k}^\varphi)\!+\!\textsf{qe}_{i,k}^\varphi}{\Delta_K}(\textsf{Dcd}_{\Delta_K}(\textsf{ee}_{i,k}^\varphi\!+\!\breve{\textsf{re}}_{i,k})\!\notag\\
	&+\textsf{qe}_{i,k}^\varphi)^\top\hat{\theta}_k,\E(\sum_{i=1}^n\frac{\varphi_{i,k}}{\Delta_K}(\textsf{Dcd}_{\Delta_K}(\textsf{ee}_{i,k+1}^y)\!+\!\textsf{qe}_{i,k+1}^y)\notag\\
	&+\sum_{i=1}^n(\varphi_{i,k}\!+\!\frac{\textsf{Dcd}_{\Delta_K}(\textsf{ee}_{i,k}^\varphi)\!+\!\textsf{qe}_{i,k}^\varphi}{\Delta_K})w_{i,k+1}\notag\\
	&+\sum_{i=1}^n\textsf{Dcd}_{\Delta_K}(Q(\frac{\textsf{ct}_{i,k}}{\mathfrak{q}_{2,K}})-\frac{\textsf{ct}_{i,k}}{\mathfrak{q}_{2,K}})+\sum_{i\neq i_0}\textsf{Dcd}_{\Delta_K}(\textsf{re}_{i,k})\notag\\
	&+\sum_{i=1}^n\textsf{Dcd}_{\Delta_K}(\textsf{me}_{i,k})|\mathcal{H}_k)\rangle.
\end{align}
{\vskip -20pt}\noindent Note that by Assumption \ref{asm4}, $w_{i,k+1}$ is independent of $\mathcal{H}_k$ for any $i=1,\dots,n,k=0,\dots,K$. Then, it can be seen that\vspace{-1em}
\begin{align}\label{t2 3}
	\E(w_{i,k+1}|\mathcal{H}_k)=\E w_{i,k+1}=0.
\end{align}
{\vskip -20pt}\noindent Moreover, note that $\textsf{Dcd}_{\Delta_K}(\textsf{ee}_{i,k+1}^y)$, $\textsf{qe}_{i,k+1}^y$, $\textsf{Dcd}_{\Delta_K}(Q($ $\frac{\textsf{ct}_{i,k}}{\mathfrak{q}_{2,K}})-$$\frac{\textsf{ct}_{i,k}}{\mathfrak{q}_{2,K}})$, $\textsf{Dcd}_{\Delta_K}(\textsf{me}_{i,k}),\textsf{Dcd}_{\Delta_K}(\textsf{re}_{i,k})$ are independent of $\mathcal{H}_k$. Then, we have\vspace{-1em}
\begin{align}\label{t2 4}
	&\E(\textsf{Dcd}_{\Delta_K}(Q(\frac{\textsf{ct}_{i,k}}{\mathfrak{q}_{2,K}})-\frac{\textsf{ct}_{i,k}}{\mathfrak{q}_{2,K}})|\mathcal{H}_k)=0,\notag\\
	&\E(\textsf{Dcd}_{\Delta_K}(\textsf{ee}_{i,k+1}^y)|\mathcal{H}_k)=0,\notag\\
	&\E(\textsf{Dcd}_{\Delta_K}(\textsf{me}_{i,k})|\mathcal{H}_k)=0,\notag\\
	&\E(\textsf{Dcd}_{\Delta_K}(\textsf{re}_{i,k})|\mathcal{H}_k)=0,\notag\\
	&\E(\textsf{qe}_{i,k+1}^y|\mathcal{H}_k)=0.
\end{align}
{\vskip -17pt}\noindent Thus, by \eqref{t2 3} and \eqref{t2 4}, it can be seen that\vspace{-1em}
\begin{align}\label{t2 6}
	&2\alpha_K\langle(I_{p+q}\!-\!\alpha_K\sum_{i=1}^n(\varphi_{i,k}\!+\!\frac{\textsf{Dcd}_{\Delta_K}(\textsf{ee}_{i,k}^\varphi)\!+\!\textsf{qe}_{i,k}^\varphi}{\Delta_K})\varphi_{i,k}^\top)\cdot\notag\\
	&(\hat{\theta}_k\!-\!\theta)-\alpha_K\sum_{i=1}^n\frac{\varphi_{i,k}}{\Delta_K}(\textsf{Dcd}_{\Delta_K}(\textsf{ee}_{i,k}^\varphi+\breve{\textsf{re}}_{i,k})+\textsf{qe}_{i,k}^\varphi)^\top\hat{\theta}_k\notag\\
	&-\alpha_K\sum_{i=1}^n\frac{\textsf{Dcd}_{\Delta_K}(\textsf{ee}_{i,k}^\varphi)\!+\!\textsf{qe}_{i,k}^\varphi}{\Delta_K}(\textsf{Dcd}_{\Delta_K}(\textsf{ee}_{i,k}^\varphi\!+\!\breve{\textsf{re}}_{i,k})\!\notag\\
	&+\textsf{qe}_{i,k}^\varphi)^\top\hat{\theta}_k,\E(\sum_{i=1}^n\frac{\varphi_{i,k}}{\Delta_K}(\textsf{Dcd}_{\Delta_K}(\textsf{ee}_{i,k+1}^y)\!+\!\textsf{qe}_{i,k+1}^y)\notag\\
	&+\sum_{i=1}^n(\varphi_{i,k}\!+\!\frac{\textsf{Dcd}_{\Delta_K}(\textsf{ee}_{i,k}^\varphi)\!+\!\textsf{qe}_{i,k}^\varphi}{\Delta_K})w_{i,k+1}\notag\\
	&+\sum_{i=1}^n\textsf{Dcd}_{\Delta_K}(Q(\frac{\textsf{ct}_{i,k}}{\mathfrak{q}_{2,K}})-\frac{\textsf{ct}_{i,k}}{\mathfrak{q}_{2,K}})+\sum_{i\neq i_0}\textsf{Dcd}_{\Delta_K}(\textsf{re}_{i,k})\notag\\
	&+\sum_{i=1}^n\textsf{Dcd}_{\Delta_K}(\textsf{me}_{i,k})|\mathcal{H}_k)\rangle\notag\\
	=&0.
\end{align}
{\vskip -17pt}\noindent Substituting \eqref{t2 6} into \eqref{t2 2} implies\vspace{-0.7em}
\begin{align}\label{t2 7}
	&\E(\|\hat{\theta}_{i,k+1}-\theta\|^2|\mathcal{H}_k)\notag\\
	\hspace{-0.5em}=&\|(I_{p+q}\!-\!\alpha_K\sum_{i=1}^n(\varphi_{i,k}\!+\!\frac{\textsf{Dcd}_{\Delta_K}(\textsf{ee}_{i,k}^\varphi)\!+\!\textsf{qe}_{i,k}^\varphi}{\Delta_K})\varphi_{i,k}^\top)(\hat{\theta}_k\!-\!\theta)\notag\\
	&-\alpha_K\sum_{i=1}^n\frac{\varphi_{i,k}}{\Delta_K}(\textsf{Dcd}_{\Delta_K}(\textsf{ee}_{i,k}^\varphi+\breve{\textsf{re}}_{i,k})+\textsf{qe}_{i,k}^\varphi)^\top\hat{\theta}_k\notag\\
	&-\alpha_K\sum_{i=1}^n\frac{\textsf{Dcd}_{\Delta_K}(\textsf{ee}_{i,k}^\varphi)\!+\!\textsf{qe}_{i,k}^\varphi}{\Delta_K}(\textsf{Dcd}_{\Delta_K}(\textsf{ee}_{i,k}^\varphi\!+\!\breve{\textsf{re}}_{i,k})\notag\\
	&+\textsf{qe}_{i,k}^\varphi)^\top\hat{\theta}_k\|^2+\alpha_K^2\E(\|\sum_{i=1}^n\frac{\varphi_{i,k}}{\Delta_K}(\textsf{Dcd}_{\Delta_K}(\textsf{ee}_{i,k+1}^y)\notag\\
	&+\textsf{qe}_{i,k+1}^y)+\sum_{i=1}^n(\varphi_{i,k}\!+\!\frac{\textsf{Dcd}_{\Delta_K}(\textsf{ee}_{i,k}^\varphi)\!+\!\textsf{qe}_{i,k}^\varphi}{\Delta_K})w_{i,k+1}\notag\\
	\noalign{\vskip -4pt}
	&+\sum_{i=1}^n\textsf{Dcd}_{\Delta_K}(Q(\frac{\textsf{ct}_{i,k}}{\mathfrak{q}_{2,K}})\!-\!\frac{\textsf{ct}_{i,k}}{\mathfrak{q}_{2,K}})\notag\\
	\noalign{\vskip -4pt}
	&+\sum_{i\neq i_0}\textsf{Dcd}_{\Delta_K}(\textsf{re}_{i,k})\!+\!\sum_{i=1}^n\textsf{Dcd}_{\Delta_K}(\textsf{me}_{i,k})\|^2|\mathcal{H}_k).
\end{align}

{\vskip -9pt}\indent Note that for any $\mathbf{a}_1$, $\mathbf{a}_2$, $\dots$, $\mathbf{a}_m\in\mathbb{R}^{p+q}$, the following inequality holds:\vspace{-0.7em}
\begin{align}\label{t2 8}
	\smash{\|\sum_{i=1}^{m}\mathbf{a}_i\|^2\leq m\sum_{i=1}^{m}\|\mathbf{a}_i\|^2.}
\end{align}
{\vskip -15pt}\noindent Moreover, $\textsf{Dcd}_{\Delta_K}(\textsf{ee}_{i,k+1}^y),\textsf{qe}_{i,k+1}^y,\textsf{Dcd}_{\Delta_K}(Q(\frac{\textsf{ct}_{i,k}}{\mathfrak{q}_{2,K}})$$-$$\frac{\textsf{ct}_{i,k}}{\mathfrak{q}_{2,K}})$, $\textsf{Dcd}_{\Delta_K}(\textsf{me}_{i,k}),\textsf{Dcd}_{\Delta_K}(\textsf{re}_{i,k})$ are independent of $\mathcal{H}_k$. Then, by \eqref{t2 8}, we have\vspace{-0.7em}
\begin{align}\label{t2 9.1}
	&\alpha_K^2\E(\|\sum_{i=1}^n\frac{\varphi_{i,k}}{\Delta_K}(\textsf{Dcd}_{\Delta_K}(\textsf{ee}_{i,k+1}^y)+\textsf{qe}_{i,k+1}^y)\notag\\
	\noalign{\vskip -3pt}
	&+\sum_{i=1}^n(\varphi_{i,k}\!+\!\frac{\textsf{Dcd}_{\Delta_K}(\textsf{ee}_{i,k}^\varphi)\!+\!\textsf{qe}_{i,k}^\varphi}{\Delta_K})w_{i,k+1}\notag\\
	\noalign{\vskip -3pt}
	&+\sum_{i=1}^n\textsf{Dcd}_{\Delta_K}(Q(\frac{\textsf{ct}_{i,k}}{\mathfrak{q}_{2,K}})\!-\!\frac{\textsf{ct}_{i,k}}{\mathfrak{q}_{2,K}})\notag\\
	\noalign{\vskip -3pt}
	&+\sum_{i\neq i_0}\textsf{Dcd}_{\Delta_K}(\textsf{re}_{i,k})\!+\!\sum_{i=1}^n\textsf{Dcd}_{\Delta_K}(\textsf{me}_{i,k})\|^2|\mathcal{H}_k)\notag\\
	\noalign{\vskip -3pt}
	\leq&10n\alpha_K^2\sum_{i=1}^n\E(\frac{\|\varphi_{i,k}\|^2(\|\textsf{Dcd}_{\Delta_K}(\textsf{ee}_{i,k+1}^y)\|^2+\|\textsf{qe}_{i,k+1}^y\|^2)}{\Delta_K^2})\notag\\
	\noalign{\vskip -3pt}
	&+15n\alpha_K^2\sum_{i=1}^n\E((\|\varphi_{i,k}\|^2\!+\!\frac{\|\textsf{Dcd}_{\Delta_K}(\textsf{ee}_{i,k}^\varphi)\|^2+\|\textsf{qe}_{i,k}^\varphi\|^2}{\Delta_K^2})\cdot\notag\\
	\noalign{\vskip -3pt}
	&\|w_{i,k+1}\|^2)+5n\alpha_K^2\sum_{i=1}^n\E\|\textsf{Dcd}_{\Delta_K}(Q(\frac{\textsf{ct}_{i,k}}{\mathfrak{q}_{2,K}})\!-\!\frac{\textsf{ct}_{i,k}}{\mathfrak{q}_{2,K}})\|^2\notag\\
	\noalign{\vskip -3pt}
	&+5(n-1)\alpha_K^2\sum_{i\neq i_0}\E\|\textsf{Dcd}_{\Delta_K}(\textsf{re}_{i,k})\|^2\notag\\
	\noalign{\vskip -3pt}
	&+5n\alpha_K^2\sum_{i=1}^n\E\|\textsf{Dcd}_{\Delta_K}(\textsf{me}_{i,k})\|^2.
\end{align}

{\vskip -15pt}\indent By Lemmas \ref{lemma 2}, \ref{lemma 3} and Assumption \ref{asm4}, \eqref{t2 9.1} can be rewritten as\vspace{-1.2em}
\begin{align}\label{t2 9.2}
	&\alpha_K^2\E(\|\sum_{i=1}^n\frac{\varphi_{i,k}}{\Delta_K}(\textsf{Dcd}_{\Delta_K}(\textsf{ee}_{i,k+1}^y)+\textsf{qe}_{i,k+1}^y)\notag\\
	\noalign{\vskip -4pt}
	&+\sum_{i=1}^n(\varphi_{i,k}\!+\!\frac{\textsf{Dcd}_{\Delta_K}(\textsf{ee}_{i,k}^\varphi)\!+\!\textsf{qe}_{i,k}^\varphi}{\Delta_K})w_{i,k+1}\notag\\
	\noalign{\vskip -4pt}
	&+\sum_{i=1}^n\textsf{Dcd}_{\Delta_K}(Q(\frac{\textsf{ct}_{i,k}}{\mathfrak{q}_{2,K}})\!-\!\frac{\textsf{ct}_{i,k}}{\mathfrak{q}_{2,K}})\notag\\
	\noalign{\vskip -4pt}
	&+\sum_{i\neq i_0}\textsf{Dcd}_{\Delta_K}(\textsf{re}_{i,k})\!+\!\sum_{i=1}^n\textsf{Dcd}_{\Delta_K}(\textsf{me}_{i,k})\|^2|\mathcal{H}_k)\notag\\
	=&O(\alpha_K^2).
\end{align}
{\vskip -17pt}\noindent Then, substituting \eqref{t2 9.2} into \eqref{t2 7} proves \eqref{t2 0}.

{\vskip -6pt}{\bf Step 2}. In this step, we prove that\vspace{-1em}
\begin{align*}
	&\E(\|\hat{\theta}_{k+1}-\theta\|^2|\mathcal{F}_k)\notag\\
	\leq&\|(I_{p+q}-\alpha_K\sum_{i=1}^n\varphi_{i,k}\varphi_{i,k}^\top)(\hat{\theta}_k-\theta)\|^2\notag\\
	\noalign{\vskip -5pt}
	&-2\alpha_K\langle(I_{p+q}\!-\!\alpha_K\sum_{i=1}^n\varphi_{i,k}\varphi_{i,k}^\top)(\hat{\theta}_k\!-\!\theta),
\end{align*}
\begin{align}\label{t2 10}
%	\noalign{\vskip -5pt}
	&\frac{1}{\Delta_K}\sum_{i=1}^n\E(\textsf{Dcd}_{\Delta_K}(\textsf{ee}_{i,k}^\varphi)\textsf{Dcd}_{\Delta_K}(\textsf{ee}_{i,k}^\varphi)^\top|\mathcal{F}_k)\hat{\theta}_k\notag\\
	&+\frac{1}{\Delta_K}\sum_{i=1}^n\E(\textsf{qe}_{i,k}^\varphi\textsf{qe}_{i,k}^{\varphi\top}|\mathcal{F}_k)\hat{\theta}_k\rangle\notag\\
	&+O(\alpha_K^2)\|\hat{\theta}_k-\theta\|^2+O(\alpha_K^2).
\end{align}

{\vskip -15pt}\indent Note that $\mathcal{F}_k\subset\mathcal{H}_k$ for any $k=0,\dots,K$. Then, by the tower property of the conditional mathematical expectation (Theorem 6.5.10(a) in \cite{ash2014real}), taking the conditional mathematical expectation on \eqref{t2 0} with respect to $\mathcal{F}_k$ implies\vspace{-1em}
\begin{align}\label{t2 11}
	&\E(\|\hat{\theta}_{k+1}-\theta\|^2|\mathcal{F}_k)\notag\\
	\leq&\|(I_{p+q}-\alpha_K\sum_{i=1}^n\varphi_{i,k}\varphi_{i,k}^\top)(\hat{\theta}_k-\theta)\|^2\notag\\
	&+\alpha_K^2\E(\|\sum_{i=1}^n(\frac{\textsf{Dcd}_{\Delta_K}(\textsf{ee}_{i,k}^\varphi)\!+\!\textsf{qe}_{i,k}^\varphi}{\Delta_K})\varphi_{i,k}^\top(\hat{\theta}_k-\theta)\notag\\
	\noalign{\vskip -5pt}
	&+\sum_{i=1}^n\frac{\varphi_{i,k}}{\Delta_K}(\textsf{Dcd}_{\Delta_K}(\textsf{ee}_{i,k}^\varphi+\breve{\textsf{re}}_{i,k})+\textsf{qe}_{i,k}^\varphi)^\top\hat{\theta}_k\notag\\
	&+\sum_{i=1}^n\frac{\textsf{Dcd}_{\Delta_K}(\textsf{ee}_{i,k}^\varphi)\!+\!\textsf{qe}_{i,k}^\varphi}{\Delta_K}(\textsf{Dcd}_{\Delta_K}(\textsf{ee}_{i,k}^\varphi\!+\!\breve{\textsf{re}}_{i,k})\notag\\
	\noalign{\vskip -5pt}
	&+\textsf{qe}_{i,k}^\varphi)^{\!\!\top}\!\hat{\theta}_k\|^2|\mathcal{F}_k)\!-\!2\alpha_K\langle(I_{p+q}\!-\!\alpha_K\!\sum_{i=1}^n\!\varphi_{i,k}\varphi_{i,k}^\top)(\hat{\theta}_k\!-\!\theta),\notag\\
	\noalign{\vskip -5pt}
	&\E(\sum_{i=1}^n(\frac{\textsf{Dcd}_{\Delta_K}(\textsf{ee}_{i,k}^\varphi)\!+\!\textsf{qe}_{i,k}^\varphi}{\Delta_K})\varphi_{i,k}^\top(\hat{\theta}_k-\theta)\notag\\
	\noalign{\vskip -5pt}
	&+\sum_{i=1}^n\frac{\varphi_{i,k}}{\Delta_K}(\textsf{Dcd}_{\Delta_K}(\textsf{ee}_{i,k}^\varphi+\breve{\textsf{re}}_{i,k})+\textsf{qe}_{i,k}^\varphi)^\top\hat{\theta}_k\notag\\
	\noalign{\vskip -5pt}
	&+\sum_{i=1}^n\frac{\textsf{Dcd}_{\Delta_K}(\textsf{ee}_{i,k}^\varphi)\!+\!\textsf{qe}_{i,k}^\varphi}{\Delta_K}(\textsf{Dcd}_{\Delta_K}(\textsf{ee}_{i,k}^\varphi\!+\!\breve{\textsf{re}}_{i,k})\notag\\
	&+\textsf{qe}_{i,k}^\varphi)^\top\hat{\theta}_k|\mathcal{F}_k)\rangle+O(\alpha_K^2).
\end{align}
{\vskip -15pt}\indent Moreover, note that $\textsf{Dcd}_{\Delta_K}(\textsf{ee}_{i,k}^\varphi),\textsf{qe}_{i,k}^\varphi,\textsf{Dcd}_{\Delta_K}(\breve{\textsf{re}}_{i,k})$ are independent of $\mathcal{F}_k$. Then, by Lemmas \ref{lemma 2} and \ref{lemma 3}, we have\vspace{-1em}
\begin{align*}
	&\alpha_K^2\E(\|\sum_{i=1}^n(\frac{\textsf{Dcd}_{\Delta_K}(\textsf{ee}_{i,k}^\varphi)\!+\!\textsf{qe}_{i,k}^\varphi}{\Delta_K})\varphi_{i,k}^\top(\hat{\theta}_k-\theta)\notag\\
	&+\sum_{i=1}^n\frac{\varphi_{i,k}}{\Delta_K}(\textsf{Dcd}_{\Delta_K}(\textsf{ee}_{i,k}^\varphi+\breve{\textsf{re}}_{i,k})+\textsf{qe}_{i,k}^\varphi)^\top\hat{\theta}_k\notag\\
	&+\sum_{i=1}^n\frac{\textsf{Dcd}_{\Delta_K}(\textsf{ee}_{i,k}^\varphi)\!+\!\textsf{qe}_{i,k}^\varphi}{\Delta_K}(\textsf{Dcd}_{\Delta_K}(\textsf{ee}_{i,k}^\varphi\!+\!\breve{\textsf{re}}_{i,k})\notag\\
	&+\textsf{qe}_{i,k}^\varphi)^\top\hat{\theta}_k\|^2|\mathcal{F}_k)\notag\\
	\leq&6n\alpha_K^2\sum_{i=1}^n\E(\frac{\|\textsf{Dcd}_{\Delta_K}(\textsf{ee}_{i,k}^\varphi)\|^2+\|\textsf{qe}_{i,k}^\varphi\|^2}{\Delta_K^2}\|\varphi_{i,k}\|^2\cdot\notag\\
	&\|\hat{\theta}_k-\theta\|^2|\mathcal{F}_k)+9n\alpha_K^2\sum_{i=1}^n\E((\|\textsf{Dcd}_{\Delta_K}(\textsf{ee}_{i,k}^\varphi)\|^2
\end{align*}
\begin{align}\label{t2 12}
	&+\|\textsf{Dcd}_{\Delta_K}(\breve{\textsf{re}}_{i,k})\|^2+\|\textsf{qe}_{i,k}^\varphi\|^2) \|\varphi_{i,k}\|^2\|\hat{\theta}_k\|^2|\mathcal{F}_k)\notag\\
	&+18n\alpha_K^2\sum_{i=1}^n\E((\|\textsf{Dcd}_{\Delta_K}(\textsf{ee}_{i,k}^\varphi)\|^2+\|\textsf{qe}_{i,k}^\varphi\|^2)\cdot\notag\\
	&(\|\textsf{Dcd}_{\Delta_K}(\textsf{ee}_{i,k}^\varphi)\|^2\!+\!\|\textsf{Dcd}_{\Delta_K}(\breve{\textsf{re}}_{i,k})\|^2\notag\\
	&+\|\textsf{qe}_{i,k}^\varphi\|^2)\|\hat{\theta}_k\|^2|\mathcal{F}_k).
\end{align}
{\vskip -17pt}\noindent Since $\|\hat{\theta}_k\|^2\leq2\|\hat{\theta}_k-\theta\|^2+2\|\theta\|^2$, \eqref{t2 12} can be rewritten as\vspace{-1em}
\begin{align}\label{t2 13}
	&\hspace{-1em}\alpha_K^2\E(\|\sum_{i=1}^n(\frac{\textsf{Dcd}_{\Delta_K}(\textsf{ee}_{i,k}^\varphi)\!+\!\textsf{qe}_{i,k}^\varphi}{\Delta_K})\varphi_{i,k}^\top(\hat{\theta}_k-\theta)\notag\\
	&\hspace{-1em}+\sum_{i=1}^n\frac{\varphi_{i,k}}{\Delta_K}(\textsf{Dcd}_{\Delta_K}(\textsf{ee}_{i,k}^\varphi+\breve{\textsf{re}}_{i,k})+\textsf{qe}_{i,k}^\varphi)^\top\hat{\theta}_k\notag\\
	&\hspace{-1em}+\sum_{i=1}^n\frac{\textsf{Dcd}_{\Delta_K}(\textsf{ee}_{i,k}^\varphi)\!+\!\textsf{qe}_{i,k}^\varphi}{\Delta_K}(\textsf{Dcd}_{\Delta_K}(\textsf{ee}_{i,k}^\varphi\!+\!\breve{\textsf{re}}_{i,k})\notag\\
	&\hspace{-1em}+\textsf{qe}_{i,k}^\varphi)^\top\hat{\theta}_k\|^2|\mathcal{F}_k)\leq O(\alpha_K^2)\|\hat{\theta}_k-\theta\|^2+O(\alpha_K^2).
\end{align}
{\vskip -17pt}\noindent Then, substituting \eqref{t2 13} into \eqref{t2 11} implies\vspace{-1em}
\begin{align}\label{t2 14}
	&\E(\|\hat{\theta}_{k+1}-\theta\|^2|\mathcal{F}_k)\cr
	\hspace{-1em}\leq&\|(I_{p+q}-\alpha_K\sum_{i=1}^n\varphi_{i,k}\varphi_{i,k}^\top)(\hat{\theta}_k-\theta)\|^2\notag\\
	\noalign{\vskip -3pt}
	&-2\alpha_K\langle(I_{p+q}-\alpha_K\sum_{i=1}^n\varphi_{i,k}\varphi_{i,k}^\top)(\hat{\theta}_k-\theta),\notag\\
	\noalign{\vskip -3pt}
	&\E(\sum_{i=1}^n(\frac{\textsf{Dcd}_{\Delta_K}(\textsf{ee}_{i,k}^\varphi)\!+\!\textsf{qe}_{i,k}^\varphi}{\Delta_K})\varphi_{i,k}^\top(\hat{\theta}_k-\theta)\notag\\
	\noalign{\vskip -3pt}
	&+\sum_{i=1}^n\frac{\varphi_{i,k}}{\Delta_K}(\textsf{Dcd}_{\Delta_K}(\textsf{ee}_{i,k}^\varphi+\breve{\textsf{re}}_{i,k})+\textsf{qe}_{i,k}^\varphi)^\top\hat{\theta}_k\notag\\
	\noalign{\vskip -3pt}
	&+\sum_{i=1}^n\frac{\textsf{Dcd}_{\Delta_K}(\textsf{ee}_{i,k}^\varphi)\!+\!\textsf{qe}_{i,k}^\varphi}{\Delta_K}(\textsf{Dcd}_{\Delta_K}(\textsf{ee}_{i,k}^\varphi\!+\!\breve{\textsf{re}}_{i,k})\notag\\
	\noalign{\vskip -3pt}
	&+\textsf{qe}_{i,k}^\varphi)^\top\hat{\theta}_k|\mathcal{F}_k)\rangle\!+\!O(\alpha_K^2)\|\hat{\theta}_k\!-\!\theta\|^2\!+\!O(\alpha_K^2).
	\raisetag{13pt}
\end{align}
Note that $\textsf{Dcd}_{\Delta_K}(\textsf{ee}_{i,k}^\varphi),\textsf{qe}_{i,k}^\varphi,\textsf{Dcd}_{\Delta_K}(\breve{\textsf{re}}_{i,k})$ are independent of $\mathcal{F}_k$. Then, we have
\begin{align}\label{t2 15}
	&\E(\textsf{Dcd}_{\Delta_K}(\textsf{ee}_{i,k}^\varphi)|\mathcal{F}_k)=0,\cr
	&\E(\textsf{Dcd}_{\Delta_K}(\breve{\textsf{re}}_{i,k})|\mathcal{F}_k)=0,\cr
	&\E(\textsf{qe}_{i,k}^\varphi|\mathcal{F}_k)=0.
\end{align}
Moreover, since $\textsf{Dcd}_{\Delta_K}(\textsf{ee}_{i,k}^\varphi),\textsf{qe}_{i,k}^\varphi,\textsf{Dcd}_{\Delta_K}(\breve{\textsf{re}}_{i,k})$ are mutually independent, by \eqref{t2 15} we have
\begin{align}\label{t2 16}
	&\E(\textsf{Dcd}_{\Delta_K}(\textsf{ee}_{i,k}^\varphi)\textsf{Dcd}_{\Delta_K}(\breve{\textsf{re}}_{i,k})^\top|\mathcal{F}_k)=0,\cr
	&\E(\textsf{qe}_{i,k}^\varphi\textsf{Dcd}_{\Delta_K}(\breve{\textsf{re}}_{i,k})^\top|\mathcal{F}_k)=0,\cr
	&\E(\textsf{Dcd}_{\Delta_K}(\textsf{ee}_{i,k}^\varphi)\textsf{qe}_{i,k}^{\varphi\top}|\mathcal{F}_k)=0,\cr
	&\E(\textsf{qe}_{i,k}^\varphi\textsf{Dcd}_{\Delta_K}(\textsf{ee}_{i,k}^\varphi)^\top|\mathcal{F}_k)=0.
\end{align}
Then, by \eqref{t2 16} we have
\begin{align}\label{t2 17}
	&-2\alpha_K\langle(I_{p+q}-\alpha_K\sum_{i=1}^n\varphi_{i,k}\varphi_{i,k}^\top)(\hat{\theta}_k-\theta),\notag\\
	&\E(\sum_{i=1}^n(\frac{\textsf{Dcd}_{\Delta_K}(\textsf{ee}_{i,k}^\varphi)\!+\!\textsf{qe}_{i,k}^\varphi}{\Delta_K})\varphi_{i,k}^\top(\hat{\theta}_k-\theta)\notag\\
	&+\sum_{i=1}^n\frac{\varphi_{i,k}}{\Delta_K}(\textsf{Dcd}_{\Delta_K}(\textsf{ee}_{i,k}^\varphi+\breve{\textsf{re}}_{i,k})+\textsf{qe}_{i,k}^\varphi)^\top\hat{\theta}_k\notag\\
	&+\sum_{i=1}^n\frac{\textsf{Dcd}_{\Delta_K}(\textsf{ee}_{i,k}^\varphi)\!+\!\textsf{qe}_{i,k}^\varphi}{\Delta_K}(\textsf{Dcd}_{\Delta_K}(\textsf{ee}_{i,k}^\varphi\!+\!\breve{\textsf{re}}_{i,k})\notag\\
	&+\textsf{qe}_{i,k}^\varphi)^\top\hat{\theta}_k|\mathcal{F}_k)\rangle\notag\\
	=&-2\alpha_K\langle(I_{p+q}\!-\!\alpha_K\sum_{i=1}^n\varphi_{i,k}\varphi_{i,k}^\top)(\hat{\theta}_k\!-\!\theta),\notag\\
	&\frac{1}{\Delta_K}\sum_{i=1}^n\E(\textsf{Dcd}_{\Delta_K}(\textsf{ee}_{i,k}^\varphi)\textsf{Dcd}_{\Delta_K}(\textsf{ee}_{i,k}^\varphi)^\top|\mathcal{F}_k)\hat{\theta}_k\notag\\
	&+\frac{1}{\Delta_K}\sum_{i=1}^n\E(\textsf{qe}_{i,k}^\varphi\textsf{qe}_{i,k}^{\varphi\top}|\mathcal{F}_k)\hat{\theta}_k\rangle.
\end{align}
Thus, substituting \eqref{t2 17} into \eqref{t2 14} proves \eqref{t2 10}.

{\bf Step 3}. In this step, we prove that $\E(\textsf{Dcd}_{\Delta_K}(\textsf{ee}_{i,k}^\varphi)\cdot$ $\textsf{Dcd}_{\Delta_K}(\textsf{ee}_{i,k}^\varphi)^\top|\mathcal{F}_k)=0$, $\E(\textsf{qe}_{i,k}^\varphi\textsf{qe}_{i,k}^{\varphi\top}|\mathcal{F}_k)\leq\frac{1}{\Delta_K^2}I_{p+q}$ for any $k=0,\dots,K$.

Let $(\textsf{ee}_{i,k}^{(1)},\dots,\textsf{ee}_{i,k}^{(N)})$, $(\textsf{e}_{i,k}^{(1)},\dots,\textsf{e}_{i,k}^{(N)}),(\textsf{e}_{i,k,1}^{(1)},\dots,\textsf{e}_{i,k,1}^{(N)})$, $(\textsf{e}_{i,k,2}^{(1)}, \dots,\textsf{e}_{i,k,2}^{(N)})$, $(\textsf{v}_{i,k}^{(1)},\dots$, $\textsf{v}_{i,k}^{(N)})$, $(\textsf{s}_{i,k}^{(1)},\dots,\textsf{s}_{i,k}^{(N)})$ be the coefficients of $\textsf{ee}_{i,k}^\varphi$, $\textsf{e}_{i,k}$, $\textsf{e}_{i,k,1}$, $\textsf{e}_{i,k,2}$, $\textsf{v}_{i,k}$, $\textsf{s}_{i,k}$, respectively. Then, by Lemma~\ref{lemma 3}, it can be seen that for any $l=1,\dots,N$,
\begin{align}\label{t2 19}
	&\textsf{ee}_{i,k}^{(l)}\notag\\
	=&\sum_{m=1}^l\! \textsf{v}_{i,k}^{(m)}\textsf{e}_{i,k}^{(l-m+1)}\!-\!\!\!\!\sum_{m=l+1}^N \!\!\textsf{v}_{i,k}^{(m)}\textsf{e}_{i,k}^{(N+l-m+1)}\!+\!\textsf{e}_{i,k,1}^{(l)}\notag\\
	&+\sum_{m=1}^l \!\textsf{s}_{i,k}^{(m)}\textsf{e}_{i,k,2}^{(l-m+1)}\!-\!\!\!\!\sum_{m=l+1}^N \!\!\textsf{s}_{i,k}^{(m)}\textsf{e}_{i,k,2}^{(N+l-m+1)}\bmod\!\mathfrak{q}_K\notag\\
	=&\sum_{m=1}^l \textsf{v}_{i,k}^{(m)}\textsf{e}_{i,k}^{(l-m+1)}\!-\!\!\!\!\sum_{m=l+1}^N \textsf{v}_{i,k}^{(m)}\textsf{e}_{i,k}^{(N+l-m+1)}\!+\!\textsf{e}_{i,k,1}^{(l)}\notag\\
	&+\sum_{m=1}^l \textsf{s}_{i,k}^{(m)}\textsf{e}_{i,k,2}^{(l-m+1)}\!-\!\!\!\!\sum_{m=l+1}^N \textsf{s}_{i,k}^{(m)}\textsf{e}_{i,k,2}^{(N+l-m+1)}.
\end{align}
Note that $(\textsf{ee}_{i,k}^{(1)},\dots,\textsf{ee}_{i,k}^{(N)})$, $(\textsf{e}_{i,k}^{(1)},\dots,\textsf{e}_{i,k}^{(N)})$, $(\textsf{e}_{i,k,1}^{(1)},\dots$, $\textsf{e}_{i,k,1}^{(N)})$, $(\textsf{e}_{i,k,2}^{(1)}, \dots,\textsf{e}_{i,k,2}^{(N)})$, $(\textsf{v}_{i,k}^{(1)},\dots$, $\textsf{v}_{i,k}^{(N)})$, $(\textsf{s}_{i,k}^{(1)},\dots,\textsf{s}_{i,k}^{(N)})$ are mutually independent and independent of $\mathcal{F}_k$. Then, taking the conditional mathematical expectation of $\textsf{ee}_{i,k}^{(l)}$ with respect to $\mathcal{F}_k$ gives
\begin{align}\label{t2 20}
	&\E(\textsf{ee}_{i,k}^{(l)}|\mathcal{F}_k)\cr
	=&\sum_{m=1}^l \E \textsf{v}_{i,k}^{(m)} \E \textsf{e}_{i,k}^{(l-m+1)}\!-\!\!\!\!\sum_{m=l+1}^N \E \textsf{v}_{i,k}^{(m)} \E \textsf{e}_{i,k}^{(N+l-m+1)}\!+\!\E \textsf{e}_{i,k,1}^{(l)}\cr
	&+\sum_{m=1}^l \E \textsf{s}_{i,k}^{(m)} \E \textsf{e}_{i,k,2}^{(l-m+1)}\!-\!\!\!\!\sum_{m=l+1}^N \E \textsf{s}_{i,k}^{(m)} \E \textsf{e}_{i,k,2}^{(N+l-m+1)}\cr
	=&0.
\end{align}
By \eqref{t2 20}, it can be seen that for any $l_1=1,\dots,N$, $l_2=1,\dots,N$, $l_1\neq l_2$, 
\begin{align}\label{t2 22}
	\E(\textsf{ee}_{i,k}^{(l_1)}\textsf{ee}_{i,k}^{(l_2)}|\mathcal{F}_k)=\E(\textsf{ee}_{i,k}^{(l_1)}\textsf{ee}_{i,k}^{(l_2)})=0.
\end{align}
Note that by \cite[Lemma 2.1]{banaszczyk1995inequalities}, $\E \textsf{e}_{i,k}^{(l)}\leq \sigma^2$ holds for any $i=1,\dots,n$, $k=0,\dots,K$, $l=1,\dots,N$. Then, it can be seen that
\begin{align}\label{t2 21}
	&\E((\textsf{ee}_{i,k}^{(l)})^2|\mathcal{F}_k)\cr
	=&\E (\textsf{ee}_{i,k}^{(l)})^2\cr
	=&\E(\sum_{m=1}^l  \textsf{v}_{i,k}^{(m)} \textsf{e}_{i,k}^{(l-m+1)}\!-\!\!\!\!\sum_{m=l+1}^N \textsf{v}_{i,k}^{(m)} \textsf{e}_{i,k}^{(N+l-m+1)})^2\!+\!\E (\textsf{e}_{i,k,1}^{(l)})^2\cr
	&+\E(\sum_{m=1}^l \textsf{s}_{i,k}^{(m)} \textsf{e}_{i,k,2}^{(l-m+1)}\!-\!\!\!\!\sum_{m=l+1}^N \textsf{s}_{i,k}^{(m)} \textsf{e}_{i,k,2}^{(N+l-m+1)})^2\cr
	\leq&\frac{(N+h+2)\sigma^2}{2}.
\end{align}
{\vskip -7pt}\noindent Note that $\{\textsf{e}_{i,k}^{(l)}|i$$=$$1,\dots,n,k$$=$$0,\dots,K,l$$=$$1,\dots,N\}$ are identically distributed. Then, by \eqref{t2 21}, there exists $\sigma_e>0$ such that $\E(\textsf{e}_{i,k}^{(l)})^2=\sigma_e^2$. Let the matrix $V$ be defined as \vspace{-0.6em}
\begin{align*}
	V=\left[\begin{matrix}
		1&\zeta&\zeta^2&\dots&\zeta^{N-1}\\
		1&\zeta^3&\zeta^6&\dots&\zeta^{3(N-1)}\\
		\vdots&\vdots&\vdots&\dots&\vdots\\
		1&\zeta^{N-1}&\zeta^{2(N-1)}&\dots&\zeta^{(N-1)^2}
	\end{matrix}\right].
\end{align*}
{\vskip -15pt}\noindent Then, we have \vspace{-0.6em}
\begin{align}\label{t2 23}
	&\E(\textsf{Dcd}_{\Delta_K}(\textsf{ee}_{i,k}^\varphi)\textsf{Dcd}_{\Delta_K}(\textsf{ee}_{i,k}^\varphi)^\top|\mathcal{F}_k)\notag\\
	=&\frac{1}{\Delta_K^2}V\cdot\left[\begin{matrix}
		\E((\textsf{e}_{i,k}^{(1)})^2|\mathcal{F}_k)&\dots&\E(\textsf{e}_{i,k}^{(1)}\textsf{e}_{i,k}^{(N)}|\mathcal{F}_k)\\
		\vdots&\dots&\vdots\\
		\E(\textsf{e}_{i,k}^{(1)}\textsf{e}_{i,k}^{(N)}|\mathcal{F}_k)&\dots&\E((\textsf{e}_{i,k}^{(N)})^2|\mathcal{F}_k)\\
	\end{matrix}\right]\cdot V^\top\notag\\
	=&\frac{\sigma_e^2}{\Delta_K^2}V\cdot V^\top.
\end{align}
{\vskip -13pt}\noindent Denote the entry in the $l_1$-th row and $l_2$-th column of $V\cdot V^\top$ as $(V\cdot V^\top)_{l_1,l_2}$ for any $l_1=1,\dots,\frac{N}{2}$, $l_2=1,\dots,\frac{N}{2}$. Then, we have \vspace{-0.8em}
\begin{align*}
	(V\cdot V^\top)_{l_1,l_2}=\sum_{m=0}^{N-1}\zeta^{m(2l_1-1)+m(2l_2-1)}=\frac{1-\zeta^{2N(l_1+l_2-1)}}{1-\zeta^{2(l_1+l_2-1)}}.
\end{align*}
{\vskip -16pt}\noindent Since $\zeta=\exp(\frac{2\pi \iu}{2N})$, we have $\zeta^{2N(l_1+l_2-1)}$$=$$(\zeta^{2N})^{l_1+l_2-1}$$=$$1$. Then, $(V\cdot V^\top)_{l_1,l_2}=0$, and thus, \eqref{t2 23} can be rewritten as \vspace{-0.6em}
\begin{align}\label{t2 24}
	\E(\textsf{Dcd}_{\Delta_K}(\textsf{ee}_{i,k}^\varphi)\textsf{Dcd}_{\Delta_K}(\textsf{ee}_{i,k}^\varphi)^\top|\mathcal{F}_k)=0.
\end{align}

{\vskip -18pt}\indent Let $\textsf{qe}_{i,k}^{\varphi(l)}$ be the $l$-th component of $\textsf{qe}_{i,k}^\varphi$ for any $l=1,\dots,p+q$. Moreover, note that $\textsf{qe}_{i,k}^\varphi$ is independent of $\mathcal{F}_k$. Then, by \eqref{pq}, we have
\begin{align}
	&\E(\textsf{qe}_{i,k}^{\varphi(l)}|\mathcal{F}_k)=\E\textsf{qe}_{i,k}^{\varphi(l)}=0,\label{t2 25a}\\
	&\E((\textsf{qe}_{i,k}^{\varphi(l)})^2|\mathcal{F}_k)=\E(\textsf{qe}_{i,k}^{\varphi(l)})^2\leq\frac{1}{\Delta_K^2}.\label{t2 25b}
\end{align}
{\vskip -16pt}\noindent Thus, for any $l_1=1,\dots,p+q$, $l_2=1,\dots,p+q$, $l_1\neq l_2$, by \eqref{t2 25a} we have\vspace{-0.4em}
\begin{align}\label{t2 26}
	\E(\textsf{qe}_{i,k}^{\varphi(l_1)}\textsf{qe}_{i,k}^{\varphi(l_2)}|\mathcal{F}_k)=\E\textsf{qe}_{i,k}^{\varphi(l_1)}\textsf{qe}_{i,k}^{\varphi(l_2)}=0.
\end{align}
By \eqref{t2 25b} and \eqref{t2 26}, we have\vspace{-1em}
\begin{align}\label{t2 27}
	&\E(\textsf{qe}_{i,k}^\varphi\textsf{qe}_{i,k}^{\varphi\top}|\mathcal{F}_k)\cr
	=&\left[\begin{matrix}
		\E(\textsf{qe}_{i,k}^{\varphi(1)})^2&\dots&\E(\textsf{qe}_{i,k}^{\varphi(1)}\textsf{qe}_{i,k}^{\varphi(p+q)})\\
		\vdots&\dots&\vdots\\
		\E(\textsf{qe}_{i,k}^{\varphi(1)}\textsf{qe}_{i,k}^{\varphi(p+q)})&\dots&\E(\textsf{qe}_{i,k}^{\varphi(p+q)})^2
	\end{matrix}\right]\cr
	\leq&\frac{1}{\Delta_K^2}I_{p+q}.
\end{align}

{\vskip -15pt}\indent{\bf Step 4}. In this step, we prove that \vspace{-0.7em}
\begin{align}\label{t2 28}
	&\E(\|\hat{\theta}_{k+1}-\theta\|^2|\mathcal{F}_k)\notag\\
	\leq&\|(I_{p+q}-\alpha_K\sum_{i=1}^n\varphi_{i,k}\varphi_{i,k}^\top)(\hat{\theta}_k-\theta)\|^2\notag\\
	&+O(\alpha_K^2)\|\hat{\theta}_k-\theta\|^2+O(\alpha_K^2).
\end{align}

{\vskip -15pt}\indent Note that $\E(\textsf{Dcd}_{\Delta_K}(\textsf{ee}_{i,k}^\varphi)\textsf{Dcd}_{\Delta_K}(\textsf{ee}_{i,k}^\varphi)^\top|\mathcal{F}_k)=0$, $\E(\textsf{qe}_{i,k}^\varphi\textsf{qe}_{i,k}^{\varphi\top}|\mathcal{F}_k)\leq\frac{1}{\Delta_K^2}I_{p+q}$. Then, by Cauchy-Schwarz inequality (Example 4(b) in \cite{zorich2015analysis}), we have\vspace{-1em}
\begin{align*}
	&-2\alpha_K\langle(I_{p+q}\!-\!\alpha_K\sum_{i=1}^n\varphi_{i,k}\varphi_{i,k}^\top)(\hat{\theta}_k\!-\!\theta),\notag\\
	\noalign{\vskip -3pt}
	&\frac{1}{\Delta_K}\sum_{i=1}^n\E(\textsf{Dcd}_{\Delta_K}(\textsf{ee}_{i,k}^\varphi)\textsf{Dcd}_{\Delta_K}(\textsf{ee}_{i,k}^\varphi)^\top|\mathcal{F}_k)\hat{\theta}_k\notag\\
	\noalign{\vskip -3pt}
	&+\frac{1}{\Delta_K}\sum_{i=1}^n\E(\textsf{qe}_{i,k}^\varphi\textsf{qe}_{i,k}^{\varphi\top}|\mathcal{F}_k)\hat{\theta}_k\rangle\notag\\
	\noalign{\vskip -3pt}
	=&-2\alpha_K\langle(I_{p+q}\!-\!\alpha_K\sum_{i=1}^n\varphi_{i,k}\varphi_{i,k}^\top)(\hat{\theta}_k\!-\!\theta),
\end{align*}
\begin{align}\label{t2 30}
%	\noalign{\vskip -3pt}
	&\frac{1}{\Delta_K}\sum_{i=1}^n\E(\textsf{qe}_{i,k}^\varphi\textsf{qe}_{i,k}^{\varphi\top}|\mathcal{F}_k)\hat{\theta}_k\rangle\notag\\
	\noalign{\vskip -3pt}
	\leq&\frac{2\alpha_K}{\Delta_K}\|(I_{p+q}\!-\!\alpha_K\sum_{i=1}^n\varphi_{i,k}\varphi_{i,k}^\top)(\hat{\theta}_k\!-\!\theta)\|\cdot\notag\\
	\noalign{\vskip -3pt}
	&\|\E(\sum_{i=1}^n\textsf{qe}_{i,k}^\varphi\textsf{qe}_{i,k}^{\varphi\top}|\mathcal{F}_k)\hat{\theta}_k\|.
\end{align}
{\vskip -12pt}\noindent By $c_1$$\leq$$\frac{1}{ntG_1^2}$ in Assumption \ref{asm6} and Lemma \ref{lemma 2}, we have $\|I_{p+q}$$-$$\alpha_K$$\sum_{i=1}^n\varphi_{i,k}$$\varphi_{i,k}^\top\|$$\leq$$1$. Then, \eqref{t2 30} can be rewritten as\vspace{-1em}
\begin{align}\label{t2 31}
	&-2\alpha_K\langle(I_{p+q}\!-\!\alpha_K\sum_{i=1}^n\varphi_{i,k}\varphi_{i,k}^\top)(\hat{\theta}_k\!-\!\theta),\notag\\
	&\frac{1}{\Delta_K}\sum_{i=1}^n\E(\textsf{Dcd}_{\Delta_K}(\textsf{ee}_{i,k}^\varphi)\textsf{Dcd}_{\Delta_K}(\textsf{ee}_{i,k}^\varphi)^\top|\mathcal{F}_k)\hat{\theta}_k\notag\\
	&+\frac{1}{\Delta_K}\sum_{i=1}^n\E(\textsf{qe}_{i,k}^\varphi\textsf{qe}_{i,k}^{\varphi\top}|\mathcal{F}_k)\hat{\theta}_k\rangle\notag\\
	\leq&\frac{3n\alpha_K}{\Delta_K^3}\|\hat{\theta}_k\!-\!\theta\|^2+\frac{n\alpha_K}{\Delta_K^3}\|\theta\|^2.
\end{align}
{\vskip -17pt}\noindent By $3p_3\geq p_1$ in Assumption \ref{asm6}, we have $\frac{\alpha_K}{\Delta_K^3}=O(\alpha_K^2)$. Thus, \eqref{t2 31} can be rewritten as\vspace{-1em}
\begin{align}\label{t2 32}
	&-2\alpha_K\langle(I_{p+q}\!-\!\alpha_K\sum_{i=1}^n\varphi_{i,k}\varphi_{i,k}^\top)(\hat{\theta}_k\!-\!\theta),\notag\\
	\noalign{\vskip -3pt}
	&\frac{1}{\Delta_K}\sum_{i=1}^n\E(\textsf{Dcd}_{\Delta_K}(\textsf{ee}_{i,k}^\varphi)\textsf{Dcd}_{\Delta_K}(\textsf{ee}_{i,k}^\varphi)^\top|\mathcal{F}_k)\hat{\theta}_k\notag\\
	&+\frac{1}{\Delta_K}\sum_{i=1}^n\E(\textsf{qe}_{i,k}^\varphi\textsf{qe}_{i,k}^{\varphi\top}|\mathcal{F}_k)\hat{\theta}_k\rangle\notag\\
	=&O(\alpha_K^2)\|\hat{\theta}_k\!-\!\theta\|^2+O(\alpha_K^2).
\end{align}
Then, substituting \eqref{t2 32} into \eqref{t2 10} proves \eqref{t2 28}.

{\bf Step 5}. In this step, we prove that $\E\|\hat{\theta}_{K+1}-\theta\|^2=O(\frac{1}{(K+1)^{p_1}})$ for any $K=t-1,t,\dots$. \vspace{-0.7em}

By \eqref{t2 28}, we have\vspace{-0.7em}
\begin{align}\label{t2 33}
	&\E(\|\hat{\theta}_{k+1}-\theta\|^2|\mathcal{F}_k)\cr
	\leq&(1+O(\alpha_K^2))\|\hat{\theta}_k-\theta\|^2-2\alpha_K\sum_{i=1}^n\|\varphi_{i,k}^\top(\hat{\theta}_k-\theta)\|^2\cr
	\noalign{\vskip -4pt}
	&+\alpha_K^2\|\sum_{i=1}^n\varphi_{i,k}\varphi_{i,k}^\top(\hat{\theta}_k-\theta)\|^2+O(\alpha_K^2).
\end{align}
By \eqref{t2 8} and Lemma \ref{lemma 2}, it can be seen that\vspace{-0.7em}
\begin{align}\label{t2 35}
	\alpha_K^2\|\!\sum_{i=1}^n\varphi_{i,k}\varphi_{i,k}^\top(\hat{\theta}_k\!-\!\theta)\|^2\leq&n^2G_1^4\alpha_K^2\|\hat{\theta}_k-\theta\|^2\cr
	=&O(\alpha_K^2)\|\hat{\theta}_k-\theta\|^2.
\end{align}
Then, substituting \eqref{t2 35} into \eqref{t2 33} implies\vspace{-0.7em}
\begin{align}\label{t2 36}
	&\E(\|\hat{\theta}_{k+1}-\theta\|^2|\mathcal{F}_k)\cr
	\leq&(1+O(\alpha_K^2))\|\hat{\theta}_k-\theta\|^2-2\alpha_K\sum_{i=1}^n\|\varphi_{i,k}^\top(\hat{\theta}_k-\theta)\|^2\cr
	&+O(\alpha_K^2).
\end{align}
By the law of total expectation (Theorem 6.5.4 in \cite{ash2014real}), taking the mathematical expectation on both sides of \eqref{t2 36} leads to\vspace{-0.7em}
\begin{align}\label{t2 37}
	&\E\|\hat{\theta}_{k+1}-\theta\|^2\cr
	\leq&(1+O(\alpha_K^2))\E\|\hat{\theta}_k-\theta\|^2-2\alpha_K\sum_{i=1}^n\E\|\varphi_{i,k}^\top(\hat{\theta}_k-\theta)\|^2\cr
	&+O(\alpha_K^2).
\end{align}
Recursively computing \eqref{t2 37} gives
\begin{align}\label{t2 37.1}
	&\E\|\hat{\theta}_{K+1}-\theta\|^2\cr
	\hspace{-1em}\leq&(1+O(\alpha_K^2))\E\|\hat{\theta}_{K-t+1}-\theta\|^2\cr
	&-2\alpha_K\!\!\sum_{l=K-t+1}^{K}\sum_{i=1}^n\E\|\varphi_{i,l}^\top(\hat{\theta}_l\!-\!\theta)\|^2\!+\!O(\alpha_K^2).
\end{align}

By Lemma \ref{lemma 3}, we have $\|\hat{\theta}_{k+1}-\hat{\theta}_k\|=O(\alpha_K)$. Then, for any $l=K-t+2,\dots,K$, we have
\begin{align}\label{t2 39}
	\|\hat{\theta}_l-\hat{\theta}_{K-t+1}\|\leq&\sum_{m=K-t+1}^{l-1}\|\hat{\theta}_{m+1}-\hat{\theta}_m\|\cr
	=&O(\alpha_K).
\end{align}
{\vskip -13pt}\indent Note that 
\begin{align}\label{t2 38}
	&\sum_{l=K\!-\!t\!+\!1}^{K}\sum_{i=1}^n\|\varphi_{i,l}^\top(\hat{\theta}_l\!-\!\theta)\|^2\cr
	\hspace{-1em}=&\sum_{l=K\!-\!t\!+\!1}^{K}\sum_{i=1}^n(\hat{\theta}_{K\!-\!t\!+\!1}\!-\!\theta)^\top\varphi_{i,l}\varphi_{i,l}^\top(\hat{\theta}_{K\!-\!t\!+\!1}\!-\!\theta)\cr
	&+2\!\!\sum_{l=K\!-\!t\!+\!1}^{K}\sum_{i=1}^n(\hat{\theta}_l\!-\!\hat{\theta}_{K\!-\!t\!+\!1})^\top\varphi_{i,l}\varphi_{i,l}^\top(\hat{\theta}_{K\!-\!t\!+\!1}\!-\!\theta)\cr
	&+\!\!\sum_{l=K\!-\!t\!+\!1}^{K}\sum_{i=1}^n(\hat{\theta}_l\!-\!\hat{\theta}_{K\!-\!t\!+\!1})^\top\!\varphi_{i,l}\varphi_{i,l}^\top(\hat{\theta}_l\!-\!\hat{\theta}_{K\!-\!t\!+\!1}).
\end{align}
{\vskip -13pt}\noindent Then, by \eqref{t2 39}, \eqref{t2 38}, and Lemma \ref{lemma 2}, it can be seen that
\begin{align}\label{t2 40}
	&-2\alpha_K\sum_{l=K-t+1}^{K}\sum_{i=1}^n\E\|\varphi_{i,l}^\top(\hat{\theta}_l-\theta)\|^2\cr
	=&-2\alpha_K(\hat{\theta}_{K-t+1}\!-\!\theta)^\top(\sum_{l=K-t+1}^{K}\sum_{i=1}^n\varphi_{i,l}\varphi_{i,l}^\top)(\hat{\theta}_{K-t+1}\!-\!\theta)\cr
	&+O(\alpha_K^2)\|\hat{\theta}_{K-t+1}-\theta\|+O(\alpha_K^3).
\end{align}
{\vskip -13pt}\noindent By Cauchy-Schwarz inequality, $\|\hat{\theta}_{K\!-\!t\!+\!1}$$-$$\theta\|$$\leq$$\frac{\|\hat{\theta}_{K\!-\!t\!+\!1}-\theta\|^2+1}{2}$. Then, \eqref{t2 40} can be rewritten as
\begin{align*}
	&-2\alpha_K\sum_{l=K-t+1}^{K}\sum_{i=1}^n\E\|\varphi_{i,l}^\top(\hat{\theta}_l-\theta)\|^2\cr
	\leq&-2\alpha_K(\hat{\theta}_{K-t+1}\!-\!\theta)^\top(\sum_{l=K-t+1}^{K}\sum_{i=1}^n\varphi_{i,l}\varphi_{i,l}^\top)(\hat{\theta}_{K-t+1}\!-\!\theta)\cr
	&+O(\alpha_K^2)\|\hat{\theta}_{K-t+1}-\theta\|^2+O(\alpha_K^2).
\end{align*}
{\vskip -13pt}\noindent Thus, by Assumption \ref{asm5}, we have
\begin{align}\label{t2 41}
	&-2\alpha_K\sum_{l=K-t+1}^{K}\sum_{i=1}^n\E\|\varphi_{i,l}^\top(\hat{\theta}_l-\theta)\|^2\cr
	\leq&-2t\varpi\alpha_K\|\hat{\theta}_{K-t+1}\!-\!\theta\|^2\cr
	&+O(\alpha_K^2)\|\hat{\theta}_{K-t+1}-\theta\|^2+O(\alpha_K^2).
\end{align}
{\vskip -13pt}\noindent Substituting \eqref{t2 41} into \eqref{t2 37.1} implies
\begin{align}\label{t2 42}
	\E\|\hat{\theta}_{K+1}-\theta\|^2\leq&(1-2t\varpi\alpha_K+O(\alpha_K^2))\E\|\hat{\theta}_{K-t+1}-\theta\|^2\notag\\
	&+O(\alpha_K^2).
\end{align}
{\vskip -13pt}\noindent Recursively computing \eqref{t2 42} leads to\vspace{-0.7em}
\begin{align}\label{t2 43}
	&\E\|\hat{\theta}_{K+1}-\theta\|^2\notag\\
	\leq&(1-2t\varpi\alpha_K+O(\alpha_K^2))^{\lfloor\frac{K+1}{t}\rfloor}\E\|\hat{\theta}_{K+1-t\lfloor\frac{K+1}{t}\rfloor}-\theta\|^2\notag\\
	&+O(\alpha_K^2)\sum_{l=0}^{\lfloor\frac{K+1}{t}\rfloor-1}(1-2t\varpi\alpha_K+O(\alpha_K^2))^l.
\end{align}

{\vskip -17pt}\indent Note that
\begin{align*}
	&(1-2t\varpi\alpha_K+O(\alpha_K^2))^{\lfloor\frac{K+1}{t}\rfloor}\cr
	=&\exp(\lfloor\frac{K+1}{t}\rfloor\ln(1-2t\varpi\alpha_K+O(\alpha_K^2))).
\end{align*}
{\vskip -7pt}\noindent Then, by $c_1<\frac{1}{2t\varpi}$ in Assumption \ref{asm6} and $\ln(1+x)\leq x$ for any $x>-1$, we have\vspace{-0.7em}
\begin{align}\label{t2 44}
	&(1-2t\varpi\alpha_K+O(\alpha_K^2))^{\lfloor\frac{K+1}{t}\rfloor}\cr
	\leq&\exp(-\frac{2t\varpi(K\!+\!1)\alpha_K}{t\!+\!1}\!+\!O(\alpha_K^2)\cdot\lfloor\frac{K\!+\!1}{t}\rfloor).
\end{align}
{\vskip -7pt}\noindent Thus, by $\alpha_K=\frac{c_1}{(K+1)^{p_1}}$ and $\frac{1}{2}<p_1<1$ in Assumption \ref{asm6}, \eqref{t2 44} can be rewritten as\vspace{-0.7em}
\begin{align}\label{t2 45}
	&(1-2t\varpi\alpha_K+O(\alpha_K^2))^{\lfloor\frac{K+1}{t}\rfloor}\cr
	=&O(\exp(-\frac{2t\varpi c_1}{t+1}(K+1)^{1-p_1})).
\end{align}
{\vskip -7pt}\noindent Moreover, by \eqref{t2 45}, we have\vspace{-0.7em}
\begin{align}\label{t2 46}
	&\sum_{l=0}^{\lfloor\frac{K+1}{t}\rfloor-1}(1-2t\varpi\alpha_K+O(\alpha_K^2))^l\cr
	=&\frac{(1-2t\varpi\alpha_K+O(\alpha_K^2))^{\lfloor\frac{K+1}{t}\rfloor}}{1-(1-2t\varpi\alpha_K+O(\alpha_K^2))}\cr
	=&O(\frac{1}{\alpha_K}).
\end{align}
{\vskip -17pt}\noindent Then, substituting \eqref{t2 45} and \eqref{t2 46} into \eqref{t2 43} implies\vspace{-0.7em}
\begin{align}\label{t2 47}
	&\E\|\hat{\theta}_{K+1}-\theta\|^2\notag\\
	\leq&O(\exp(-\frac{2t\varpi c_1}{t+1}(K+1)^{1-p_1}))+O(\alpha_K^2\cdot\frac{1}{\alpha_K})\notag\\
	=&O(\frac{1}{(K+1)^{p_1}}).
\end{align}

{\vskip -17pt}\indent {\bf Step 6}. In this step, we prove that $\E\|\hat{\theta}_{K+1}-\theta\|^2=O(\frac{1}{(K+1)^{p_1}})$ for any $K=0,1,\dots$.

By \eqref{t2 47}, there exists $G_5>0$ such that $\E\|\hat{\theta}_{K+1}-\theta\|^2\leq\frac{G_5}{(K+1)^{p_1}}$ for any $K=t-1,t,\dots$. Let $G_6$$=$$\max\{\|\hat{\theta}_0-\theta\|^2$, $2^{p_1}\E\|\hat{\theta}_1-\theta\|^2,\dots,t^{p_1}\E\|\hat{\theta}_{t-1}-\theta\|^2,G_5\}$. Then, for any $K=0,1,\dots$, we have
\begin{align}\label{t2 48}
	\E\|\hat{\theta}_{K+1}-\theta\|^2\leq\frac{G_6}{(K+1)^{p_1}}=O(\frac{1}{(K+1)^{p_1}}).
\end{align}
Thus, this theorem is proved. $\hfill\blacksquare$

\bibliographystyle{elsarticle-harv}

\end{document}